# Spatiotemporal Emotional Synchrony in Dyadic Interactions: *The Role of Speech Conditions in Facial and Vocal Affective Alignment*

Von Ralph Dane Herbuela
*International Research Center for Neurointelligence*
*The University of Tokyo*
Tokyo, Japan

Yukie Nagai
*International Research Center for Neurointelligence*
*The University of Tokyo*
Tokyo, Japan

***Abstract*—Understanding how humans express and synchronize emotions across multiple communication channels—particularly facial expressions and speech—has significant implications for emotion recognition systems and human-computer interaction. Motivated by the notion that non-overlapping speech promotes clearer emotional coordination, while overlapping speech disrupts synchrony, this study examines how these conversational dynamics shape the spatial and temporal alignment of arousal and valence across facial and vocal modalities. Using dyadic interactions from the IEMOCAP dataset, we extracted continuous emotion estimates via EmoNet (facial video) and a Wav2Vec2-based model (speech audio). Segments were categorized based on speech overlap, and emotional alignment was assessed using Pearson correlation, lag-adjusted analysis, and Dynamic Time Warping (DTW). Across analyses, non-overlapping speech was associated with more stable and predictable emotional synchrony than overlapping speech. While zero-lag correlations were low and not statistically different, non-overlapping speech showed reduced variability, especially for arousal. Lag-adjusted correlations and best-lag distributions revealed clearer, more consistent temporal alignment in these segments. In contrast, overlapping speech exhibited higher variability and flatter lag profiles, though DTW indicated unexpectedly tighter alignment—suggesting distinct coordination strategies. Notably, directionality patterns showed that facial expressions more often preceded speech during turn-taking, while speech led during simultaneous vocalizations. These findings underscore the importance of conversational structure in regulating emotional communication and provide new insight into the spatial and temporal dynamics of multimodal affective alignment in real-world interaction.**

## I. Introduction

Understanding how humans express and perceive emotions across multiple communication channels—such as facial expressions, vocal prosody, gesture, and body movement—is a foundational question in affective science and human-centered computing [1] [2]. These nonverbal signals are not merely passive reflections of internal states; they are active components of social communication that serve to convey affect, regulate interaction, and foster interpersonal understanding. Among the multimodal repertoire, facial expressions and vocal prosody have emerged as the two most salient and extensively researched channels, largely due to their universality, expressiveness, and ease of measurement [3] [4].

Facial cues offer rich, spatially encoded emotional information through changes in muscle activation and facial morphology, while vocal cues—such as pitch, intensity, and rhythm—convey temporal dynamics of arousal and valence in speech [5]. Arousal and valence are fundamental dimensions in emotion theory, capturing the intensity (arousal) and the pleasantness (valence) of emotional states, respectively. These dimensions provide a useful framework for understanding emotional expression across modalities and offer a common ground for aligning facial and vocal cues in real-time interactions [6]. Importantly, these modalities are not independent; they function as tightly coupled, interactive systems that co-evolve during social interaction, contributing to emotional synchrony and mutual regulation [7]. In real-life dyadic conversations, however, these signals rarely occur in perfect unison. Instead, they unfold continuously and often asynchronously, influenced by contextual variables such as speaker intent, response timing, and conversational dynamics like turn-taking and overlap [8] [9].

While prior research has advanced emotion recognition using unimodal approaches—leveraging either speech-based [10] [11] or vision-based features [12]—relatively few studies have examined how these signals align at a fine-grained, frame-level timescale in real-world interpersonal interactions. This gap is particularly critical for understanding emotion communication as a dynamic, multimodal process rather than a static classification problem. To build emotionally intelligent systems that can respond appropriately in real-time, it is crucial to investigate not only what emotions are present in each modality but also how they interact—spatially (are they conveying similar signals?) and temporally (are they unfolding in sync?).

Multimodal emotion analysis has gained momentum with the advancement of deep learning models capable of extracting affective features from raw audio and video data [13]. Research has shown that cross-modal fusion of speech and facial features can enhance emotion recognition accuracy [14] [15], yet the nature of their alignment—particularly in dynamic interpersonal settings—remains underexplored. Previous work has demonstrated that cross-modal congruence can be disrupted by contextual factors, such as simultaneous speech or high cognitive load [16] [17], suggesting that emotional synchrony is sensitive to the structure of conversation. Although many cross-modal studies focus on correlation or classification performance, fewer investigate the fine-grained temporal dynamics, such as the lag or directionality of emotional alignment between modalities [18] [19].

A key but underexplored factor in emotional alignment is the structure of speech itself. In dyadic interactions, non-overlapping speech (turn-taking) and overlapping speech (simultaneous vocalization) represent two distinct communicative contexts [20]. Non-overlapping speech is associated with organized exchanges, mutual regulation, and clearer affective signaling, whereas overlapping speech often emerges in emotionally charged moments, interruptions, or misunderstandings—conditions that may disrupt emotional coordination [16]. Overlapping speech is also known to increase cognitive load, which may reduce attentional synchrony between speakers, potentially degrading emotional alignment between modalities [21]. This disruption is especially noticeable when switching between speakers or dealing with divergent speech patterns, which demands more cognitive resources for processing and coordinating emotional signals. Temporally misaligned emotional utterances, often due to natural asynchrony in multimodal expression (e.g., a smile may precede speech), reflect this challenge [22]. This misalignment, however, may indicate a shared latent affective process across modalities, rather than one modality driving the other, especially in naturalistic settings [14] [23]. Thus, examining the role of speech structure in emotional synchrony offers valuable insights into the dynamic interplay between facial and vocal expressions.

The present study aims to investigate the alignment between facial and vocal emotional signals in dyadic interactions, focusing on the role of speech structure in emotional synchrony. Specifically, we examine two key types of speech conditions: non-overlapping speech (turn-taking, where one speaker speaks while the other listens) and overlapping speech (simultaneous vocalizations, where both speakers speak at the same time). These speech conditions represent two distinct communicative dynamics: non-overlapping speech, which generally supports organized conversation and clearer emotional signals, and overlapping speech, which may introduce disruptions due to interruptions or misunderstandings, potentially affecting emotional coordination. To examine how these speech conditions impact the alignment of facial and vocal emotions, we use the IEMOCAP dataset [10], which provides high-quality audiovisual recordings of naturalistic dyadic conversations. The study focuses on both spatial alignment (whether facial and vocal emotional signals align) and temporal alignment (whether facial and vocal emotional signals unfold synchronously or with time lags) in terms of arousal and valence emotion dimensions.

Thus, we explore two key hypotheses regarding the mechanisms of emotional synchrony in arousal and valence:

Hypothesis 1 (Spatial Alignment): Facial and vocal emotional signals will show greater spatial alignment (i.e., higher correlation) in non-overlapping speech conditions compared to overlapping speech. In non-overlapping speech, turn-taking should promote clearer emotional signals and better synchrony between facial and vocal modalities.

Hypothesis 2 (Temporal Alignment): Temporal alignment (i.e., the extent to which facial and vocal emotional signals align across time) will be stronger in non-overlapping speech than in overlapping speech. In non-overlapping speech, facial and vocal signals are expected to unfold more predictably, whereas simultaneous speech may lead to asynchrony due to the competing nature of overlapping vocalizations.

Overall, this study provides valuable insights into the dynamic nature of emotional synchrony in dyadic interactions, emphasizing the role of speech structure in aligning facial and vocal emotional signals. By examining both spatial and temporal alignment, we contribute to a deeper understanding of how emotional expressions are coordinated across modalities during dyadic conversations. The findings highlight the complex interplay between turn-taking and overlapping speech in shaping emotional synchrony, offering a more nuanced perspective on how individuals regulate and synchronize their emotional expressions in real-time interactions. This work advances our understanding of multimodal communication and lays the groundwork for more effective emotionally intelligent systems that can interpret and respond to human emotions in interactive settings.

## II. Audiovisual dataset: IEMOCAP

To investigate emotional synchrony in dyadic interactions, we focused on the IEMOCAP (Interactive Emotional Dyadic Motion Capture) dataset, a well-established dataset for multimodal emotion recognition [10]. It consists of approximately 12 hours of audiovisual recordings collected from 10 professional actors (5 male-female dyads) across five sessions. Each session includes both scripted and improvised dialogues designed to capture a range of emotional expressions. The corpus provides synchronized stereo audio (captured with individual head-mounted microphones for each speaker) and side-by-side frontal video recordings (captured using high-resolution cameras). Each session is annotated with both categorical emotion labels (e.g., anger, happiness, sadness) and dimensional ratings of arousal and valence on a continuous scale, which enables time-resolved analysis of emotional expression across speech and facial modalities. [10].

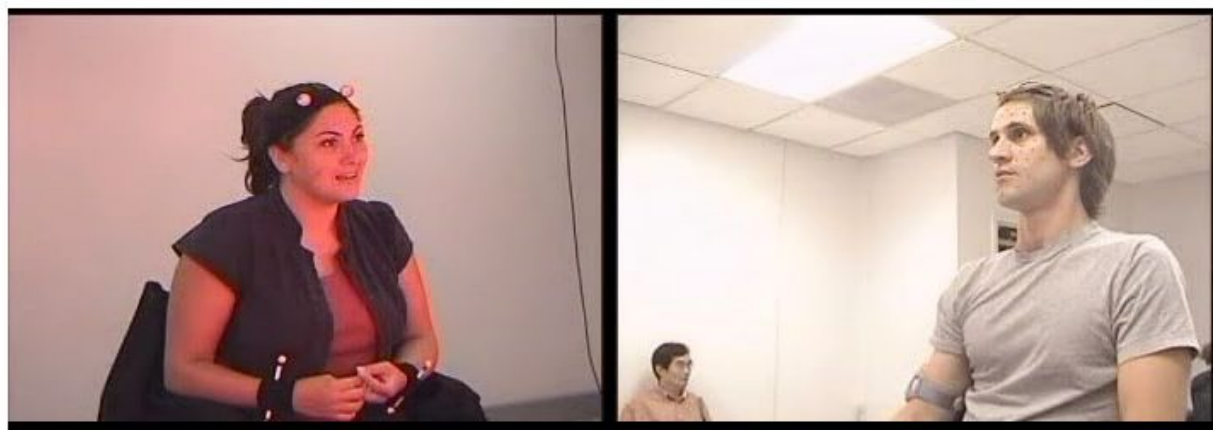

Fig. 1. IEMOCAP Data Overview. Example frame showing the dyadic interaction during the session. The left speaker is the target main speaker, whose facial and vocal emotional expressions are analyzed in this study.

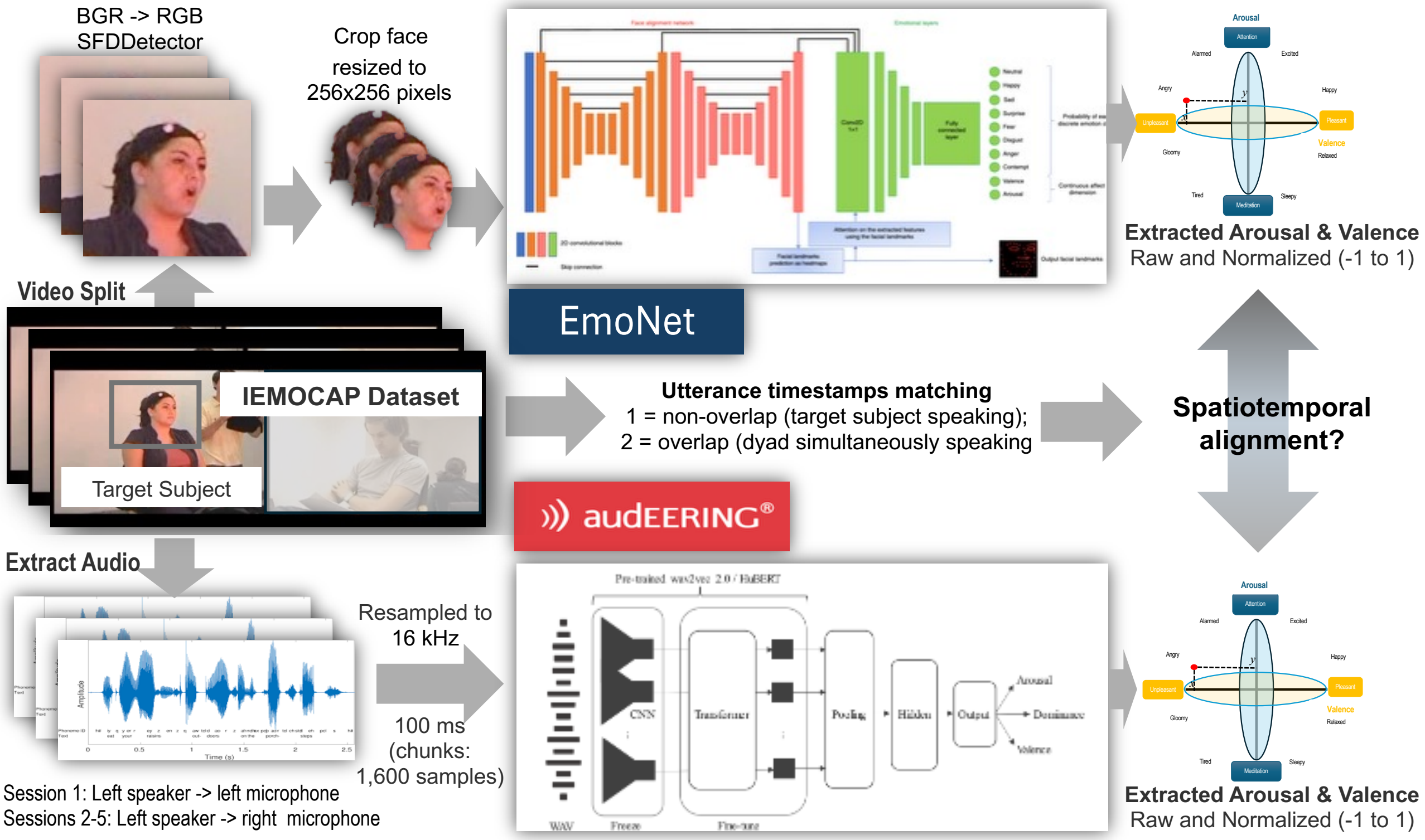

Fig. 2. Overview of the audiovisual emotion processing (EmoNet [24] and audEERING [25]) and spatiotemporal analysis pipeline

## III. METHOD

To investigate how speech-based and facial-based emotional signals align during dyadic interaction, we developed a multimodal analysis pipeline combining audiovisual feature extraction, speaker-specific annotation, and condition-based alignment analysis (Figure 1). Using the IEMOCAP dataset, we extracted frame-level arousal and valence estimates from synchronized audio and video data, isolated the target speaker's signals, and segmented each recording into non-overlapping and overlapping speech conditions. We then computed spatial (correlation) and temporal (lag adjusted and DTW-based) alignment metrics across modalities to assess the consistency and timing of emotional expressions under different conversational dynamics.

### *A. Pre-Processing*

Each .avi video in the IEMOCAP corpus contains stereo audio and side-by-side visual footage of two speakers. As our analysis focused exclusively on one speaker, we extracted only the left speaker's audio and visual streams**.** Audio separation was conducted using FFmpeg's channelsplit filter, leveraging the dataset's known microphone configuration: the Front Left (FL) channel was extracted for Session 1, and the Front Right (FR) channel was extracted for Sessions 2–5, corresponding to the left-positioned speaker in each session. In parallel, the left half of each video frame was cropped to isolate the facial region of the primary speaker. The resulting audio (.wav) and video (.mp4) files were saved with synchronized timestamps to ensure temporal correspondence. To eliminate non-task-related or preparatory behaviors typically present at the beginning of recordings, we trimmed the first 4 seconds from each cropped video using FFmpeg's stream copy mode. This ensured that subsequent analyses focused on segments with higher likelihood of meaningful emotional expression.

### *B. Facial and Audio Emotion Feature Extraction*

For facial emotion analysis, we employed EmoNet, a deep convolutional neural network developed for real-time, frame-level emotion recognition [24]. EmoNet utilizes an advanced architecture based on a combination of landmark detection and emotion recognition, making it a unique and powerful solution for dynamic facial emotion recognition. The network operates by first detecting facial landmarks using the Facial Action Network (FAN), which predicts the location of key facial points such as the eyes, eyebrows, and mouth. EmoNet integrates these features into a multi-layered architecture where the initial layers capture low-level facial features (e.g., edges and textures) while the deeper layers focus on high-level morphological features (e.g., overall facial shape). The Hourglass Network, a key component of EmoNet, is responsible for extracting multi-resolution features that help in capturing both local and global facial patterns [24]. EmoNet processes each input image by first isolating the face using the Single Shot Scale-invariant Face Detector (SFD) and cropping the facial region to a 256x256 pixel size. Additionally, EmoNet incorporates an attention mechanism that prioritizes key facial regions relevant for emotion recognition, effectively increasing the model's accuracy in distinguishing subtle emotional expressions. The architecture is trained using a hybrid loss function that combines cross-entropy for classification tasks and Concordance Correlation Coefficient (CCC) for continuous emotion regression (>.60) [24]. This architecture is further enhanced by temporal smoothing and knowledge distillation, which improves robustness against noisy data and ensures accurate prediction across a wide range of emotional expressions. The network then outputs continuous valence and arousal values (normalized to -1, 1 range), categorical emotion labels (e.g., Happy, Sad, Anger), and 68 facial landmarks. The original audio, extracted during preprocessing, is then reattached to the annotated video using FFmpeg to generate a fully synchronized audiovisual file for each trial.

For audio processing, we utilized audEERING's wav2vec2-large-robust-12-ft-emotion-msp-dim model for speech emotion recognition. This model is based on the Wav2Vec2 architecture, a self-supervised speech representation learning model, and was specifically fine-tuned on the MSP-Podcast dataset to predict continuous emotional dimensions like arousal and valence [25]. Wav2Vec2 uses a transformer-based encoder to process the raw speech waveform and predict emotion-related features. In this model, the speech signal is passed through a feature encoder consisting of convolutional layers, which project the raw waveform into a feature space. This feature set is then input into a series of transformer layers designed for temporal feature extraction, utilizing multi-head self-attention mechanisms [26]. This model, trained on a variety of speech datasets including LibriSpeech, CommonVoice, and others, demonstrated state-of-the-art performance on valence recognition, with a CCC of 0.638 on the MSP-Podcast dataset [25]. The transformer-based architecture enables robust emotion recognition with high generalization across datasets and robust handling of noisy environments, outperforming traditional convolutional neural networks (CNNs) in the dimensional speech emotion recognition task. The model outputs valence and arousal predictions on a frame-level basis, with the final emotion score rescaled to the range [-1, 1]. For temporal alignment, predictions from the speech modality were extracted using overlapping 100 ms windows, synchronized with the visual modality's time base to align speech and facial signals for emotional analysis.

### C. *Temporal alignment and Utterance-Level Annotation of Speech Conditions*

To segment emotional signals based on interaction dynamics, we temporally aligned frame-level audiovisual emotion features with utterance transcripts from the IEMOCAP dataset. Each transcription file was parsed into a structured format containing onset/offset times, speaker identity, and utterance content. Using the utterance intervals of both interlocutors, we generated a binary speech state label for each frame: speech_state = 1 for non-overlapping speech (only the target speaker was speaking) and speech_state = 2 for overlapping speech (both speakers spoke simultaneously). Frames with no active speech were excluded. These frame-level annotations were used to segment the emotion time series into non-overlap and overlap conditions. To enable this, we first created a unified multimodal dataset by merging audio and visual emotion features extracted from the audEERING and EmoNet models. For each video, frame-level arousal and valence values were aligned across modalities using the shared timeframe column (in seconds), and only frames with matching timestamps were retained via an inner join. The resulting merged files contained synchronized multimodal features per frame and were organized into session-wise folders for downstream spatial and temporal alignment analyses.

### D. *Analysis*

To evaluate the spatial alignment between speech-based and facial-based emotional signals, we computed the Pearson correlation coefficient (r) for arousal and valence values at each frame, without applying any temporal shift. The Pearson correlation coefficient (r) measures the linear relationship between the emotional expressions from speech and facial signals, providing insight into how similarly these modalities are expressing emotions. This analysis was conducted separately for non-overlapping and overlapping speech conditions, with correlation coefficients computed per dyad by pooling data across all utterances. The resulting dyad-level Pearson correlation values for arousal and valence were then compared using paired t-tests to evaluate whether the strength of alignment differed significantly between non-overlapping and overlapping speech conditions.

For temporal alignment analysis, we first applied a 5-frame moving average to smooth the arousal and valence time series. This step helped reduce short-term fluctuations and provided a clearer representation of emotional dynamics over time. Subsequently, we performed lag correlation analysis, shifting one modality's emotional signals against the other within a ±10-frame window. This process enabled us to determine the best lag (the time offset at which the emotional signals from speech and facial expressions were most aligned). The lag directionality (whether speech or facial expression leads or lags) was also identified through this analysis. In addition, we used Dynamic Time Warping (DTW) to assess the nonlinear misalignment between the emotional signals of speech and facial expressions. DTW is a robust technique for comparing two time series by finding the optimal alignment between them. It works by stretching or compressing the time series to minimize the distance between corresponding points. All temporal metrics were calculated per dyad and paired t-tests were used to compare condition-level differences (non-overlapping vs. overlapping speech) at the global level. This approach allowed us to retain dyad-level variability and provided a statistically robust means of assessing how interaction structure influences cross-modal emotional alignment across the entire dataset.

## IV. RESULTS

### A. *Spatial analyses*

***Pearson Correlation.*** To evaluate the degree of emotional synchrony between speech-based and facial-based signals, we computed Pearson correlation coefficients for arousal and valence at the dyad level, separately for non-overlapping and overlapping speech conditions. Correlation magnitudes were generally low across both affective dimensions but varied by speech condition and emotional signal (Figure 3). For arousal, the average correlation in non-overlapping speech was r = 0.005 (SD = 0.041), while overlapping speech showed a slightly more dispersed alignment with $r$ = –0.01 (SD = 0.10,). Valence followed a similar trend: $r$ = 0.01 (SD = 0.043,) in non-overlapping speech, dropping to $r$ = –0.003 (SD = 0.101,) under overlapping conditions. Paired-sample t-tests showed that the differences in correlation strength between conditions were not

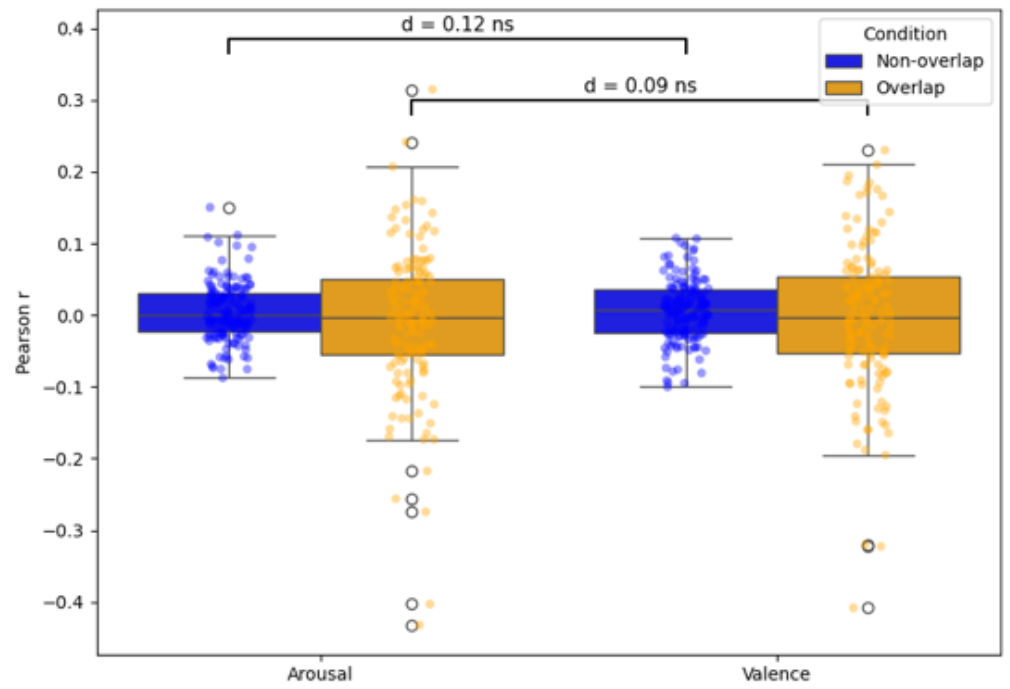

Fig 3. Dyad-level Pearson correlations between speech-based and facial-based arousal (left) and valence (right) signals, compared across non-overlapping and overlapping speech conditions.

statistically significant for either arousal ($t = 1.53$, $d = 0.12$, $p = .128$) or valence ($t = 1.08$, $d = 0.09$, $p = .283$). Notably, arousal exhibited slightly stronger and more consistent synchrony than valence across both conditions. However, the higher standard deviations during overlapping speech suggest that synchrony—especially for valence—is more susceptible to disruption during simultaneous vocalization, likely due to reduced cue clarity, disfluency, or turn-taking conflicts. Overall, while both modalities show modest cross-modal alignment, arousal appears more robust than valence, and non-overlapping speech supports more stable synchrony than overlapping interactions (Figure 4). (see Supplementary Table 1 for the dyad-level correlation across speech conditions and sample cases in Figure 1)

***Arousal-Valence (AV) Space.*** To further visualize how emotional signals unfold across modalities, we plotted the global pooled dyad-level distribution of speech- and facial-derived emotion estimates in AV space, separately for non-overlapping and overlapping speech segments.

Figure 5 shows that facial expressions are more concentrated in the high arousal, low valence quadrant (top-left), but also extend into the high arousal, high valence (top-right) and low arousal, low valence (bottom-left) quadrants, suggesting that facial expressions capture a variety of emotional states, including positive emotions with high energy, as well as negative emotions or neutral expressions in both non-overlapping and non-overlapping speech conditions. However, there is a clear emphasis on the high arousal, low valence region, indicating that facial expressions often convey more enhanced, negative emotional states.

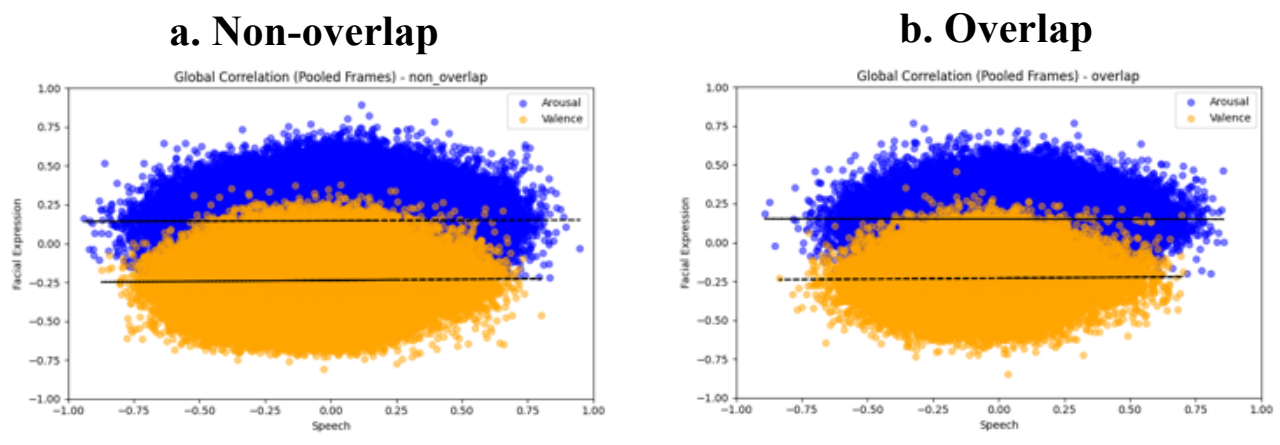

Fig 4. Pooled Pearson correlations between speech-derived and facial-derived arousal (blue) and valence (orange) signals at the dyad level for non-overlapping and overlapping speech conditions.

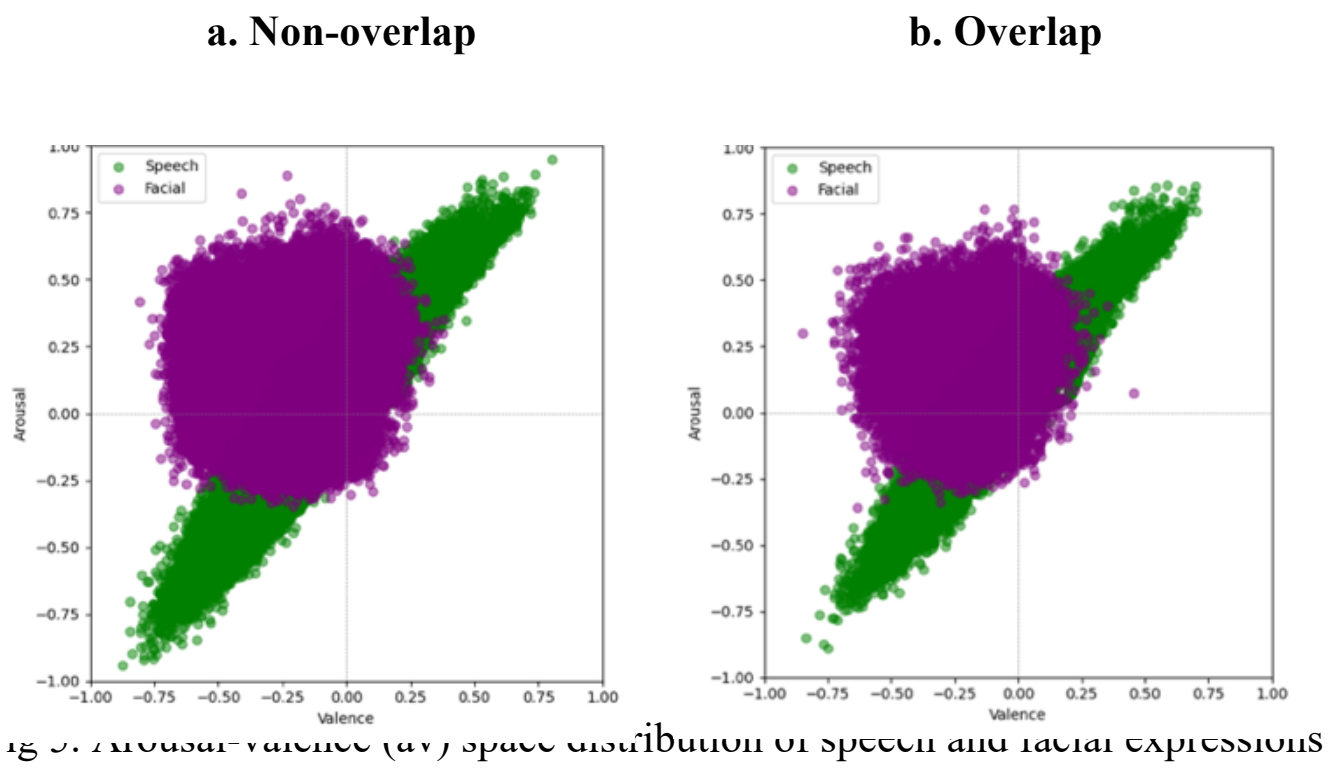

Fig 5. Arousal-valence (av) space distribution of speech and facial expressions in non-overlapping and overlapping speech conditions.

On the other hand, speech primarily covers the high arousal, high valence quadrant (top-right) and the low arousal, low valence quadrant (bottom-left), with much less coverage in the low arousal-high valence and high arousal-low valence regions. This suggests that speech tends to be more focused on either intense positive emotions or low-energy, negative/neutral states. The difference in coverage between speech and facial expressions indicates that facial expressions offer a more diverse emotional representation, whereas speech tends to focus on more extreme or neutral emotional expressions.

## B. *Temporal analyses*

***Lag Correlation Analysis***. To evaluate the temporal alignment between speech and facial expressions, we conducted a lag-adjusted Pearson correlation analysis (Figure 6). The lag-adjusted correlations showed improvements in synchrony compared to the zero-lag correlations. Specifically, for non-overlapping speech, the arousal correlation increased from r = 0.01 ± 0.06 to r = 0.03 ± 0.07, with a medium effect size (d = –0.44, p < .001). Valence correlation increased from r = 0.01 ± 0.07 to r = 0.09 ± 0.15, showing a large effect (d = –0.67, p < .001). For overlapping speech, the arousal correlation improved from r = -0.01 ± 0.15 to r = 0.09 ± 0.14, again with a large effect (d = –0.64, p < .001). Valence increased from r = -0.00 ± 0.15 to r = 0.09 ± 0.15, showing a moderate-to-large effect (d = –0.46, p < .001). These results demonstrate that accounting for temporal shifts (lag adjustments) improves the alignment between emotional signals, especially in non-overlapping speech conditions, with arousal and valence correlations showing clear positive shifts. Notably, while lag-adjusted synchrony improved for both arousal and valence across conditions, the enhancement was more pronounced for valence in non-overlapping speech and for arousal in overlapping speech, suggesting modality-specific sensitivities to interaction structure.

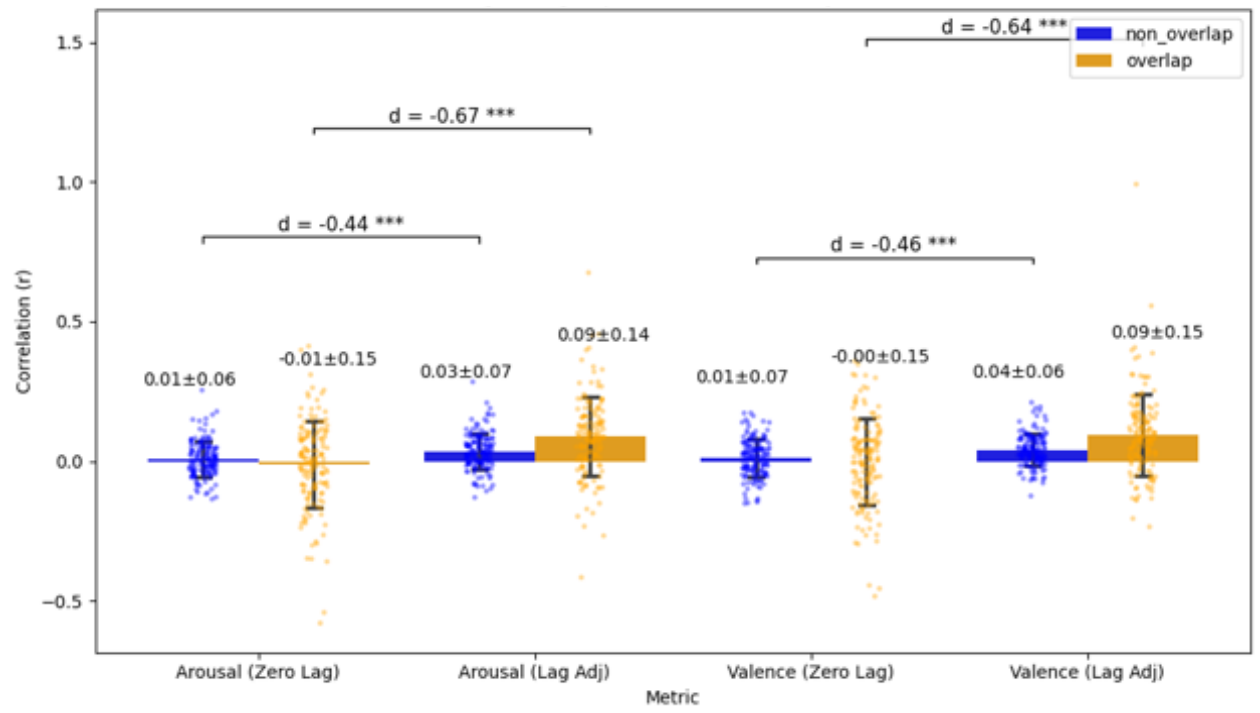

Fig 6. Zero-lag vs lag-adjusted correlation (smoothed) between speech and facial expressions in aroudal and valence before and after lag adjustment across different conditions.

Figure 7 shows the distribution of the best lag values (per dyad) for arousal and valence across both non-overlapping and overlapping speech conditions with kernel density estimates (KDE). The distributions for both signals are bimodal, with clear peaks around –10 and +10 frames, indicating that maximal emotional synchrony between vocal and facial expressions often occurs at the temporal extremes—suggesting substantial lead-lag relationships across modalities. For arousal (left plot), non-overlapping speech exhibits a more structured distribution, with defined peaks at both ends and a subtle rise near zero lag, reflecting more consistent alignment. In contrast, the overlapping speech shows a flatter distribution, especially in the central region, and a more dispersed KDE curve, suggesting weaker or more variable synchrony. Although the means are similar (non-overlap = -0.1 ± 7.4; overlap = 0.1 ± 7.6, t = -0.20, d = -0.02, p = .84), the wider spread in overlapping speech reflects less temporal stability during emotional coordination. A similar trend is observed for valence (right plot), where non-overlapping speech retains strong boundary peaks and a tighter distribution, while overlapping speech displays a flatter, more irregular profile. Notably, the mean lag in overlapping speech shifts slightly negative (-0.6 ± 7.7 vs. 0.6 ± 7.3, t = -0.02, d = -0.003, p = .98), implying that facial expressions may precede vocal valence cues more frequently when multiple speakers are vocalizing simultaneously. These results highlight that non-overlapping speech supports clearer and more predictable temporal alignment between modalities, while overlapping speech introduces diffuse, inconsistent lag structures, reflecting disrupted or delayed emotional synchrony under more complex

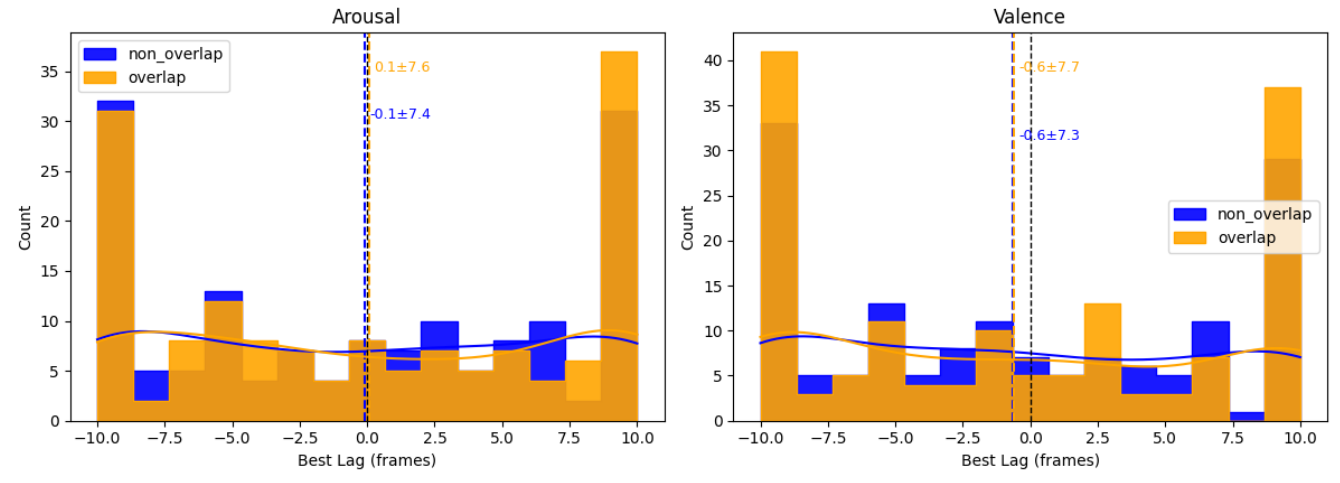

Fig 7. Distributions of best lag values for arousal and valence across non-overlapping and overlapping speech conditions. Histograms are overlaid with kernel density estimates (KDE) to visualize the shape of the distributions. Dashed vertical lines represent the group means, with accompanying labels denoting the mean ± standard deviation.

turn-taking conditions. (see Supplementary Table 2 and Figure 4 for the dyad-level temporal correlation (smoothed) results and sample cases, respectively; Table 3 for dyad-level lag correlations (smoothed) [best lags, correlations, and directionality], and; Figure 4 for the temporal correlation sample cases)

***Dynamic Time Warping.*** To complement the correlation-based analyses, we applied DTW to quantify the temporal alignment between speech- and facial-derived emotional signals. Contrary to our initial hypothesis, DTW scores were significantly lower during overlapping speech, suggesting enhanced temporal alignment of emotional expressions when interlocutors speak simultaneously (Figure 8). Specifically, the average DTW score for arousal decreased from 322.1 ± 212.2 in the non-overlapping condition 77.6 ± 69.7 during overlaps; for valence, scores dropped from 427.8 ± 297.5 to 108.2 ± 80.0. These differences were robust (arousal: $t(116) = 13.45$, $d = 1.55$, $p < 0.001$; valence: $t(116) = 12.75$, $d = 1.47$, $p < 0.001$), indicating a large effect of interaction structure on cross-modal alignment. This finding challenges the assumption that structured turn-taking facilitates better emotional synchrony. (see Supplementary Table 4, for the dyad-level DTW misalignment scores, lag direction, mean shift; DTW misalignment heatmaps of sample cases in Figure 5).

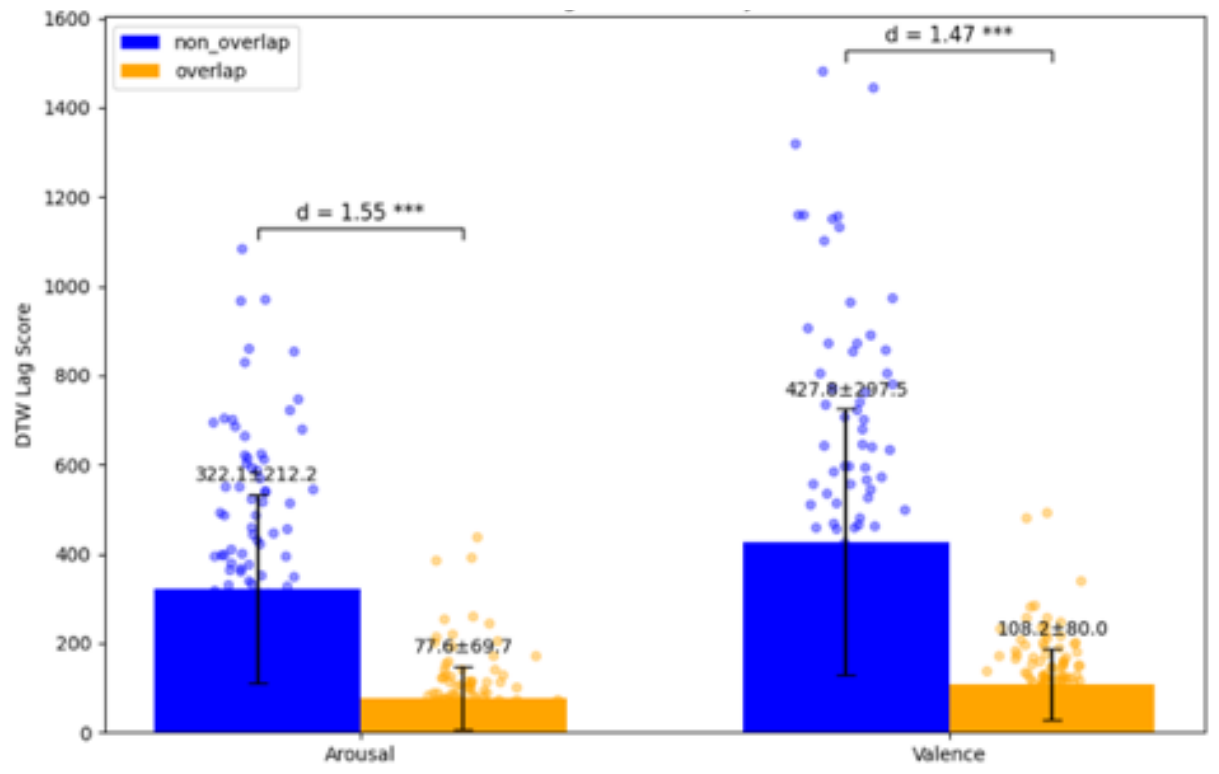

Fig 8. DTW misalignment scores by speech condition for arousal and valence across non-overlapping and overlapping speech conditions.

## V. Disccusion and Conclusion

We investigated the spatial and temporal alignment of arousal and valence in facial and vocal emotional expressions during dyadic interactions, comparing non-overlapping and overlapping speech conditions. Our findings underscore the significant role of conversational structure, particularly the presence or absence of turn-taking, in shaping emotional synchrony between modalities.

Across analyses, we observed that non-overlapping speech was associated with slightly stronger and more stable cross-modal emotional synchrony compared to overlapping speech. Although zero-lag correlations were low overall and not statistically different between conditions, non-overlapping speech exhibited reduced variability and more consistent alignment patterns, particularly for arousal. Temporal analyses further reinforced these findings: lag-adjusted correlations improved significantly in non-overlapping segments, and best-lag distributions showed clearer bimodal structures, indicating more predictable temporal coupling. In contrast, overlapping speech was characterized by higher variability in synchrony and flatter lag distributions, despite unexpectedly tighter DTW-based alignment, implying that different interaction structures recruit distinct emotional coordination mechanisms.

Our spatial analysis revealed that zero-lag correlations between facial- and speech-based emotional signals were generally low across conditions. Although non-overlapping speech yielded slightly stronger average correlations and lower variability—particularly for arousal—these differences were not statistically significant. Still, the lower standard deviations in the non-overlapping condition point to more stable synchrony, likely due to clearer segmentation and predictability of expressive turns. In contrast, the greater variability observed in overlapping speech, especially for valence, suggests that emotional synchrony is more vulnerable to disruption when vocal streams overlap. These disruptions may stem from reduced clarity in acoustic and visual cues, potential asynchrony in expressive timing, or increased ambiguity when affective signals from both interlocutors compete temporally. The simultaneity of overlapping speech can blur the temporal boundaries necessary for cross-modal alignment, making it harder to decode and integrate affective cues across modalities.

The AV space distributions further support this dissociation by showing facial expressions spanning a range of arousal and valence states, especially in the high arousal–low valence quadrant, suggesting stronger expression of intense or negative emotions. In contrast, speech estimates were more narrowly concentrated in polarized regions, particularly high arousal–high valence and low arousal–low valence, reflecting potentially limited emotional range. This difference may be attributed in part to the characteristics of the EmoNet and audEERING models, which were trained on distinct datasets and may encode emotional information with varying sensitivity to subtle or context-dependent cues. Additionally, the scripted nature of some IEMOCAP segments may constrain vocal expressivity more than facial expressivity, leading to reduced diversity in speech-derived emotion estimates.

Temporal alignment analyses provided stronger support for the influence of interaction structure. Lag-adjusted Pearson correlations improved significantly compared to zero-lag correlations, particularly for valence in non-overlapping speech and arousal in overlapping speech. These improvements suggest that multimodal emotional signals are often temporally misaligned and benefit from slight temporal adjustments to reveal underlying synchrony. The stronger enhancement in non-overlapping speech indicates that turn-taking facilitates temporal predictability and regularity, which may enhance the perceptual coherence of affective cues. The distribution of best lag values further reflected these dynamics. In non-overlapping speech, lag distributions showed clear bimodal patterns and central peaks, indicating more structured and predictable alignment patterns across dyads. In contrast, overlapping speech was associated with flatter and more dispersed lag distributions, suggesting greater variability in the timing of emotional signals. Although mean lags did not differ significantly, the increased dispersion in overlapping speech reflects more unstable temporal coupling between modalities.

The DTW analysis also revealed an unexpected pattern: overlapping speech yielded significantly lower misalignment scores than non-overlapping speech, indicating tighter temporal coupling. One possible interpretation is that overlapping speech imposes greater demands on the coordination of expressive behavior, leading to tighter, albeit more reactive, multimodal coupling. This may reflect a compensatory mechanism, wherein individuals adjust facial and vocal expressions more tightly in time to maintain communicative coherence under less structured conditions. The nature of the IEMOCAP dataset—which includes both scripted and improvised emotional expressions—may also contribute to this result, as overlapping speech segments often emerge in more emotionally expressive or less constrained improvisational contexts. These segments may require heightened attentional and expressive synchronization, resulting in tighter but more constrained alignment.

Finally, DTW analyses of directionality revealed that in non-overlapping speech, facial expressions tended to precede speech-based emotional cues, suggesting a leading role for facial signals in guiding vocal affect. In contrast, overlapping speech showed more variable directionality, including cases where speech preceded facial expression. This pattern suggests that facial expressions serve as the primary channel for emotional communication in structured, turn-taking contexts, often initiating the emotional response while speech follows. In overlapping speech, this leading pattern is less consistent, possibly due to cognitive load, divided attention, or the inherently chaotic nature of simultaneous vocalizations. When both speakers vocalize at the same time, the emotional cues from facial expressions may be delayed or less coordinated, as the attention and processing resources are split between the two modalities. In these cases, speech tends to dominate, influencing the emotional tone before facial expressions can fully reflect or align with it. These findings emphasize that directionality is not fixed but dynamically shaped by interaction context, reinforcing the importance of examining both alignment strength and expressive sequencing.

In sum, our findings emphasize that interaction structure—particularly turn-taking—plays a critical role in shaping emotional synchrony across facial and vocal modalities. Non-overlapping speech supports more stable and predictable alignment, as evidenced by lower variability, clearer lead-lag distributions, and stronger lag-adjusted correlations. These results suggest that structured exchanges enhance the integration of multimodal affective cues by enabling sequential decoding and reducing cross-modal ambiguity. In contrast, overlapping speech introduces greater temporal complexity and expressive interference, but it can also elicit tighter moment-to-moment coupling, as reflected in lower DTW misalignment scores. This pattern implies that even under less organized conditions, interlocutors may engage in real-time adaptive mechanisms to sustain emotional attunement. Together, these findings point to distinct coordination strategies shaped by conversational context, with implications for understanding how emotional signals are modulated and maintained across modalities in naturalistic social interaction.

The findings from this study provide important insights into real-world emotional communication. Non-overlapping speech, which mimics natural turn-taking behavior in everyday conversations, tends to facilitate better synchrony between emotional expressions in speech and facial signals. This suggests that conversations with clear speaking turns allow individuals to better align their emotional cues, leading to more coherent and coordinated emotional exchanges. In contrast, overlapping speech—common in more rapid, emotionally charged, or competitive interactions—introduces more variability and misalignment, which could hinder effective emotional communication, but under certain conditions, they may also elicit compensatory adjustments that foster tighter, albeit less predictable, temporal coupling. Understanding these patterns can help inform how individuals navigate emotional expression in different social contexts, highlighting the importance of conversational structure in promoting emotional synchrony. Future studies could explore strategies to improve emotional alignment in overlapping speech situations, such as through training or using assistive technologies to help regulate emotional expression during simultaneous speaking.

While our study provides valuable insights into the dynamics of emotional synchrony, there are several limitations to consider. First, our emotion estimates relied on pre-trained models, which may introduce biases or constrain sensitivity to subtle, context-dependent cues, particularly in spontaneous interactions or overlapping speech. These models were trained on different datasets with potentially incompatible labeling schemes, which may affect cross-modal comparability. Additionally, the analysis relied on segment-level aggregation, which may obscure finer-grained fluctuations—particularly during transitions between turn-taking and overlap—might be better captured using higher temporal resolution or dynamic modeling approaches. The IEMOCAP dataset includes interactions in a controlled lab setting which may limit the ecological validity of the findings, as emotional communication in naturalistic environments may involve different patterns of synchrony and disruption. Moreover, the dataset featured a relatively homogeneous group of actors, limiting demographic diversity of cultures, age groups, and neurodiverse populations, which should be explored in future studies to ensure generalizability. Lastly, while DTW provides a flexible and informative measure of temporal alignment, it does not distinguish whether tighter coupling reflects deliberate coordination, reactive mimicry, or compensatory mechanisms. The interpretation of alignment strength—particularly under overlapping speech—remains speculative without complementary behavioral or neural data.

## Ethical Impact Statement

This study analyzed the publicly available IEMOCAP dataset, which was originally collected under approved institutional review protocols. Therefore, no new human subjects were recruited, and IRB approval was not required for this specific analysis. The original dataset includes informed consent from participants for the use and public distribution of their audiovisual recordings for research purposes. No personally identifiable information was used or inferred during our analyses. Because this study involved secondary analysis of existing data, no instructions were given to participants by the authors. However, original participants were instructed to perform scripted and improvised dialogues in emotionally expressive ways. These conditions were systematically labeled and have been preserved in our analysis. Participants in the original IEMOCAP study were compensated by the collecting institution. Our study did not involve any new participant recruitment or compensation. This research contributes to the development of multimodal emotional alignment models that could benefit neurodiverse populations by improving mutual understanding in social interactions. However, such technologies could also be misused in applications such as emotion-based surveillance or behavioral profiling. To mitigate this risk, we explicitly frame our results within assistive and educational contexts and avoid normative assumptions about emotional "correctness". We also emphasize that alignment should be interpreted as contextual and not prescriptive. There is also a risk of perpetuating biases due to the demographic composition of the IEMOCAP dataset, which consists primarily of English-speaking actors and lacks cultural and neurodiversity. We caution against generalizing these results to broader populations and call for future validation on more diverse and naturalistic datasets. Findings may not generalize to non-English speakers, non-acted interactions, or individuals with divergent emotional expression patterns (e.g., autistic individuals). While our results highlight condition-specific patterns (e.g., overlap vs. non-overlap), the absolute values of synchrony should not be treated as normative benchmarks. We do not claim that our system fully captures emotional understanding. Rather, we investigate temporal alignment as one aspect of affective interaction. Our results should be interpreted as insights into patterns of cross-modal coordination under structured conditions, not as generalizable indicators of emotional intelligence or social competence. Our results are sensitive to speaking conditions (overlap vs. non-overlap), signal quality, and time-series preprocessing choices (e.g., smoothing window). The alignment patterns we report are contextual and may vary significantly in spontaneous or multilingual settings. Future work should explore robustness to diverse interactional and cultural contexts.

## Acknowledgment

**Supplementary Material:**

## 1. Spatial analysis

**Correlation Analysis.** To evaluate the spatial alignment between speech-based and facial-based emotional signals, we computed the Pearson correlation coefficient (r) for arousal and valence values at each frame, without applying any temporal shift. The Pearson correlation coefficient (r) measures the linear relationship between the emotional expressions from speech and facial signals, providing insight into how similarly these modalities are expressing emotions. This analysis was conducted separately for non-overlapping and overlapping speech conditions, with correlation coefficients computed per dyad by pooling data across all utterances. The resulting dyad-level Pearson correlation values for arousal and valence were then compared using paired t-tests to evaluate whether the strength of alignment differed significantly between non-overlapping and overlapping speech conditions.

**Table 1.** Dyad-Level Correlation and Significance Across Speech Conditions

| | | Non-overlap | | | | Overlap | | | |
|---|---|---|---|---|---|---|---|---|---|
| Session | Dyads | Arousal (*r*) | *p-value* | Valence (*r*) | *p-value* | Arousal (*r*) | *p-value* | Valence (*r*) | *p-value* |
| Session1 | Ses01F_impro01 | -0.009 | 0.750 | -0.033 | 0.236 | 0.153 | 0.001 | 0.013 | 0.793 |
| Session1 | Ses01F_impro02 | 0.109 | 0.000 | 0.014 | 0.464 | 0.315 | 0.001 | 0.116 | 0.229 |
| Session1 | Ses01F_impro03 | 0.042 | 0.051 | -0.046 | 0.034 | 0.022 | 0.624 | 0.028 | 0.536 |
| Session1 | Ses01F_impro04 | -0.014 | 0.465 | 0.107 | 0.000 | -0.060 | 0.076 | 0.070 | 0.039 |
| Session1 | Ses01F_impro05 | -0.058 | 0.008 | 0.083 | 0.000 | 0.035 | 0.191 | -0.050 | 0.061 |
| Session1 | Ses01F_impro06 | -0.034 | 0.014 | 0.051 | 0.000 | 0.119 | 0.000 | 0.144 | 0.000 |
| Session1 | Ses01F_impro07 | -0.028 | 0.217 | 0.053 | 0.020 | -0.062 | 0.035 | 0.040 | 0.172 |
| Session1 | Ses01F_script01_1 | 0.004 | 0.760 | 0.062 | 0.000 | 0.032 | 0.441 | 0.178 | 0.000 |
| Session1 | Ses01F_script01_2 | 0.035 | 0.117 | -0.001 | 0.977 | 0.012 | 0.731 | 0.138 | 0.000 |
| Session1 | Ses01F_script01_3 | 0.014 | 0.332 | 0.052 | 0.000 | 0.124 | 0.091 | -0.160 | 0.029 |
| Session1 | Ses01F_script02_1 | -0.068 | 0.003 | -0.100 | 0.000 | -0.081 | 0.135 | -0.056 | 0.299 |
| Session1 | Ses01F_script02_2 | -0.012 | 0.312 | -0.011 | 0.363 | -0.024 | 0.554 | 0.185 | 0.000 |
| Session1 | Ses01F_script03_1 | 0.059 | 0.001 | -0.046 | 0.008 | 0.121 | 0.001 | -0.044 | 0.229 |
| Session1 | Ses01F_script03_2 | 0.040 | 0.013 | -0.001 | 0.971 | 0.047 | 0.086 | 0.115 | 0.000 |
| Session1 | Ses01M_impro01 | -0.013 | 0.521 | 0.037 | 0.059 | 0.078 | 0.030 | 0.044 | 0.223 |
| Session1 | Ses01M_impro02 | 0.051 | 0.013 | -0.028 | 0.171 | 0.115 | 0.020 | -0.121 | 0.014 |
| Session1 | Ses01M_impro03 | -0.030 | 0.201 | -0.074 | 0.002 | 0.068 | 0.004 | 0.078 | 0.001 |
| Session1 | Ses01M_impro04 | -0.005 | 0.807 | -0.010 | 0.599 | -0.137 | 0.000 | -0.081 | 0.015 |
| Session1 | Ses01M_impro05 | -0.021 | 0.242 | 0.065 | 0.000 | -0.005 | 0.873 | -0.006 | 0.841 |
| Session1 | Ses01M_impro06 | -0.028 | 0.105 | -0.039 | 0.028 | -0.014 | 0.686 | 0.073 | 0.035 |
| Session1 | Ses01M_impro07 | -0.022 | 0.261 | 0.031 | 0.105 | 0.018 | 0.415 | 0.047 | 0.036 |
| Session1 | Ses01M_script01_1 | 0.054 | 0.000 | -0.019 | 0.150 | -0.053 | 0.151 | -0.091 | 0.014 |
| Session1 | Ses01M_script01_2 | 0.150 | 0.000 | 0.016 | 0.517 | -0.052 | 0.426 | -0.141 | 0.030 |
| Session1 | Ses01M_script01_3 | 0.028 | 0.011 | 0.050 | 0.000 | -0.036 | 0.542 | 0.110 | 0.060 |
| Session1 | Ses01M_script02_1 | -0.008 | 0.482 | 0.028 | 0.010 | 0.046 | 0.325 | -0.038 | 0.413 |
| Session1 | Ses01M_script02_2 | -0.036 | 0.002 | 0.101 | 0.000 | -0.080 | 0.070 | -0.016 | 0.713 |
| Session1 | Ses01M_script03_1 | 0.036 | 0.017 | 0.023 | 0.138 | 0.023 | 0.441 | 0.042 | 0.152 |
| Session1 | Ses01M_script03_2 | 0.018 | 0.210 | -0.003 | 0.814 | 0.078 | 0.022 | 0.134 | 0.000 |
| Session2 | Ses02F_impro01 | 0.006 | 0.760 | -0.051 | 0.005 | -0.256 | 0.000 | -0.015 | 0.728 |
| Session2 | Ses02F_impro02 | -0.088 | 0.000 | -0.046 | 0.013 | -0.432 | 0.000 | -0.409 | 0.000 |
| Session2 | Ses02F_impro03 | 0.102 | 0.000 | 0.048 | 0.001 | -0.077 | 0.046 | -0.048 | 0.212 |
| Session2 | Ses02F_impro04 | 0.052 | 0.001 | -0.003 | 0.822 | -0.144 | 0.009 | -0.049 | 0.378 |
| Session2 | Ses02F_impro05 | 0.023 | 0.185 | -0.095 | 0.000 | 0.018 | 0.694 | 0.019 | 0.682 |
| Session2 | Ses02F_impro06 | -0.071 | 0.003 | -0.062 | 0.010 | -0.403 | 0.153 | -0.320 | 0.264 |
| Session2 | Ses02F_impro07 | -0.002 | 0.927 | -0.032 | 0.065 | -0.048 | 0.051 | 0.063 | 0.011 |
| Session2 | Ses02F_impro08 | 0.000 | 0.993 | -0.005 | 0.740 | -0.093 | 0.036 | -0.057 | 0.199 |
| Session2 | Ses02F_script01_1 | 0.007 | 0.635 | 0.022 | 0.150 | 0.076 | 0.130 | -0.030 | 0.551 |
| Session2 | Ses02F_script01_2 | -0.059 | 0.005 | -0.014 | 0.492 | 0.147 | 0.030 | -0.188 | 0.005 |
| Session2 | Ses02F_script01_3 | 0.033 | 0.056 | -0.004 | 0.838 | -0.118 | 0.411 | -0.097 | 0.498 |
| Session2 | Ses02F_script02_1 | -0.058 | 0.016 | -0.034 | 0.157 | 0.242 | 0.044 | -0.096 | 0.430 |
| Session2 | Ses02F_script02_2 | -0.024 | 0.058 | 0.022 | 0.070 | -0.065 | 0.260 | -0.124 | 0.031 |
| Session2 | Ses02F_script03_1 | -0.032 | 0.099 | 0.009 | 0.652 | -0.046 | 0.376 | 0.098 | 0.059 |
| Session2 | Ses02F_script03_2 | 0.033 | 0.051 | -0.028 | 0.102 | 0.060 | 0.109 | -0.133 | 0.000 |
| Session2 | Ses02M_impro01 | 0.043 | 0.044 | -0.021 | 0.327 | 0.132 | 0.002 | -0.023 | 0.594 |
| Session2 | Ses02M_impro02 | 0.008 | 0.630 | 0.039 | 0.026 | 0.207 | 0.122 | 0.230 | 0.085 |
| Session2 | Ses02M_impro03 | -0.062 | 0.001 | 0.081 | 0.000 | -0.017 | 0.648 | -0.012 | 0.753 |
| Session2 | Ses02M_impro04 | -0.017 | 0.287 | -0.026 | 0.114 | 0.045 | 0.569 | 0.166 | 0.033 |
| Session2 | Ses02M_impro05 | -0.032 | 0.086 | -0.045 | 0.016 | -0.056 | 0.209 | -0.020 | 0.645 |
| Session2 | Ses02M_impro06 | -0.013 | 0.358 | 0.081 | 0.000 | -0.116 | 0.110 | 0.195 | 0.007 |
| Session2 | Ses02M_impro07 | -0.028 | 0.154 | 0.071 | 0.000 | 0.060 | 0.018 | -0.011 | 0.676 |
| Session2 | Ses02M_impro08 | -0.032 | 0.063 | 0.036 | 0.037 | 0.019 | 0.604 | -0.018 | 0.624 |
| Session2 | Ses02M_script01_1 | 0.001 | 0.932 | 0.020 | 0.139 | -0.098 | 0.033 | 0.119 | 0.010 |
| Session2 | Ses02M_script01_2 | -0.011 | 0.703 | 0.013 | 0.660 | 0.019 | 0.727 | 0.021 | 0.707 |
| Session2 | Ses02M_script01_3 | 0.044 | 0.000 | -0.027 | 0.020 | -0.067 | 0.076 | 0.031 | 0.410 |
| Session2 | Ses02M_script02_1 | -0.017 | 0.156 | 0.039 | 0.001 | -0.112 | 0.058 | -0.051 | 0.390 |
| Session2 | Ses02M_script02_2 | 0.031 | 0.027 | 0.008 | 0.561 | -0.042 | 0.286 | -0.003 | 0.929 |
| Session2 | Ses02M_script03_1 | 0.024 | 0.165 | 0.015 | 0.388 | -0.005 | 0.891 | 0.004 | 0.914 |
| Session2 | Ses02M_script03_2 | 0.038 | 0.018 | 0.062 | 0.000 | 0.055 | 0.066 | 0.120 | 0.000 |
| Session3 | Ses03F_impro01 | -0.040 | 0.193 | 0.020 | 0.518 | 0.058 | 0.343 | -0.094 | 0.123 |
| Session3 | Ses03F_impro02 | 0.042 | 0.003 | -0.030 | 0.035 | 0.046 | 0.294 | 0.019 | 0.658 |
| Session3 | Ses03F_impro03 | -0.008 | 0.665 | 0.017 | 0.382 | 0.046 | 0.080 | -0.026 | 0.317 |
| Session3 | Ses03F_impro04 | 0.017 | 0.332 | -0.028 | 0.110 | -0.006 | 0.896 | -0.028 | 0.558 |
| Session3 | Ses03F_impro05 | -0.001 | 0.954 | -0.028 | 0.176 | -0.141 | 0.001 | 0.043 | 0.339 |
| Session3 | Ses03F_impro06 | 0.002 | 0.889 | 0.011 | 0.441 | -0.170 | 0.003 | 0.210 | 0.000 |
| Session3 | Ses03F_impro07 | -0.001 | 0.969 | 0.049 | 0.014 | -0.043 | 0.082 | -0.041 | 0.102 |
| Session3 | Ses03F_impro08 | 0.039 | 0.054 | 0.011 | 0.575 | -0.026 | 0.419 | -0.035 | 0.280 |
| Session3 | Ses03F_script01_1 | -0.026 | 0.047 | 0.039 | 0.003 | 0.066 | 0.026 | -0.044 | 0.140 |
| Session3 | Ses03F_script01_2 | 0.025 | 0.247 | -0.017 | 0.416 | 0.003 | 0.955 | 0.036 | 0.533 |
| Session3 | Ses03F_script01_3 | 0.039 | 0.017 | 0.011 | 0.514 | -0.110 | 0.023 | 0.008 | 0.865 |
| Session3 | Ses03F_script02_1 | 0.060 | 0.007 | 0.040 | 0.079 | -0.041 | 0.448 | 0.171 | 0.001 |
| Session3 | Ses03F_script02_2 | 0.026 | 0.034 | -0.007 | 0.568 | 0.066 | 0.068 | 0.036 | 0.325 |
| Session3 | Ses03F_script03_1 | 0.028 | 0.187 | 0.016 | 0.461 | -0.014 | 0.624 | 0.039 | 0.178 |
| Session3 | Ses03F_script03_2 | -0.033 | 0.037 | -0.079 | 0.000 | -0.071 | 0.088 | 0.025 | 0.542 |
| Session3 | Ses03M_impro01 | -0.028 | 0.195 | -0.005 | 0.808 | -0.050 | 0.100 | -0.008 | 0.791 |
| Session3 | Ses03M_impro02 | -0.074 | 0.000 | -0.081 | 0.000 | -0.008 | 0.741 | 0.036 | 0.157 |
| Session3 | Ses03M_impro03 | -0.006 | 0.711 | 0.019 | 0.248 | -0.008 | 0.730 | 0.030 | 0.177 |
| Session3 | Ses03M_impro04 | -0.026 | 0.080 | -0.076 | 0.000 | 0.001 | 0.972 | -0.054 | 0.140 |
| Session3 | Ses03M_impro05a | -0.012 | 0.569 | 0.033 | 0.109 | -0.144 | 0.001 | -0.069 | 0.114 |
| Session3 | Ses03M_impro05b | 0.013 | 0.422 | 0.097 | 0.000 | -0.027 | 0.367 | 0.057 | 0.060 |
| Session3 | Ses03M_impro06 | 0.019 | 0.235 | 0.039 | 0.014 | -0.274 | 0.000 | -0.195 | 0.002 |
| Session3 | Ses03M_impro07 | 0.031 | 0.143 | -0.017 | 0.416 | 0.012 | 0.784 | -0.020 | 0.652 |
| Session3 | Ses03M_impro08a | -0.043 | 0.020 | -0.023 | 0.206 | 0.066 | 0.021 | 0.026 | 0.365 |
| Session3 | Ses03M_impro08b | 0.024 | 0.214 | -0.022 | 0.251 | -0.072 | 0.030 | 0.031 | 0.359 |
| Session3 | Ses03M_script01_1 | -0.006 | 0.624 | 0.004 | 0.757 | 0.014 | 0.626 | 0.068 | 0.021 |
| Session3 | Ses03M_script01_2 | -0.048 | 0.041 | -0.045 | 0.059 | 0.161 | 0.023 | -0.165 | 0.020 |
| Session3 | Ses03M_script01_3 | 0.025 | 0.038 | 0.017 | 0.150 | 0.093 | 0.004 | -0.003 | 0.929 |
| Session3 | Ses03M_script02_1 | -0.012 | 0.274 | 0.006 | 0.601 | -0.081 | 0.077 | -0.032 | 0.486 |
| Session3 | Ses03M_script02_2 | 0.022 | 0.091 | -0.027 | 0.038 | -0.151 | 0.000 | 0.047 | 0.123 |
| Session3 | Ses03M_script03_1 | 0.011 | 0.453 | -0.039 | 0.009 | -0.015 | 0.762 | -0.060 | 0.217 |
| Session3 | Ses03M_script03_2 | -0.031 | 0.039 | 0.037 | 0.014 | -0.048 | 0.053 | -0.082 | 0.001 |
| Session4 | Ses04F_impro01 | 0.048 | 0.015 | 0.108 | 0.000 | 0.159 | 0.000 | 0.029 | 0.398 |
| Session4 | Ses04F_impro02 | -0.008 | 0.657 | 0.000 | 0.979 | 0.009 | 0.813 | -0.180 | 0.000 |
| Session4 | Ses04F_impro03 | 0.019 | 0.238 | -0.050 | 0.002 | -0.043 | 0.168 | 0.006 | 0.852 |
| Session4 | Ses04F_impro04 | 0.028 | 0.082 | 0.045 | 0.005 | 0.054 | 0.034 | 0.062 | 0.015 |
| Session4 | Ses04F_impro05 | -0.010 | 0.574 | -0.023 | 0.204 | 0.005 | 0.862 | -0.079 | 0.008 |
| Session4 | Ses04F_impro06 | -0.039 | 0.052 | -0.032 | 0.117 | 0.075 | 0.312 | -0.069 | 0.353 |
| Session4 | Ses04F_impro07 | -0.048 | 0.001 | -0.049 | 0.001 | -0.056 | 0.009 | -0.009 | 0.681 |
| Session4 | Ses04F_impro08 | -0.026 | 0.179 | 0.015 | 0.455 | 0.067 | 0.090 | -0.154 | 0.000 |
| Session4 | Ses04F_script01_1 | 0.030 | 0.041 | 0.026 | 0.076 | 0.109 | 0.005 | -0.023 | 0.547 |
| Session4 | Ses04F_script01_2 | 0.017 | 0.398 | -0.006 | 0.783 | -0.016 | 0.776 | -0.102 | 0.062 |
| Session4 | Ses04F_script01_3 | 0.057 | 0.001 | 0.002 | 0.917 | 0.074 | 0.177 | -0.004 | 0.945 |
| Session4 | Ses04F_script02_1 | 0.079 | 0.001 | -0.095 | 0.000 | 0.113 | 0.038 | 0.080 | 0.145 |
| Session4 | Ses04F_script02_2 | 0.053 | 0.000 | 0.063 | 0.000 | 0.003 | 0.945 | -0.005 | 0.895 |
| Session4 | Ses04F_script03_1 | -0.010 | 0.628 | -0.043 | 0.037 | 0.024 | 0.565 | 0.061 | 0.146 |
| Session4 | Ses04F_script03_2 | 0.037 | 0.037 | 0.037 | 0.040 | 0.137 | 0.000 | 0.038 | 0.198 |
| Session4 | Ses04M_impro01 | -0.006 | 0.761 | 0.029 | 0.145 | -0.003 | 0.927 | 0.023 | 0.527 |
| Session4 | Ses04M_impro02 | 0.001 | 0.948 | -0.012 | 0.412 | 0.142 | 0.001 | -0.026 | 0.549 |
| Session4 | Ses04M_impro03 | -0.025 | 0.055 | -0.024 | 0.063 | 0.031 | 0.378 | 0.066 | 0.063 |
| Session4 | Ses04M_impro04 | 0.006 | 0.696 | 0.028 | 0.071 | 0.077 | 0.118 | -0.048 | 0.326 |
| Session4 | Ses04M_impro05 | -0.010 | 0.475 | 0.036 | 0.007 | 0.036 | 0.128 | 0.000 | 0.993 |
| Session4 | Ses04M_impro06 | 0.017 | 0.287 | 0.024 | 0.149 | -0.218 | 0.004 | -0.148 | 0.053 |
| Session4 | Ses04M_impro07 | -0.013 | 0.363 | 0.047 | 0.001 | 0.066 | 0.015 | 0.052 | 0.054 |
| Session4 | Ses04M_impro08 | -0.012 | 0.509 | -0.058 | 0.001 | -0.054 | 0.049 | 0.006 | 0.814 |
| Session4 | Ses04M_script01_1 | -0.004 | 0.793 | -0.010 | 0.466 | -0.004 | 0.919 | -0.075 | 0.054 |
| Session4 | Ses04M_script01_2 | 0.095 | 0.000 | 0.093 | 0.000 | 0.020 | 0.738 | 0.176 | 0.003 |
| Session4 | Ses04M_script01_3 | -0.008 | 0.502 | -0.043 | 0.001 | 0.040 | 0.471 | 0.062 | 0.260 |
| Session4 | Ses04M_script02_1 | 0.018 | 0.132 | 0.010 | 0.380 | -0.174 | 0.004 | -0.051 | 0.407 |
| Session4 | Ses04M_script02_2 | -0.007 | 0.652 | 0.009 | 0.539 | -0.067 | 0.119 | 0.129 | 0.003 |
| Session4 | Ses04M_script03_1 | 0.062 | 0.000 | 0.061 | 0.000 | 0.025 | 0.601 | -0.055 | 0.244 |
| Session4 | Ses04M_script03_2 | -0.031 | 0.048 | 0.019 | 0.222 | 0.002 | 0.951 | -0.043 | 0.171 |
| Session5 | Ses05F_impro01 | -0.032 | 0.133 | 0.076 | 0.000 | -0.173 | 0.000 | 0.187 | 0.000 |
| Session5 | Ses05F_impro02 | -0.017 | 0.266 | 0.001 | 0.939 | -0.042 | 0.370 | 0.030 | 0.513 |
| Session5 | Ses05F_impro03 | 0.043 | 0.016 | 0.020 | 0.265 | -0.001 | 0.969 | 0.074 | 0.005 |
| Session5 | Ses05F_impro04 | 0.008 | 0.554 | 0.006 | 0.663 | 0.031 | 0.377 | 0.062 | 0.076 |
| Session5 | Ses05F_impro05 | 0.016 | 0.292 | 0.002 | 0.888 | 0.034 | 0.195 | 0.054 | 0.038 |
| Session5 | Ses05F_impro06 | -0.060 | 0.001 | 0.010 | 0.577 | -0.124 | 0.100 | -0.323 | 0.000 |
| Session5 | Ses05F_impro07 | 0.019 | 0.384 | 0.012 | 0.584 | 0.003 | 0.936 | 0.020 | 0.606 |
| Session5 | Ses05F_impro08 | 0.076 | 0.000 | 0.002 | 0.885 | 0.117 | 0.002 | 0.168 | 0.000 |
| Session5 | Ses05F_script01_1 | 0.014 | 0.347 | -0.022 | 0.135 | 0.036 | 0.375 | 0.014 | 0.734 |
| Session5 | Ses05F_script01_2 | -0.042 | 0.059 | -0.091 | 0.000 | -0.165 | 0.020 | -0.095 | 0.184 |
| Session5 | Ses05F_script01_3 | -0.001 | 0.950 | -0.026 | 0.132 | -0.159 | 0.116 | -0.026 | 0.798 |
| Session5 | Ses05F_script02_1 | -0.075 | 0.002 | 0.071 | 0.005 | -0.015 | 0.811 | 0.102 | 0.098 |
| Session5 | Ses05F_script02_2 | 0.009 | 0.489 | 0.043 | 0.001 | -0.008 | 0.859 | -0.085 | 0.065 |
| Session5 | Ses05F_script03_1 | 0.029 | 0.144 | -0.002 | 0.901 | -0.004 | 0.926 | -0.149 | 0.000 |
| Session5 | Ses05F_script03_2 | 0.112 | 0.000 | 0.090 | 0.000 | -0.011 | 0.766 | 0.046 | 0.203 |
| Session5 | Ses05M_impro01 | 0.037 | 0.097 | -0.030 | 0.178 | -0.008 | 0.784 | -0.033 | 0.271 |
| Session5 | Ses05M_impro02 | -0.009 | 0.566 | -0.044 | 0.004 | 0.035 | 0.369 | -0.011 | 0.782 |
| Session5 | Ses05M_impro03 | -0.075 | 0.000 | -0.027 | 0.072 | -0.042 | 0.118 | 0.000 | 0.987 |
| Session5 | Ses05M_impro04 | 0.054 | 0.001 | 0.024 | 0.135 | 0.009 | 0.712 | 0.056 | 0.028 |
| Session5 | Ses05M_impro05 | -0.039 | 0.070 | -0.009 | 0.668 | -0.022 | 0.623 | -0.127 | 0.004 |
| Session5 | Ses05M_impro06 | -0.012 | 0.455 | 0.009 | 0.562 | -0.031 | 0.447 | 0.010 | 0.810 |
| Session5 | Ses05M_impro07 | -0.024 | 0.177 | 0.017 | 0.334 | -0.038 | 0.127 | -0.005 | 0.835 |
| Session5 | Ses05M_impro08 | 0.040 | 0.029 | 0.034 | 0.059 | 0.047 | 0.152 | 0.096 | 0.003 |
| Session5 | Ses05M_script01_1 | 0.007 | 0.600 | 0.001 | 0.928 | -0.040 | 0.412 | 0.104 | 0.033 |
| Session5 | Ses05M_script01_1b | -0.016 | 0.217 | 0.016 | 0.191 | -0.096 | 0.054 | -0.026 | 0.596 |
| Session5 | Ses05M_script01_2 | 0.038 | 0.135 | 0.057 | 0.025 | -0.091 | 0.188 | -0.070 | 0.316 |
| Session5 | Ses05M_script01_3 | -0.060 | 0.000 | 0.006 | 0.586 | -0.012 | 0.856 | -0.160 | 0.015 |
| Session5 | Ses05M_script02_1 | 0.010 | 0.373 | -0.006 | 0.591 | 0.064 | 0.247 | 0.022 | 0.694 |
| Session5 | Ses05M_script02_2 | 0.054 | 0.000 | 0.017 | 0.189 | -0.113 | 0.009 | -0.069 | 0.114 |
| Session5 | Ses05M_script03_1 | -0.008 | 0.584 | -0.020 | 0.191 | 0.079 | 0.069 | -0.130 | 0.003 |
| Session5 | Ses05M_script03_2 | 0.097 | 0.000 | 0.046 | 0.004 | 0.054 | 0.100 | 0.042 | 0.195 |

This table presents Pearson correlation coefficients (r) and corresponding p-values for arousal and valence alignment between speech and facial expressions across dyads and sessions, under both non-overlapping and overlapping speech conditions. Each row represents a unique interaction, while columns represent correlation strength and statistical significance. Values are color-coded for emphasis: Lighter colors indicate stronger positive correlations; Darker orange highlights strong negative correlations; and intermediate shades represent weaker or less significant relationships.

At the dyad-level, there was substantial variability across dyads. In the **non-overlapping** condition, for example, **Ses01M_script01_2** showed the **highest arousal correlation** (**r = 0.15,** p < .001), indicating statistically significant alignment (Figure 1a). In contrast, **Ses02F_impro02** showed a **nonsignificant negative correlation** (**r = -0.09,** p < .001), suggesting reduced or absent synchrony (Figure 1b). For **valence**, the strongest alignment was found in **Ses04F_impro01** (**r = 0.11,** p < .001) (Figure 1c), while the weakest occurred in **Ses01F_script02_1** (**r = -0.10, p = 0.003**) (Figure 1d). These findings highlight the influence of dyadic dynamics and conversational context on emotional alignment. In the **overlapping speech** condition, emotional synchrony became more volatile. Notably, in Figure 1e, **Ses01F_impro02** showed **high arousal synchrony** (**r = 0.31, p = 0.001**), suggesting that in some cases, simultaneous speaking may reflect **mutual engagement** and **shared emotional intensity**—possibly functioning as an **affiliative overlap**. Conversely, **Ses02F_impro02** (Figure 1f) exhibited **strong desynchrony** in both **arousal** (**r = -0.43**) and **valence** (**r = -0.41**), indicating potential **emotional misalignment**, perhaps due to **interruptions**, **competition**, or **expressive conflict**. Several scripted sessions showed nonsignificant correlations, likely due to **low emotional variability** or **focus on task execution** rather than spontaneous expression. Thus, the absence of synchrony should not necessarily be interpreted as conflict—it may reflect **emotional regulation**, **neutrality**, or **reduced expressiveness**. Collectively, these dyad-level patterns emphasize the **dual role of overlapping speech**: while typically disruptive to synchrony, it can, under emotionally charged conditions, signal **heightened connection**. These findings underscore the value of integrating global trends with dyad-level nuance and demonstrate how **interaction context, speaker dynamics, and structural speech features** shape emotional alignment in naturalistic conversations.

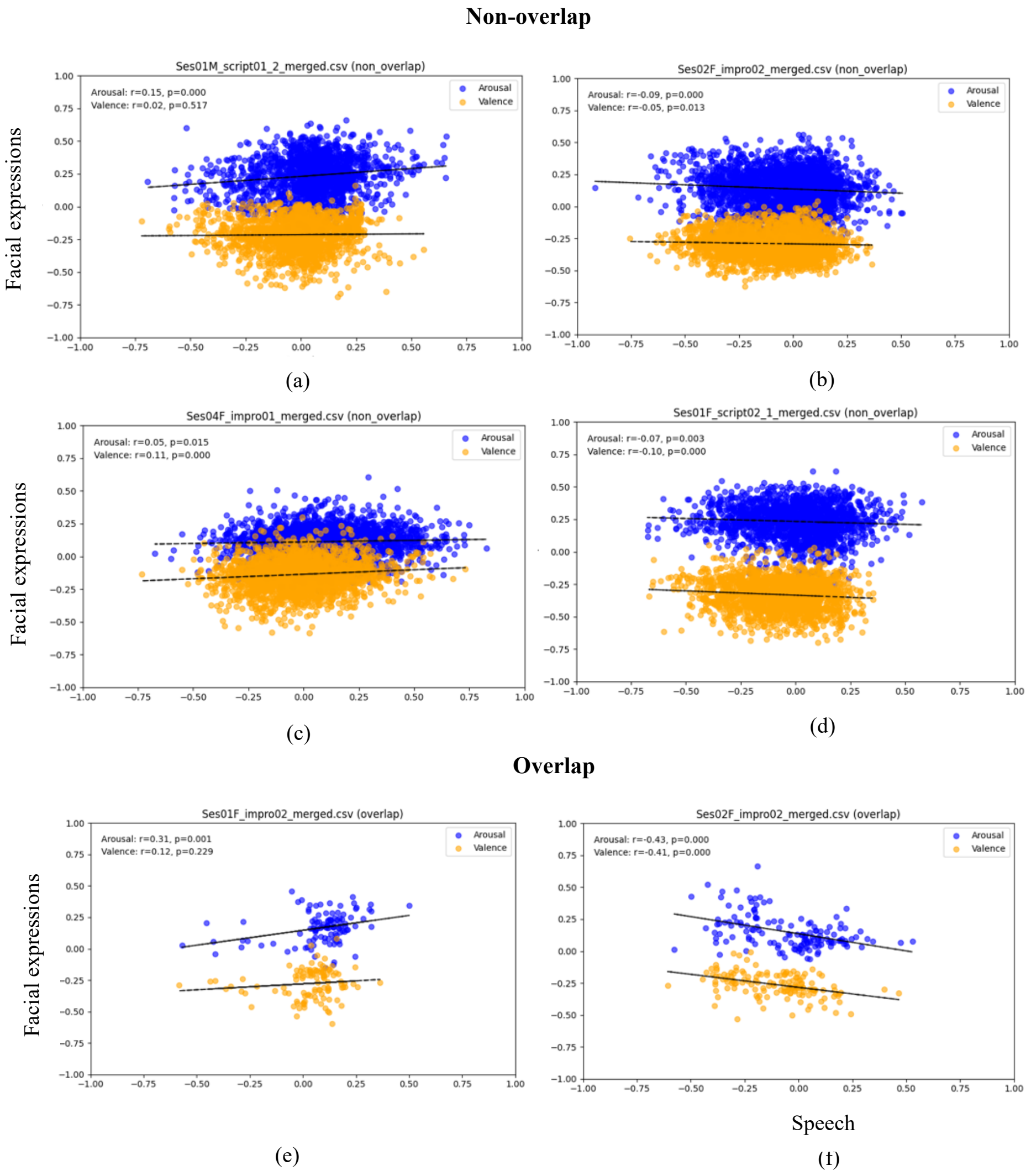

Speech

**Figure 1.** Dyad-level correlation plots illustrating the strongest and weakest Pearson correlations (r) examples between speech-derived and facial-derived arousal and valence signals in both non-overlapping and overlapping speech conditions. Each plot presents a linear trend line fitted to the respective arousal or valence data. The blue and orange points represent speech and facial emotion signals, respectively.

## 2. Temporal analyses

For temporal alignment analysis, we first applied a 5-frame moving average to smooth the arousal and valence time series. This step helped reduce short-term fluctuations and provided a clearer representation of emotional dynamics over time. Subsequently, we performed lag correlation analysis, shifting one modality's emotional signals against the other within a ±10-frame window. This process enabled us to determine the best lag (the time offset at which the emotional signals from speech and facial expressions were most aligned). The lag directionality (whether speech or facial expression leads or lags) was also identified through this analysis. In addition, we used Dynamic Time Warping (DTW) to assess the nonlinear misalignment between the emotional signals of speech and facial expressions. DTW is a robust technique for comparing two time series by finding the optimal alignment between them. It works by stretching or compressing the time series to minimize the distance between corresponding points. All temporal metrics were calculated per dyad and paired t-tests were used to compare condition-level differences (non-overlapping vs. overlapping speech) at the global level. This approach allowed us to retain dyad-level variability and provided a statistically robust means of assessing how interaction structure influences cross-modal emotional alignment across the entire dataset.

### a. Correlation analysis

**Table 2.** Dyad-Level Temporal Correlation and Significance (By Condition)

| Session | Dyads | Condition | Arousal | p-value | Valence | p-value |
|---|---|---|---|---|---|---|
| Session1 | Ses01M_impro06 | non_overlap | -0.047 | 0.016 | -0.050 | 0.000 |
| Session1 | Ses01M_impro06 | overlap | -0.012 | 0.233 | 0.103 | 0.137 |
| Session1 | Ses01F_script02_2 | non_overlap | -0.020 | 0.294 | -0.018 | 0.140 |
| Session1 | Ses01F_script02_2 | overlap | -0.033 | 0.173 | 0.299 | 0.510 |
| Session1 | Ses01F_impro01 | non_overlap | -0.022 | 0.786 | -0.051 | 0.374 |
| Session1 | Ses01F_impro01 | overlap | 0.263 | 0.077 | 0.007 | 0.117 |
| Session1 | Ses01M_impro03 | non_overlap | -0.055 | 0.005 | -0.141 | 0.000 |
| Session1 | Ses01M_impro03 | overlap | 0.105 | 0.021 | 0.121 | 0.000 |
| Session1 | Ses01F_impro04 | non_overlap | -0.023 | 0.471 | 0.172 | 0.000 |
| Session1 | Ses01F_impro04 | overlap | -0.115 | 0.045 | 0.124 | 0.000 |
| Session1 | Ses01F_script01_1 | non_overlap | 0.014 | 0.379 | 0.095 | 0.000 |
| Session1 | Ses01F_script01_1 | overlap | 0.125 | 0.000 | 0.219 | 0.825 |
| Session1 | Ses01F_script03_2 | non_overlap | 0.055 | 0.000 | 0.008 | 0.136 |
| Session1 | Ses01F_script03_2 | overlap | 0.090 | 0.773 | 0.152 | 0.304 |
| Session1 | Ses01F_impro03 | non_overlap | 0.046 | 0.601 | -0.048 | 0.757 |
| Session1 | Ses01F_impro03 | overlap | 0.035 | 0.616 | 0.093 | 0.520 |
| Session1 | Ses01M_script02_1 | non_overlap | -0.015 | 0.014 | 0.028 | 0.728 |
| Session1 | Ses01M_script02_1 | overlap | 0.082 | 0.012 | -0.036 | 0.519 |
| Session1 | Ses01M_impro04 | non_overlap | 0.009 | 0.325 | -0.023 | 0.931 |
| Session1 | Ses01M_impro04 | overlap | -0.239 | 0.415 | -0.086 | 0.259 |
| Session1 | Ses01F_impro06 | non_overlap | -0.054 | 0.798 | 0.076 | 0.001 |
| Session1 | Ses01F_impro06 | overlap | 0.165 | 0.685 | 0.217 | 0.000 |
| Session1 | Ses01M_impro01 | non_overlap | -0.008 | 0.003 | 0.046 | 0.998 |
| Session1 | Ses01M_impro01 | overlap | 0.149 | 0.245 | 0.064 | 0.056 |
| Session1 | Ses01M_script01_2 | non_overlap | 0.254 | 0.000 | 0.019 | 0.654 |
| Session1 | Ses01M_script01_2 | overlap | -0.083 | 0.786 | -0.242 | 0.186 |
| Session1 | Ses01M_script03_1 | non_overlap | 0.049 | 0.389 | 0.048 | 0.008 |
| Session1 | Ses01M_script03_1 | overlap | -0.019 | 0.021 | 0.074 | 0.516 |
| Session1 | Ses01F_script01_3 | non_overlap | 0.013 | 0.230 | 0.074 | 0.015 |
| Session1 | Ses01F_script01_3 | overlap | 0.150 | 0.078 | -0.185 | 0.414 |
| Session1 | Ses01M_impro02 | non_overlap | 0.068 | 0.504 | -0.026 | 0.003 |
| Session1 | Ses01M_impro02 | overlap | 0.141 | 0.003 | -0.173 | 0.374 |
| Session1 | Ses01F_impro05 | non_overlap | -0.086 | 0.033 | 0.100 | 0.169 |
| Session1 | Ses01F_impro05 | overlap | 0.051 | 0.052 | -0.150 | 0.120 |
| Session1 | Ses01M_script03_2 | non_overlap | 0.030 | 0.063 | -0.010 | 0.148 |
| Session1 | Ses01M_script03_2 | overlap | 0.115 | 0.498 | 0.172 | 0.139 |
| Session1 | Ses01M_script01_1 | non_overlap | 0.087 | 0.000 | -0.026 | 0.184 |
| Session1 | Ses01M_script01_1 | overlap | -0.164 | 0.070 | -0.182 | 0.453 |
| Session1 | Ses01M_impro07 | non_overlap | -0.034 | 0.158 | 0.055 | 0.052 |
| Session1 | Ses01M_impro07 | overlap | 0.047 | 0.001 | 0.071 | 0.103 |
| Session1 | Ses01M_script02_2 | non_overlap | -0.054 | 0.012 | 0.149 | 0.000 |
| Session1 | Ses01M_script02_2 | overlap | -0.096 | 0.234 | 0.027 | 0.973 |
| Session1 | Ses01F_impro07 | non_overlap | -0.050 | 0.349 | 0.077 | 0.064 |
| Session1 | Ses01F_impro07 | overlap | -0.114 | 0.787 | 0.045 | 0.014 |
| Session1 | Ses01F_script03_1 | non_overlap | 0.082 | 0.001 | -0.070 | 0.115 |
| Session1 | Ses01F_script03_1 | overlap | 0.159 | 0.000 | -0.038 | 0.327 |
| Session1 | Ses01M_script01_3 | non_overlap | 0.039 | 0.003 | 0.075 | 0.000 |
| Session1 | Ses01M_script01_3 | overlap | -0.039 | 0.182 | 0.206 | 0.356 |
| Session1 | Ses01F_script01_2 | non_overlap | 0.051 | 0.134 | -0.030 | 0.036 |
| Session1 | Ses01F_script01_2 | overlap | 0.033 | 0.008 | 0.266 | 0.010 |
| Session1 | Ses01F_impro02 | non_overlap | 0.152 | 0.000 | 0.027 | 0.016 |
| Session1 | Ses01F_impro02 | overlap | 0.412 | 0.345 | 0.154 | 0.023 |
| Session1 | Ses01F_script02_1 | non_overlap | -0.082 | 0.000 | -0.152 | 0.669 |
| Session1 | Ses01F_script02_1 | overlap | -0.138 | 0.000 | -0.170 | 0.400 |
| Session1 | Ses01M_impro05 | non_overlap | -0.016 | 0.225 | 0.107 | 0.026 |
| Session1 | Ses01M_impro05 | overlap | 0.015 | 0.104 | 0.037 | 0.407 |
| Session2 | Ses02F_script01_1 | non_overlap | 0.015 | 0.187 | 0.046 | 0.001 |
| Session2 | Ses02F_script01_1 | overlap | 0.098 | 0.000 | -0.078 | 0.198 |
| Session2 | Ses02F_impro01 | non_overlap | 0.013 | 0.012 | -0.074 | 0.000 |
| Session2 | Ses02F_impro01 | overlap | -0.349 | 0.000 | -0.120 | 0.480 |
| Session2 | Ses02F_script03_2 | non_overlap | 0.070 | 0.000 | -0.051 | 0.034 |
| Session2 | Ses02F_script03_2 | overlap | 0.079 | 0.927 | -0.153 | 0.002 |
| Session2 | Ses02M_impro06 | non_overlap | -0.012 | 0.894 | 0.110 | 0.000 |
| Session2 | Ses02M_impro06 | overlap | -0.199 | 0.200 | 0.307 | 0.923 |
| Session2 | Ses02F_script02_2 | non_overlap | -0.031 | 0.199 | 0.042 | 0.010 |
| Session2 | Ses02F_script02_2 | overlap | -0.084 | 0.109 | -0.203 | 0.176 |
| Session2 | Ses02F_impro04 | non_overlap | 0.070 | 0.001 | -0.002 | 0.235 |
| Session2 | Ses02F_impro04 | overlap | -0.289 | 0.191 | -0.038 | 0.284 |
| Session2 | Ses02M_impro03 | non_overlap | -0.111 | 0.000 | 0.126 | 0.002 |
| Session2 | Ses02M_impro03 | overlap | -0.031 | 0.862 | -0.040 | 0.152 |
| Session2 | Ses02M_impro04 | non_overlap | -0.037 | 0.109 | -0.026 | 0.113 |
| Session2 | Ses02M_impro04 | overlap | 0.036 | 0.089 | 0.230 | 0.001 |
| Session2 | Ses02M_script01_2 | non_overlap | 0.000 | 0.747 | 0.047 | 0.479 |
| Session2 | Ses02M_script01_2 | overlap | 0.099 | 0.178 | 0.045 | 0.091 |
| Session2 | Ses02F_script01_3 | non_overlap | 0.055 | 0.968 | 0.011 | 0.091 |
| Session2 | Ses02F_script01_3 | overlap | -0.543 | 0.831 | -0.061 | 0.462 |
| Session2 | Ses02M_script03_1 | non_overlap | 0.035 | 0.237 | 0.021 | 0.308 |
| Session2 | Ses02M_script03_1 | overlap | -0.051 | 0.964 | 0.035 | 0.840 |
| Session2 | Ses02F_impro03 | non_overlap | 0.166 | 0.000 | 0.084 | 0.000 |
| Session2 | Ses02F_impro03 | overlap | -0.169 | 0.002 | -0.077 | 0.026 |
| Session2 | Ses02M_script02_1 | non_overlap | -0.033 | 0.352 | 0.063 | 0.777 |
| Session2 | Ses02M_script02_1 | overlap | -0.097 | 0.497 | -0.182 | 0.120 |
| Session2 | Ses02M_impro01 | non_overlap | 0.066 | 0.338 | -0.035 | 0.931 |
| Session2 | Ses02M_impro01 | overlap | 0.232 | 0.000 | -0.041 | 0.118 |
| Session2 | Ses02F_impro06 | non_overlap | -0.103 | 0.000 | -0.096 | 0.001 |
| Session2 | Ses02F_impro06 | overlap | -0.134 | 0.808 | -0.457 | 0.128 |
| Session2 | Ses02F_impro08 | non_overlap | -0.027 | 0.026 | -0.020 | 0.006 |
| Session2 | Ses02F_impro08 | overlap | -0.154 | 0.018 | -0.118 | 0.013 |
| Session2 | Ses02M_script02_2 | non_overlap | 0.054 | 0.447 | 0.015 | 0.594 |
| Session2 | Ses02M_script02_2 | overlap | -0.088 | 0.616 | -0.025 | 0.426 |
| Session2 | Ses02F_impro05 | non_overlap | 0.026 | 0.000 | -0.134 | 0.210 |
| Session2 | Ses02F_impro05 | overlap | 0.037 | 0.190 | 0.037 | 0.345 |
| Session2 | Ses02M_impro02 | non_overlap | 0.010 | 0.719 | 0.064 | 0.009 |
| Session2 | Ses02M_impro02 | overlap | 0.243 | 0.117 | 0.347 | 0.925 |
| Session2 | Ses02M_script03_2 | non_overlap | 0.079 | 0.178 | 0.088 | 0.007 |
| Session2 | Ses02M_script03_2 | overlap | 0.081 | 0.312 | 0.191 | 0.005 |
| Session2 | Ses02M_impro07 | non_overlap | -0.015 | 0.958 | 0.097 | 0.339 |
| Session2 | Ses02M_impro07 | overlap | 0.084 | 0.018 | -0.006 | 0.855 |
| Session2 | Ses02M_script01_1 | non_overlap | -0.002 | 0.185 | 0.041 | 0.282 |
| Session2 | Ses02M_script01_1 | overlap | -0.183 | 0.633 | 0.252 | 0.749 |
| Session2 | Ses02F_script02_1 | non_overlap | -0.114 | 0.000 | -0.060 | 0.991 |
| Session2 | Ses02F_script02_1 | overlap | 0.398 | 0.034 | -0.188 | 0.194 |
| Session2 | Ses02F_impro07 | non_overlap | -0.019 | 0.156 | -0.073 | 0.000 |
| Session2 | Ses02F_impro07 | overlap | -0.077 | 0.277 | 0.103 | 0.114 |
| Session2 | Ses02M_script01_3 | non_overlap | 0.062 | 0.183 | -0.018 | 0.031 |
| Session2 | Ses02M_script01_3 | overlap | -0.156 | 0.021 | 0.037 | 0.409 |
| Session2 | Ses02M_impro05 | non_overlap | -0.048 | 0.122 | -0.074 | 0.001 |
| Session2 | Ses02M_impro05 | overlap | -0.080 | 0.471 | 0.003 | 0.002 |
| Session2 | Ses02F_script03_1 | non_overlap | -0.079 | 0.379 | -0.017 | 0.092 |
| Session2 | Ses02F_script03_1 | overlap | -0.127 | 0.106 | 0.074 | 0.227 |
| Session2 | Ses02F_impro02 | non_overlap | -0.136 | 0.000 | -0.083 | 0.126 |
| Session2 | Ses02F_impro02 | overlap | -0.581 | 0.000 | -0.485 | 0.009 |
| Session2 | Ses02F_script01_2 | non_overlap | -0.087 | 0.016 | -0.017 | 0.370 |
| Session2 | Ses02F_script01_2 | overlap | 0.254 | 0.189 | -0.266 | 0.027 |
| Session2 | Ses02M_impro08 | non_overlap | -0.045 | 0.007 | 0.056 | 0.353 |
| Session2 | Ses02M_impro08 | overlap | 0.012 | 0.660 | 0.020 | 0.492 |
| Session3 | Ses03M_impro06 | non_overlap | 0.024 | 0.128 | 0.067 | 0.002 |
| Session3 | Ses03M_impro06 | overlap | -0.360 | 0.003 | -0.293 | 0.022 |
| Session3 | Ses03F_impro01 | non_overlap | -0.075 | 0.766 | 0.023 | 0.065 |
| Session3 | Ses03F_impro01 | overlap | 0.162 | 0.029 | -0.248 | 0.000 |
| Session3 | Ses03M_script02_1 | non_overlap | -0.029 | 0.937 | 0.022 | 0.034 |
| Session3 | Ses03M_script02_1 | overlap | -0.147 | 0.631 | -0.057 | 0.054 |
| Session3 | Ses03M_impro03 | non_overlap | -0.019 | 0.418 | 0.028 | 0.182 |
| Session3 | Ses03M_impro03 | overlap | -0.015 | 0.494 | 0.046 | 0.998 |
| Session3 | Ses03M_script03_1 | non_overlap | 0.015 | 0.090 | -0.055 | 0.029 |
| Session3 | Ses03M_script03_1 | overlap | 0.030 | 0.183 | -0.066 | 0.048 |
| Session3 | Ses03F_script01_3 | non_overlap | 0.057 | 0.115 | 0.009 | 0.962 |
| Session3 | Ses03F_script01_3 | overlap | -0.229 | 0.067 | 0.076 | 0.032 |
| Session3 | Ses03M_script01_2 | non_overlap | -0.050 | 0.368 | -0.041 | 0.774 |
| Session3 | Ses03M_script01_2 | overlap | 0.309 | 0.050 | -0.264 | 0.612 |
| Session3 | Ses03F_impro04 | non_overlap | 0.028 | 0.000 | -0.061 | 0.000 |
| Session3 | Ses03F_impro04 | overlap | -0.092 | 0.017 | 0.004 | 0.846 |
| Session3 | Ses03M_impro05a | non_overlap | -0.011 | 0.324 | 0.064 | 0.490 |
| Session3 | Ses03M_impro05a | overlap | -0.196 | 0.041 | -0.092 | 0.030 |
| Session3 | Ses03F_impro03 | non_overlap | -0.024 | 0.797 | 0.029 | 0.888 |
| Session3 | Ses03F_impro03 | overlap | 0.066 | 0.126 | -0.055 | 0.945 |
| Session3 | Ses03M_impro08b | non_overlap | 0.006 | 0.146 | -0.018 | 0.054 |
| Session3 | Ses03M_impro08b | overlap | -0.111 | 0.144 | 0.075 | 0.001 |
| Session3 | Ses03M_impro04 | non_overlap | -0.031 | 0.906 | -0.100 | 0.000 |
| Session3 | Ses03M_impro04 | overlap | 0.025 | 0.742 | -0.114 | 0.190 |
| Session3 | Ses03F_script02_2 | non_overlap | 0.038 | 0.994 | -0.007 | 0.448 |
| Session3 | Ses03F_script02_2 | overlap | 0.100 | 0.111 | 0.102 | 0.738 |
| Session3 | Ses03F_impro06 | non_overlap | -0.013 | 0.064 | 0.019 | 0.388 |
| Session3 | Ses03F_impro06 | overlap | -0.223 | 0.629 | 0.311 | 0.390 |
| Session3 | Ses03F_script03_2 | non_overlap | -0.049 | 0.140 | -0.135 | 0.000 |
| Session3 | Ses03F_script03_2 | overlap | -0.117 | 0.027 | 0.040 | 0.003 |
| Session3 | Ses03M_impro01 | non_overlap | -0.040 | 0.013 | -0.033 | 0.123 |
| Session3 | Ses03M_impro01 | overlap | -0.077 | 0.802 | 0.040 | 0.305 |
| Session3 | Ses03F_script01_1 | non_overlap | -0.042 | 0.089 | 0.068 | 0.003 |
| Session3 | Ses03F_script01_1 | overlap | 0.112 | 0.282 | -0.064 | 0.961 |
| Session3 | Ses03F_impro08 | non_overlap | 0.051 | 0.001 | 0.033 | 0.393 |
| Session3 | Ses03F_impro08 | overlap | -0.027 | 0.533 | -0.071 | 0.716 |
| Session3 | Ses03F_script01_2 | non_overlap | 0.034 | 0.000 | -0.010 | 0.859 |
| Session3 | Ses03F_script01_2 | overlap | -0.005 | 0.026 | 0.125 | 0.797 |
| Session3 | Ses03M_impro02 | non_overlap | -0.118 | 0.000 | -0.118 | 0.068 |
| Session3 | Ses03M_impro02 | overlap | -0.019 | 0.330 | 0.052 | 0.171 |
| Session3 | Ses03F_script03_1 | non_overlap | 0.048 | 0.403 | 0.047 | 0.408 |
| Session3 | Ses03F_script03_1 | overlap | -0.025 | 0.992 | 0.028 | 0.119 |
| Session3 | Ses03M_script01_3 | non_overlap | 0.029 | 0.369 | 0.037 | 0.013 |
| Session3 | Ses03M_script01_3 | overlap | 0.142 | 0.000 | -0.005 | 0.022 |
| Session3 | Ses03F_impro05 | non_overlap | 0.002 | 0.915 | -0.041 | 0.087 |
| Session3 | Ses03F_impro05 | overlap | -0.265 | 0.000 | 0.042 | 0.086 |
| Session3 | Ses03M_impro07 | non_overlap | 0.048 | 0.894 | -0.055 | 0.334 |
| Session3 | Ses03M_impro07 | overlap | 0.081 | 0.003 | 0.033 | 0.024 |
| Session3 | Ses03M_impro08a | non_overlap | -0.054 | 0.028 | -0.019 | 0.903 |
| Session3 | Ses03M_impro08a | overlap | 0.102 | 0.383 | 0.030 | 0.203 |
| Session3 | Ses03F_script02_1 | non_overlap | 0.108 | 0.003 | 0.065 | 0.044 |
| Session3 | Ses03F_script02_1 | overlap | -0.042 | 0.418 | 0.307 | 0.887 |
| Session3 | Ses03M_script01_1 | non_overlap | 0.006 | 0.583 | 0.002 | 0.299 |
| Session3 | Ses03M_script01_1 | overlap | 0.011 | 0.532 | 0.082 | 0.056 |
| Session3 | Ses03F_impro07 | non_overlap | 0.010 | 0.135 | 0.090 | 0.733 |
| Session3 | Ses03F_impro07 | overlap | -0.082 | 0.722 | -0.056 | 0.180 |
| Session3 | Ses03M_impro05b | non_overlap | 0.009 | 0.791 | 0.143 | 0.001 |
| Session3 | Ses03M_impro05b | overlap | 0.000 | 0.231 | 0.066 | 0.101 |
| Session3 | Ses03M_script03_2 | non_overlap | -0.040 | 0.345 | 0.042 | 0.020 |
| Session3 | Ses03M_script03_2 | overlap | -0.080 | 0.008 | -0.134 | 0.000 |
| Session3 | Ses03F_impro02 | non_overlap | 0.066 | 0.062 | -0.030 | 0.463 |
| Session3 | Ses03F_impro02 | overlap | 0.076 | 0.800 | -0.049 | 0.092 |
| Session3 | Ses03M_script02_2 | non_overlap | 0.036 | 0.700 | -0.038 | 0.263 |
| Session3 | Ses03M_script02_2 | overlap | -0.214 | 0.223 | 0.075 | 0.227 |
| Session4 | Ses04M_impro03 | non_overlap | -0.034 | 0.795 | -0.046 | 0.350 |
| Session4 | Ses04M_impro03 | overlap | -0.005 | 0.044 | 0.131 | 0.252 |
| Session4 | Ses04M_script02_1 | non_overlap | 0.009 | 0.486 | 0.018 | 0.950 |
| Session4 | Ses04M_script02_1 | overlap | -0.209 | 0.443 | -0.067 | 0.348 |
| Session4 | Ses04F_impro04 | non_overlap | 0.037 | 0.059 | 0.056 | 0.995 |
| Session4 | Ses04F_impro04 | overlap | 0.082 | 0.959 | 0.079 | 0.001 |
| Session4 | Ses04M_script03_1 | non_overlap | 0.107 | 0.101 | 0.138 | 0.000 |
| Session4 | Ses04M_script03_1 | overlap | 0.012 | 0.960 | -0.130 | 0.326 |
| Session4 | Ses04F_script01_3 | non_overlap | 0.067 | 0.263 | 0.037 | 0.121 |
| Session4 | Ses04F_script01_3 | overlap | 0.073 | 0.151 | 0.041 | 0.129 |
| Session4 | Ses04M_script01_2 | non_overlap | 0.149 | 0.337 | 0.162 | 0.000 |
| Session4 | Ses04M_script01_2 | overlap | 0.109 | 0.759 | 0.213 | 0.003 |
| Session4 | Ses04M_impro06 | non_overlap | 0.023 | 0.095 | 0.039 | 0.414 |
| Session4 | Ses04M_impro06 | overlap | -0.351 | 0.085 | -0.229 | 0.940 |
| Session4 | Ses04F_impro01 | non_overlap | 0.076 | 0.858 | 0.172 | 0.000 |
| Session4 | Ses04F_impro01 | overlap | 0.243 | 0.004 | 0.015 | 0.402 |
| Session4 | Ses04F_impro06 | non_overlap | -0.063 | 0.113 | -0.051 | 0.305 |
| Session4 | Ses04F_impro06 | overlap | 0.191 | 0.011 | -0.012 | 0.363 |
| Session4 | Ses04M_impro01 | non_overlap | 0.019 | 0.315 | 0.047 | 0.047 |
| Session4 | Ses04M_impro01 | overlap | 0.013 | 0.495 | 0.032 | 0.984 |
| Session4 | Ses04F_script02_2 | non_overlap | 0.084 | 0.000 | 0.089 | 0.004 |
| Session4 | Ses04F_script02_2 | overlap | -0.004 | 0.429 | 0.076 | 0.405 |
| Session4 | Ses04F_script03_2 | non_overlap | 0.051 | 0.425 | 0.075 | 0.890 |
| Session4 | Ses04F_script03_2 | overlap | 0.187 | 0.749 | 0.057 | 0.001 |
| Session4 | Ses04F_script01_1 | non_overlap | 0.046 | 0.085 | 0.036 | 0.339 |
| Session4 | Ses04F_script01_1 | overlap | 0.122 | 0.559 | -0.056 | 0.769 |
| Session4 | Ses04F_impro03 | non_overlap | 0.017 | 0.003 | -0.057 | 0.002 |
| Session4 | Ses04F_impro03 | overlap | -0.140 | 0.292 | 0.049 | 0.491 |
| Session4 | Ses04M_impro04 | non_overlap | 0.017 | 0.722 | 0.042 | 0.504 |
| Session4 | Ses04M_impro04 | overlap | 0.141 | 0.058 | -0.116 | 0.908 |
| Session4 | Ses04F_script01_2 | non_overlap | 0.040 | 0.080 | -0.028 | 0.010 |
| Session4 | Ses04F_script01_2 | overlap | 0.048 | 0.110 | -0.167 | 0.074 |
| Session4 | Ses04F_script03_1 | non_overlap | -0.027 | 0.013 | -0.083 | 0.749 |
| Session4 | Ses04F_script03_1 | overlap | 0.073 | 0.643 | 0.066 | 0.295 |
| Session4 | Ses04M_script01_3 | non_overlap | -0.021 | 0.213 | -0.078 | 0.000 |
| Session4 | Ses04M_script01_3 | overlap | 0.108 | 0.276 | 0.082 | 0.060 |
| Session4 | Ses04M_impro07 | non_overlap | -0.031 | 0.033 | 0.079 | 0.013 |
| Session4 | Ses04M_impro07 | overlap | 0.087 | 0.376 | 0.113 | 0.023 |
| Session4 | Ses04F_script02_1 | non_overlap | 0.133 | 0.001 | -0.147 | 0.783 |
| Session4 | Ses04F_script02_1 | overlap | 0.189 | 0.817 | 0.172 | 0.000 |
| Session4 | Ses04M_impro02 | non_overlap | 0.000 | 0.008 | -0.010 | 0.203 |
| Session4 | Ses04M_impro02 | overlap | 0.219 | 0.646 | -0.019 | 0.499 |
| Session4 | Ses04F_impro05 | non_overlap | -0.014 | 0.516 | -0.049 | 0.154 |
| Session4 | Ses04F_impro05 | overlap | -0.008 | 0.281 | -0.146 | 0.029 |
| Session4 | Ses04F_impro08 | non_overlap | -0.039 | 0.647 | 0.041 | 0.021 |
| Session4 | Ses04F_impro08 | overlap | 0.108 | 0.266 | -0.297 | 0.607 |
| Session4 | Ses04M_script01_1 | non_overlap | -0.013 | 0.007 | -0.013 | 0.074 |
| Session4 | Ses04M_script01_1 | overlap | -0.008 | 0.433 | -0.142 | 0.044 |
| Session4 | Ses04M_impro08 | non_overlap | -0.005 | 0.788 | -0.092 | 0.000 |
| Session4 | Ses04M_impro08 | overlap | -0.043 | 0.211 | 0.005 | 0.684 |
| Session4 | Ses04M_script03_2 | non_overlap | -0.041 | 0.011 | 0.028 | 0.275 |
| Session4 | Ses04M_script03_2 | overlap | -0.012 | 0.061 | -0.078 | 0.012 |
| Session4 | Ses04F_impro02 | non_overlap | -0.024 | 0.491 | 0.014 | 0.971 |
| Session4 | Ses04F_impro02 | overlap | -0.018 | 0.355 | -0.288 | 0.000 |
| Session4 | Ses04M_impro05 | non_overlap | 0.003 | 0.261 | 0.074 | 0.208 |
| Session4 | Ses04M_impro05 | overlap | 0.067 | 0.118 | 0.012 | 0.857 |
| Session4 | Ses04M_script02_2 | non_overlap | -0.009 | 0.162 | 0.019 | 0.481 |
| Session4 | Ses04M_script02_2 | overlap | -0.115 | 0.047 | 0.136 | 0.139 |
| Session4 | Ses04F_impro07 | non_overlap | -0.068 | 0.007 | -0.082 | 0.719 |
| Session4 | Ses04F_impro07 | overlap | -0.099 | 0.299 | 0.000 | 0.899 |
| Session5 | Ses05F_impro04 | non_overlap | -0.011 | 0.617 | 0.021 | 0.034 |
| Session5 | Ses05F_impro04 | overlap | 0.037 | 0.295 | 0.119 | 0.202 |
| Session5 | Ses05M_impro03 | non_overlap | -0.130 | 0.000 | -0.049 | 0.166 |
| Session5 | Ses05M_impro03 | overlap | -0.085 | 0.349 | -0.003 | 0.044 |
| Session5 | Ses05F_script01_1 | non_overlap | 0.028 | 0.079 | -0.029 | 0.075 |
| Session5 | Ses05F_script01_1 | overlap | 0.043 | 0.052 | -0.003 | 0.892 |
| Session5 | Ses05F_script03_2 | non_overlap | 0.178 | 0.000 | 0.139 | 0.000 |
| Session5 | Ses05F_script03_2 | overlap | -0.043 | 0.153 | 0.061 | 0.031 |
| Session5 | Ses05F_script02_2 | non_overlap | 0.010 | 0.486 | 0.071 | 0.115 |
| Session5 | Ses05F_script02_2 | overlap | -0.010 | 0.859 | -0.108 | 0.080 |
| Session5 | Ses05F_impro01 | non_overlap | -0.061 | 0.084 | 0.111 | 0.008 |
| Session5 | Ses05F_impro01 | overlap | -0.296 | 0.028 | 0.358 | 0.004 |
| Session5 | Ses05M_impro06 | non_overlap | -0.038 | 0.295 | 0.036 | 0.098 |
| Session5 | Ses05M_impro06 | overlap | -0.034 | 0.508 | -0.045 | 0.727 |
| Session5 | Ses05M_impro01 | non_overlap | 0.043 | 0.973 | -0.040 | 0.125 |
| Session5 | Ses05M_impro01 | overlap | 0.022 | 0.733 | -0.034 | 0.122 |
| Session5 | Ses05F_impro06 | non_overlap | -0.088 | 0.002 | 0.007 | 0.175 |
| Session5 | Ses05F_impro06 | overlap | -0.171 | 0.061 | -0.445 | 0.416 |
| Session5 | Ses05M_script01_2 | non_overlap | 0.079 | 0.171 | 0.094 | 0.034 |
| Session5 | Ses05M_script01_2 | overlap | -0.104 | 0.372 | -0.114 | 0.783 |
| Session5 | Ses05F_script01_3 | non_overlap | -0.005 | 0.798 | -0.036 | 0.862 |
| Session5 | Ses05F_script01_3 | overlap | -0.223 | 0.431 | -0.017 | 0.540 |
| Session5 | Ses05M_script03_1 | non_overlap | -0.015 | 0.008 | -0.033 | 0.003 |
| Session5 | Ses05M_script03_1 | overlap | 0.128 | 0.280 | -0.211 | 0.023 |
| Session5 | Ses05M_script01_1b | non_overlap | -0.032 | 0.486 | 0.020 | 0.229 |
| Session5 | Ses05M_script01_1b | overlap | -0.053 | 0.955 | -0.013 | 0.577 |
| Session5 | Ses05M_impro04 | non_overlap | 0.087 | 0.001 | 0.047 | 0.001 |
| Session5 | Ses05M_impro04 | overlap | 0.025 | 0.017 | 0.079 | 0.066 |
| Session5 | Ses05M_script02_1 | non_overlap | -0.003 | 0.309 | -0.017 | 0.416 |
| Session5 | Ses05M_script02_1 | overlap | 0.119 | 0.247 | -0.024 | 0.725 |
| Session5 | Ses05F_impro03 | non_overlap | 0.077 | 0.319 | 0.032 | 0.635 |
| Session5 | Ses05F_impro03 | overlap | 0.018 | 0.043 | 0.158 | 0.151 |
| Session5 | Ses05M_script02_2 | non_overlap | 0.077 | 0.020 | 0.042 | 0.004 |
| Session5 | Ses05M_script02_2 | overlap | -0.191 | 0.833 | -0.132 | 0.376 |
| Session5 | Ses05M_impro07 | non_overlap | -0.051 | 0.004 | 0.034 | 0.588 |
| Session5 | Ses05M_impro07 | overlap | -0.060 | 0.070 | -0.013 | 0.420 |
| Session5 | Ses05F_impro05 | non_overlap | 0.024 | 0.108 | 0.000 | 0.923 |
| Session5 | Ses05F_impro05 | overlap | 0.027 | 0.445 | 0.068 | 0.038 |
| Session5 | Ses05M_impro02 | non_overlap | 0.004 | 0.101 | -0.072 | 0.245 |
| Session5 | Ses05M_impro02 | overlap | 0.007 | 0.483 | 0.016 | 0.682 |
| Session5 | Ses05F_impro08 | non_overlap | 0.115 | 0.006 | 0.010 | 0.719 |
| Session5 | Ses05F_impro08 | overlap | 0.177 | 0.570 | 0.261 | 0.001 |
| Session5 | Ses05M_script03_2 | non_overlap | 0.144 | 0.000 | 0.090 | 0.000 |
| Session5 | Ses05M_script03_2 | overlap | 0.091 | 0.000 | 0.072 | 0.061 |
| Session5 | Ses05M_script01_1 | non_overlap | 0.027 | 0.414 | 0.027 | 0.214 |
| Session5 | Ses05M_script01_1 | overlap | -0.038 | 0.010 | 0.206 | 0.871 |
| Session5 | Ses05M_impro08 | non_overlap | 0.075 | 0.010 | 0.073 | 0.001 |
| Session5 | Ses05M_impro08 | overlap | 0.078 | 0.201 | 0.146 | 0.125 |
| Session5 | Ses05M_impro05 | non_overlap | -0.045 | 0.829 | -0.018 | 0.005 |
| Session5 | Ses05M_impro05 | overlap | -0.036 | 0.262 | -0.160 | 0.424 |
| Session5 | Ses05F_impro02 | non_overlap | -0.018 | 0.104 | 0.004 | 0.039 |
| Session5 | Ses05F_impro02 | overlap | -0.046 | 0.011 | 0.113 | 0.078 |
| Session5 | Ses05F_script02_1 | non_overlap | -0.131 | 0.000 | 0.139 | 0.000 |
| Session5 | Ses05F_script02_1 | overlap | 0.012 | 0.267 | 0.196 | 0.804 |
| Session5 | Ses05F_impro07 | non_overlap | 0.019 | 0.027 | 0.015 | 0.856 |
| Session5 | Ses05F_impro07 | overlap | 0.026 | 0.024 | 0.057 | 0.717 |
| Session5 | Ses05M_script01_3 | non_overlap | -0.101 | 0.000 | 0.001 | 0.347 |
| Session5 | Ses05M_script01_3 | overlap | -0.029 | 0.142 | -0.144 | 0.827 |
| Session5 | Ses05F_script03_1 | non_overlap | 0.050 | 0.723 | -0.005 | 0.080 |
| Session5 | Ses05F_script03_1 | overlap | 0.002 | 0.725 | -0.238 | 0.001 |
| Session5 | Ses05F_script01_2 | non_overlap | -0.059 | 0.879 | -0.150 | 0.000 |
| Session5 | Ses05F_script01_2 | overlap | -0.302 | 0.011 | -0.189 | 0.567 |

This table presents Pearson correlation coefficients (r) and corresponding p-values for facial-speech alignment in arousal and valence signals across dyads and speech conditions (non-overlapping and overlapping). Each dyad's correlation strength is calculated between smoothed facial and speech signals, with statistical significance indicated by p-values. Stronger correlations are highlighted in lighter shades, and weaker or negative correlations in darker shades. Only dyads with sufficient temporal frames were included.

To visualize how emotional synchrony evolves over time, we also plotted time-aligned trajectories of arousal and valence for representative dyads under overlapping and non-overlapping speech conditions (Figure 2a–h). These visualizations illustrate dynamic rises and falls in emotional signals across modalities. Importantly, the plotted emotional trajectories were smoothed using a 5-frame moving average to reduce noise and highlight underlying trends in synchrony. These plots illustrate how facial and vocal signals dynamically rise and fall together—or diverge—across the course of an interaction. In **non-overlapping speech, Ses01M_script01_2** exhibits strong arousal synchrony (*arousal* r = 0.25; Figure 4a), and **Ses04F_impro01** shows moderate valence synchrony (*valence* r = 0.17; Figure 2c), highlighting emotional coordination when only one speaker is active. Meanwhile, **Ses02F_impro02** again reflects weak arousal alignment (*arousal* r = –0.14; Figure 2b), and **Ses01F_script02_1** shows weak valence synchrony (*valence* r = –0.15; Figure 2d), pointing to variability even in turn-taking segments. In **overlapping speech**, sessions with strong synchrony such as **Ses01F_impro02** (*arousal* r = 0.41; Figure 2e) and **Ses05F_impro01** (*valence* r = 0.36; Figure 2g) show facial and vocal trajectories that move in parallel, reflecting heightened emotional alignment during simultaneous vocalization. In contrast, **Ses02F_impro02** displays pronounced desynchrony, with diverging trajectories in both **arousal** (r = –0.58; Figure 2f) and **valence** (r = –0.48; Figure 2h), suggesting competing emotional cues or reduced mutual responsiveness. These examples underscore how emotional synchrony varies not only across sessions but also across communicative contexts, revealing that both **overlapping** and **non-overlapping** speech dynamics influence the unfolding of cross-modal emotional alignment.

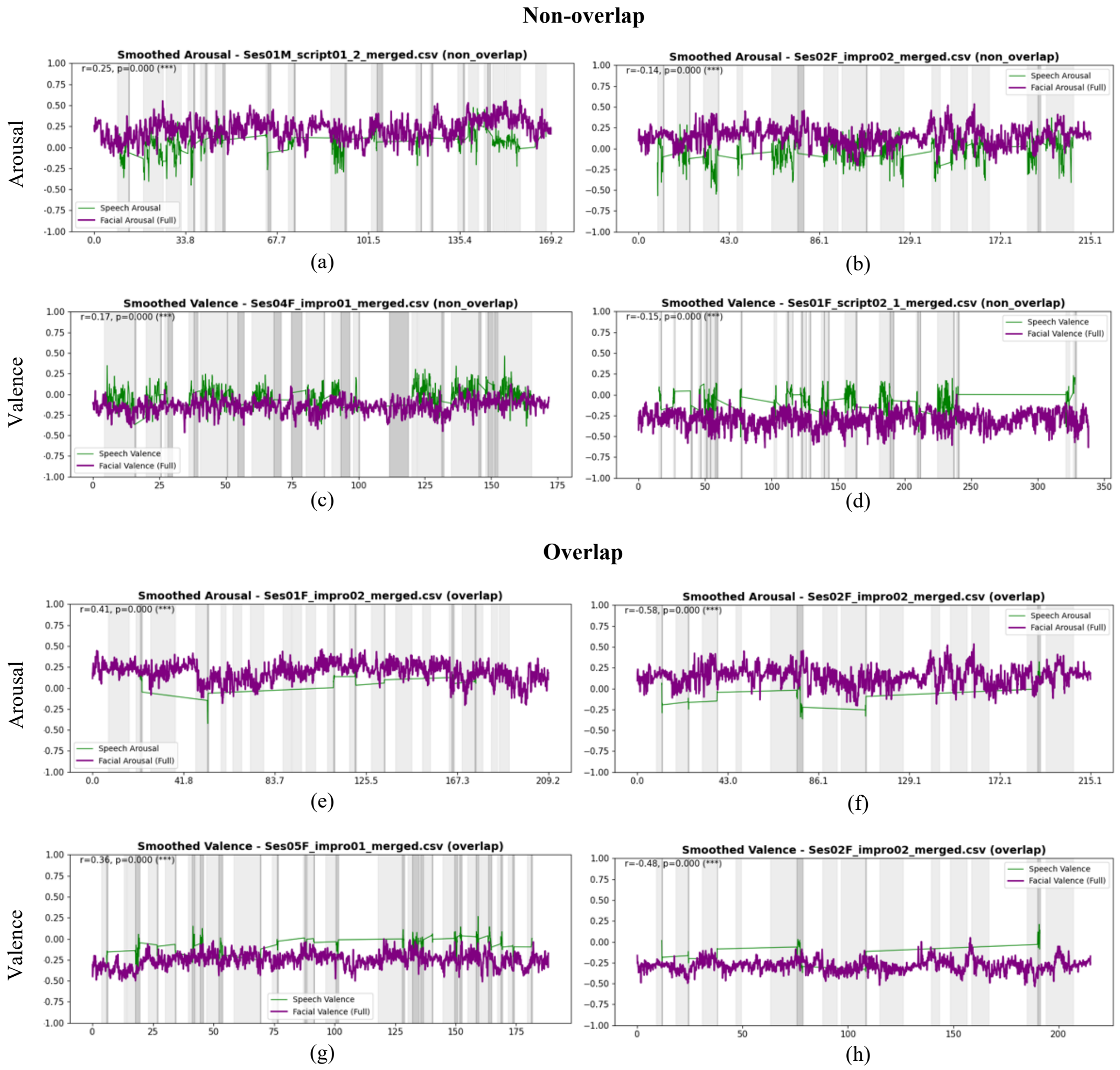

**Figure 2. Temporal dynamics of smoothed arousal and valence signals from facial and vocal modalities under non-overlapping (light grey) and overlapping speech (darker grey) conditions. The alignment and temporal unfolding of emotional signals across modalities are compared, illustrating how conversational structure (non-overlap vs. overlap) influences cross-modal emotional coordination.**

To evaluate the degree of emotional synchrony between speech-based and facial-based arousal and valence signals, we computed Pearson correlation coefficients for each dyad, analyzing them separately for non-overlapping and overlapping speech conditions. This dyad-level analysis allowed us to quantify the spatial alignment of emotional expressions across modalities under varying conversational dynamics. Overall, correlations were generally low in magnitude but varied across conditions (Figure 3). For arousal, the average correlation in non-overlapping speech was $r = 0.006$, $SD = 0.064$ ($r^2 = 0.0000$), indicating minimal shared variance between vocal and facial signals. In overlapping speech, the mean correlation was slightly lower at $r = –0.011$, $SD = 0.155$ ($r^2 = 0.0001$), suggesting even weaker and more dispersed alignment. Similarly, for valence, the mean correlation during non-overlapping speech was $r = 0.011$, $SD = 0.068$ ($r^2 = 0.0001$), while in overlapping speech it dropped to $r = –0.001$, $SD = 0.155$ ($r^2 = 0.0000$). While effect sizes remained small, the standard deviations were consistently higher during overlapping speech. This increase in dispersion suggests that emotional synchrony was less stable when both speakers were vocalizing simultaneously, which may reflect contextual challenges such as reduced clarity of emotional cues, disfluency, or interruptions during overlapping talk.

To statistically assess whether synchrony differed across conditions, paired-sample t-tests were conducted (Figure 3). For arousal, the difference between non-overlapping and overlapping speech was not statistically significant, $t(85) = 1.30$, Cohen's $d = 0.11$, $p = .194$. Similarly, for valence, the difference was also non-significant, $t(85) = 0.95$, $p = .344$, $d = 0.08$. These results suggest that while overlapping speech may introduce greater variability, its overall impact on the mean strength of emotional alignment between modalities is modest. The increased dispersion observed in overlapping segments may reflect contextual disruptions—such as emotional misalignment, reduced expressive clarity, or difficulties in simultaneous message processing—but these interpretations remain speculative and should be further tested in future studies incorporating turn-taking structure and conversational context. In contrast, non-overlapping (turn-taking) segments may provide a clearer window into emotional mirroring, though even here, average synchrony remains low.

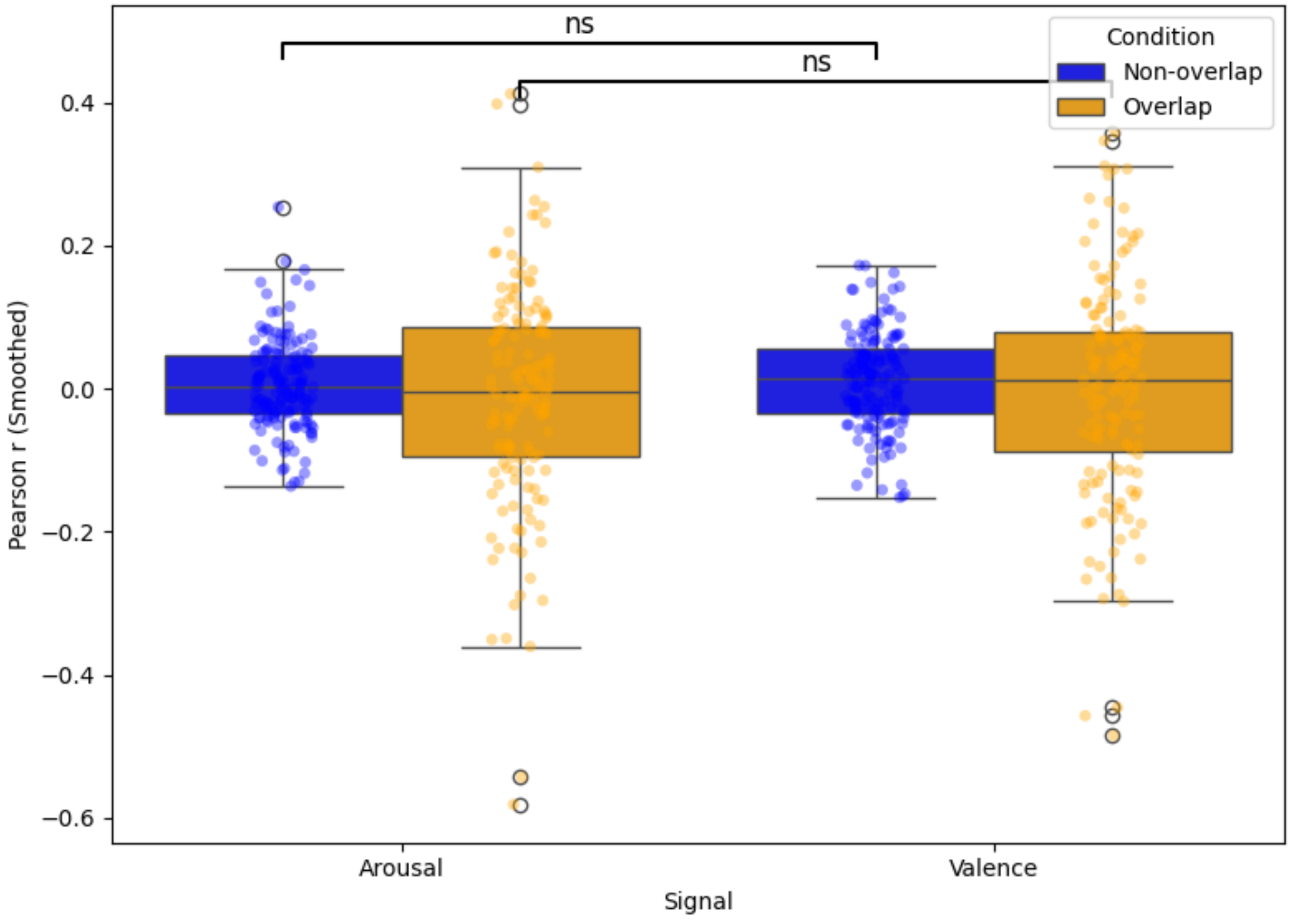

**Figure 3.** Dyad-level Pearson correlations (smoothed) between speech-based and facial-based arousal (left) and valence (right) signals, compared across non-overlapping and overlapping speech conditions. Each point represents a dyad-level correlation, with paired-sample t-tests used to assess differences between conditions.

### b. Lag correlation analysis

**Table 3. Lag-Adjusted Correlation Metrics for Arousal and Valence Across Conditions**

| | | | Arousal | | | Valence | | |
|---|---|---|---|---|---|---|---|---|
| Session | Dyads | Condition | Best lag | Best corr | p-value | Best lag | Best corr | p-value |
| Session1 | Ses01M_impro06 | non_overlap | 0 | -0.047 | 0.105 | 2 | -0.045 | 0.341 |
| Session1 | Ses01M_impro06 | overlap | -4 | 0.034 | 0.673 | -1 | 0.106 | 0.019 |
| Session1 | Ses01F_script02_2 | non_overlap | -10 | -0.005 | 0.753 | 10 | 0.026 | 0.140 |
| Session1 | Ses01F_script02_2 | overlap | 10 | 0.081 | 0.173 | 2 | 0.302 | 0.000 |
| Session1 | Ses01F_impro01 | non_overlap | 7 | 0.009 | 0.636 | -10 | 0.004 | 0.902 |
| Session1 | Ses01F_impro01 | overlap | 1 | 0.268 | 0.000 | -5 | 0.034 | 0.096 |
| Session1 | Ses01M_impro03 | non_overlap | -10 | -0.006 | 0.614 | -10 | -0.064 | 0.077 |
| Session1 | Ses01M_impro03 | overlap | 5 | 0.133 | 0.000 | 10 | 0.158 | 0.000 |
| Session1 | Ses01F_impro04 | non_overlap | -10 | 0.017 | 0.431 | 4 | 0.190 | 0.000 |
| Session1 | Ses01F_impro04 | overlap | 4 | -0.085 | 0.241 | -9 | 0.306 | 0.000 |
| Session1 | Ses01F_script01_1 | non_overlap | 3 | 0.026 | 0.097 | 1 | 0.095 | 0.000 |
| Session1 | Ses01F_script01_1 | overlap | 8 | 0.215 | 0.001 | -2 | 0.232 | 0.003 |
| Session1 | Ses01F_script03_2 | non_overlap | 10 | 0.087 | 0.000 | 9 | 0.037 | 0.072 |
| Session1 | Ses01F_script03_2 | overlap | 0 | 0.090 | 0.086 | -2 | 0.172 | 0.000 |
| Session1 | Ses01F_impro03 | non_overlap | -3 | 0.061 | 0.104 | -10 | 0.053 | 0.065 |
| Session1 | Ses01F_impro03 | overlap | 2 | 0.048 | 0.400 | -4 | 0.197 | 0.003 |
| Session1 | Ses01M_script02_1 | non_overlap | -2 | -0.009 | 0.612 | -10 | 0.036 | 0.042 |
| Session1 | Ses01M_script02_1 | overlap | -10 | 0.207 | 0.004 | -5 | 0.122 | 0.007 |
| Session1 | Ses01M_impro04 | non_overlap | 5 | 0.032 | 0.375 | 6 | 0.014 | 0.537 |
| Session1 | Ses01M_impro04 | overlap | -10 | -0.090 | 0.120 | 10 | 0.042 | 0.259 |
| Session1 | Ses01F_impro06 | non_overlap | 10 | -0.006 | 0.798 | -9 | 0.087 | 0.000 |
| Session1 | Ses01F_impro06 | overlap | -6 | 0.190 | 0.000 | 10 | 0.244 | 0.000 |
| Session1 | Ses01M_impro01 | non_overlap | -5 | 0.011 | 0.196 | -1 | 0.048 | 0.058 |
| Session1 | Ses01M_impro01 | overlap | -3 | 0.174 | 0.000 | 4 | 0.100 | 0.034 |
| Session1 | Ses01M_script01_2 | non_overlap | -6 | 0.284 | 0.000 | -10 | 0.035 | 0.337 |
| Session1 | Ses01M_script01_2 | overlap | -10 | 0.143 | 0.049 | -10 | -0.041 | 0.119 |
| Session1 | Ses01M_script03_1 | non_overlap | 1 | 0.052 | 0.005 | 5 | 0.068 | 0.001 |
| Session1 | Ses01M_script03_1 | overlap | 0 | -0.019 | 0.441 | 2 | 0.079 | 0.036 |
| Session1 | Ses01F_script01_3 | non_overlap | -7 | 0.037 | 0.014 | -5 | 0.083 | 0.000 |
| Session1 | Ses01F_script01_3 | overlap | 5 | 0.214 | 0.009 | -10 | 0.141 | 0.227 |
| Session1 | Ses01M_impro02 | non_overlap | -5 | 0.079 | 0.006 | -8 | 0.030 | 0.182 |
| Session1 | Ses01M_impro02 | overlap | 9 | 0.179 | 0.009 | 7 | -0.058 | 0.709 |
| Session1 | Ses01F_impro05 | non_overlap | 10 | -0.061 | 0.033 | 0 | 0.100 | 0.000 |
| Session1 | Ses01F_impro05 | overlap | 10 | 0.082 | 0.052 | 10 | -0.088 | 0.120 |
| Session1 | Ses01M_script03_2 | non_overlap | -10 | 0.068 | 0.001 | 0 | -0.010 | 0.814 |
| Session1 | Ses01M_script03_2 | overlap | 0 | 0.115 | 0.022 | 0 | 0.172 | 0.000 |
| Session1 | Ses01M_script01_1 | non_overlap | 5 | 0.108 | 0.000 | 6 | -0.009 | 0.797 |
| Session1 | Ses01M_script01_1 | overlap | 10 | 0.098 | 0.070 | 10 | -0.051 | 0.453 |
| Session1 | Ses01M_impro07 | non_overlap | -7 | 0.014 | 0.510 | 5 | 0.073 | 0.016 |
| Session1 | Ses01M_impro07 | overlap | 10 | 0.128 | 0.001 | 2 | 0.083 | 0.006 |
| Session1 | Ses01M_script02_2 | non_overlap | -9 | -0.024 | 0.343 | 5 | 0.162 | 0.000 |
| Session1 | Ses01M_script02_2 | overlap | 0 | -0.096 | 0.070 | -10 | 0.112 | 0.068 |
| Session1 | Ses01F_impro07 | non_overlap | 5 | -0.027 | 0.727 | 1 | 0.077 | 0.031 |
| Session1 | Ses01F_impro07 | overlap | -10 | 0.007 | 0.680 | -8 | 0.101 | 0.069 |
| Session1 | Ses01F_script03_1 | non_overlap | -8 | 0.115 | 0.000 | 10 | -0.035 | 0.115 |
| Session1 | Ses01F_script03_1 | overlap | 10 | 0.224 | 0.000 | -10 | 0.037 | 0.524 |
| Session1 | Ses01M_script01_3 | non_overlap | 10 | 0.053 | 0.003 | 7 | 0.108 | 0.000 |
| Session1 | Ses01M_script01_3 | overlap | 10 | 0.066 | 0.182 | -9 | 0.375 | 0.000 |
| Session1 | Ses01F_script01_2 | non_overlap | 10 | 0.060 | 0.134 | 10 | 0.068 | 0.036 |
| Session1 | Ses01F_script01_2 | overlap | 9 | 0.143 | 0.006 | -5 | 0.302 | 0.000 |
| Session1 | Ses01F_impro02 | non_overlap | -10 | 0.171 | 0.000 | 10 | 0.058 | 0.016 |
| Session1 | Ses01F_impro02 | overlap | -2 | 0.455 | 0.000 | 1 | 0.157 | 0.252 |
| Session1 | Ses01F_script02_1 | non_overlap | 4 | -0.058 | 0.217 | -10 | 0.016 | 0.553 |
| Session1 | Ses01F_script02_1 | overlap | -10 | 0.232 | 0.039 | 10 | -0.013 | 0.400 |
| Session1 | Ses01M_impro05 | non_overlap | -8 | 0.067 | 0.009 | -4 | 0.118 | 0.000 |
| Session1 | Ses01M_impro05 | overlap | 9 | 0.078 | 0.174 | -10 | 0.089 | 0.022 |
| Session2 | Ses02F_script01_1 | non_overlap | 10 | 0.031 | 0.187 | 10 | 0.089 | 0.001 |
| Session2 | Ses02F_script01_1 | overlap | 10 | 0.328 | 0.000 | 10 | 0.119 | 0.198 |
| Session2 | Ses02F_impro01 | non_overlap | 10 | 0.061 | 0.012 | -3 | -0.063 | 0.130 |
| Session2 | Ses02F_impro01 | overlap | -7 | -0.268 | 0.000 | 10 | -0.094 | 0.480 |
| Session2 | Ses02F_script03_2 | non_overlap | 10 | 0.102 | 0.000 | -10 | 0.010 | 0.304 |
| Session2 | Ses02F_script03_2 | overlap | 3 | 0.084 | 0.113 | -10 | -0.137 | 0.033 |
| Session2 | Ses02M_impro06 | non_overlap | -7 | 0.006 | 0.511 | -1 | 0.112 | 0.000 |
| Session2 | Ses02M_impro06 | overlap | 10 | 0.122 | 0.200 | 3 | 0.387 | 0.003 |
| Session2 | Ses02F_script02_2 | non_overlap | 10 | 0.016 | 0.199 | 3 | 0.046 | 0.020 |
| Session2 | Ses02F_script02_2 | overlap | 8 | 0.162 | 0.160 | 10 | -0.126 | 0.176 |
| Session2 | Ses02F_impro04 | non_overlap | 8 | 0.096 | 0.000 | -5 | 0.022 | 0.554 |
| Session2 | Ses02F_impro04 | overlap | 10 | -0.096 | 0.191 | 5 | 0.052 | 0.506 |
| Session2 | Ses02M_impro03 | non_overlap | -10 | -0.090 | 0.000 | -3 | 0.133 | 0.000 |
| Session2 | Ses02M_impro03 | overlap | -10 | 0.220 | 0.001 | -8 | -0.011 | 0.690 |
| Session2 | Ses02M_impro04 | non_overlap | 0 | -0.037 | 0.287 | -10 | 0.021 | 0.610 |
| Session2 | Ses02M_impro04 | overlap | -3 | 0.125 | 0.438 | 10 | 0.407 | 0.001 |
| Session2 | Ses02M_script01_2 | non_overlap | -10 | 0.150 | 0.001 | -4 | 0.086 | 0.054 |
| Session2 | Ses02M_script01_2 | overlap | -5 | 0.280 | 0.000 | -3 | 0.073 | 0.368 |
| Session2 | Ses02F_script01_3 | non_overlap | 2 | 0.063 | 0.019 | -7 | 0.034 | 0.178 |
| Session2 | Ses02F_script01_3 | overlap | -6 | 0.195 | 0.146 | -6 | 0.556 | 0.001 |
| Session2 | Ses02M_script03_1 | non_overlap | 1 | 0.037 | 0.173 | -1 | 0.023 | 0.129 |
| Session2 | Ses02M_script03_1 | overlap | -10 | 0.049 | 0.299 | -7 | 0.122 | 0.183 |
| Session2 | Ses02F_impro03 | non_overlap | 0 | 0.166 | 0.000 | 4 | 0.096 | 0.000 |
| Session2 | Ses02F_impro03 | overlap | -10 | 0.106 | 0.016 | -9 | 0.022 | 0.492 |
| Session2 | Ses02M_script02_1 | non_overlap | -7 | -0.013 | 0.852 | -1 | 0.063 | 0.008 |
| Session2 | Ses02M_script02_1 | overlap | 7 | 0.051 | 0.202 | -10 | -0.053 | 0.853 |
| Session2 | Ses02M_impro01 | non_overlap | -10 | 0.090 | 0.036 | -5 | 0.007 | 0.234 |
| Session2 | Ses02M_impro01 | overlap | 10 | 0.405 | 0.000 | -7 | 0.172 | 0.010 |
| Session2 | Ses02F_impro06 | non_overlap | -2 | -0.091 | 0.025 | -10 | -0.069 | 0.034 |
| Session2 | Ses02F_impro06 | overlap | -5 | 0.675 | 0.009 | -9 | 0.992 | 0.016 |
| Session2 | Ses02F_impro08 | non_overlap | 10 | 0.059 | 0.026 | 1 | -0.016 | 0.807 |
| Session2 | Ses02F_impro08 | overlap | -7 | -0.039 | 0.710 | -3 | -0.099 | 0.349 |
| Session2 | Ses02M_script02_2 | non_overlap | -10 | 0.084 | 0.000 | -10 | 0.040 | 0.019 |
| Session2 | Ses02M_script02_2 | overlap | 8 | 0.020 | 0.578 | -9 | 0.120 | 0.044 |
| Session2 | Ses02F_impro05 | non_overlap | 10 | 0.111 | 0.000 | 10 | -0.038 | 0.210 |
| Session2 | Ses02F_impro05 | overlap | -3 | 0.090 | 0.145 | 1 | 0.045 | 0.256 |
| Session2 | Ses02M_impro02 | non_overlap | 4 | 0.019 | 0.263 | -1 | 0.064 | 0.003 |
| Session2 | Ses02M_impro02 | overlap | -4 | 0.451 | 0.023 | 2 | 0.368 | 0.024 |
| Session2 | Ses02M_script03_2 | non_overlap | -5 | 0.096 | 0.000 | 2 | 0.090 | 0.001 |
| Session2 | Ses02M_script03_2 | overlap | -4 | 0.115 | 0.009 | 2 | 0.201 | 0.000 |
| Session2 | Ses02M_impro07 | non_overlap | -10 | -0.005 | 0.826 | 0 | 0.097 | 0.000 |
| Session2 | Ses02M_impro07 | overlap | 1 | 0.088 | 0.007 | -10 | 0.014 | 0.312 |
| Session2 | Ses02M_script01_1 | non_overlap | 10 | 0.038 | 0.185 | -2 | 0.048 | 0.003 |
| Session2 | Ses02M_script01_1 | overlap | 10 | 0.009 | 0.633 | 0 | 0.252 | 0.010 |
| Session2 | Ses02F_script02_1 | non_overlap | -8 | -0.108 | 0.011 | 10 | -0.016 | 0.991 |
| Session2 | Ses02F_script02_1 | overlap | 0 | 0.398 | 0.044 | -10 | 0.118 | 0.487 |
| Session2 | Ses02F_impro07 | non_overlap | -3 | -0.006 | 0.904 | -10 | -0.042 | 0.095 |
| Session2 | Ses02F_impro07 | overlap | 6 | -0.015 | 0.917 | -2 | 0.115 | 0.016 |
| Session2 | Ses02M_script01_3 | non_overlap | -10 | 0.081 | 0.000 | 10 | 0.043 | 0.031 |
| Session2 | Ses02M_script01_3 | overlap | -6 | -0.071 | 0.635 | 9 | 0.043 | 0.283 |
| Session2 | Ses02M_impro05 | non_overlap | 5 | -0.016 | 0.957 | -4 | -0.031 | 0.929 |
| Session2 | Ses02M_impro05 | overlap | -7 | -0.061 | 0.641 | 10 | 0.206 | 0.002 |
| Session2 | Ses02F_script03_1 | non_overlap | 10 | -0.038 | 0.379 | -5 | 0.022 | 0.101 |
| Session2 | Ses02F_script03_1 | overlap | 5 | -0.095 | 0.666 | 1 | 0.075 | 0.090 |
| Session2 | Ses02F_impro02 | non_overlap | -9 | -0.085 | 0.003 | 10 | -0.059 | 0.126 |
| Session2 | Ses02F_impro02 | overlap | -7 | -0.417 | 0.000 | -10 | -0.206 | 0.052 |
| Session2 | Ses02F_script01_2 | non_overlap | -2 | -0.083 | 0.086 | -10 | 0.014 | 0.343 |
| Session2 | Ses02F_script01_2 | overlap | -4 | 0.320 | 0.000 | -6 | -0.038 | 0.632 |
| Session2 | Ses02M_impro08 | non_overlap | -10 | 0.000 | 0.801 | 5 | 0.069 | 0.034 |
| Session2 | Ses02M_impro08 | overlap | -9 | 0.128 | 0.032 | 5 | 0.127 | 0.018 |
| Session3 | Ses03M_impro06 | non_overlap | 7 | 0.040 | 0.050 | -10 | 0.115 | 0.000 |
| Session3 | Ses03M_impro06 | overlap | -10 | -0.172 | 0.022 | -9 | -0.075 | 0.580 |
| Session3 | Ses03F_impro01 | non_overlap | 9 | 0.045 | 0.189 | -6 | 0.037 | 0.487 |
| Session3 | Ses03F_impro01 | overlap | 10 | 0.269 | 0.029 | -10 | -0.141 | 0.150 |
| Session3 | Ses03M_script02_1 | non_overlap | 10 | 0.002 | 0.937 | 6 | 0.045 | 0.005 |
| Session3 | Ses03M_script02_1 | overlap | 6 | 0.023 | 0.367 | -5 | 0.026 | 0.649 |
| Session3 | Ses03M_impro03 | non_overlap | -10 | 0.004 | 0.984 | -3 | 0.046 | 0.034 |
| Session3 | Ses03M_impro03 | overlap | 0 | -0.015 | 0.730 | 0 | 0.046 | 0.177 |
| Session3 | Ses03M_script03_1 | non_overlap | 10 | 0.048 | 0.090 | -8 | -0.010 | 0.888 |
| Session3 | Ses03M_script03_1 | overlap | -7 | 0.162 | 0.029 | 10 | 0.102 | 0.048 |
| Session3 | Ses03F_script01_3 | non_overlap | 1 | 0.059 | 0.008 | 4 | 0.026 | 0.202 |
| Session3 | Ses03F_script01_3 | overlap | 10 | -0.091 | 0.067 | -5 | 0.202 | 0.001 |
| Session3 | Ses03M_script01_2 | non_overlap | 9 | 0.019 | 0.343 | 6 | 0.019 | 0.580 |
| Session3 | Ses03M_script01_2 | overlap | 4 | 0.363 | 0.000 | -10 | 0.153 | 0.065 |
| Session3 | Ses03F_impro04 | non_overlap | 10 | 0.094 | 0.000 | -1 | -0.060 | 0.098 |
| Session3 | Ses03F_impro04 | overlap | -10 | 0.003 | 0.885 | -4 | 0.071 | 0.224 |
| Session3 | Ses03M_impro05a | non_overlap | -5 | -0.007 | 0.846 | -1 | 0.065 | 0.106 |
| Session3 | Ses03M_impro05a | overlap | -10 | -0.114 | 0.113 | -10 | 0.058 | 0.587 |
| Session3 | Ses03F_impro03 | non_overlap | 4 | -0.006 | 0.888 | -2 | 0.038 | 0.029 |
| Session3 | Ses03F_impro03 | overlap | -4 | 0.102 | 0.007 | 7 | -0.036 | 0.483 |
| Session3 | Ses03M_impro08b | non_overlap | 7 | 0.048 | 0.312 | 4 | 0.011 | 0.383 |
| Session3 | Ses03M_impro08b | overlap | -7 | -0.047 | 0.701 | 9 | 0.175 | 0.000 |
| Session3 | Ses03M_impro04 | non_overlap | 10 | -0.008 | 0.906 | -10 | -0.029 | 0.277 |
| Session3 | Ses03M_impro04 | overlap | -5 | 0.037 | 0.758 | 10 | -0.077 | 0.190 |
| Session3 | Ses03F_script02_2 | non_overlap | 1 | 0.039 | 0.026 | -10 | 0.007 | 0.223 |
| Session3 | Ses03F_script02_2 | overlap | 1 | 0.113 | 0.003 | -10 | 0.190 | 0.001 |
| Session3 | Ses03F_impro06 | non_overlap | 0 | -0.013 | 0.889 | 1 | 0.020 | 0.175 |
| Session3 | Ses03F_impro06 | overlap | 10 | -0.056 | 0.629 | 0 | 0.311 | 0.000 |
| Session3 | Ses03F_script03_2 | non_overlap | -9 | -0.030 | 0.411 | -10 | -0.125 | 0.000 |
| Session3 | Ses03F_script03_2 | overlap | -4 | -0.077 | 0.609 | 10 | 0.168 | 0.003 |
| Session3 | Ses03M_impro01 | non_overlap | 3 | -0.031 | 0.643 | -10 | -0.020 | 0.562 |
| Session3 | Ses03M_impro01 | overlap | 10 | -0.027 | 0.802 | 5 | 0.066 | 0.620 |
| Session3 | Ses03F_script01_1 | non_overlap | -4 | -0.031 | 0.268 | -8 | 0.086 | 0.000 |
| Session3 | Ses03F_script01_1 | overlap | 2 | 0.121 | 0.013 | -10 | 0.012 | 0.883 |
| Session3 | Ses03F_impro08 | non_overlap | 10 | 0.131 | 0.001 | 4 | 0.080 | 0.004 |
| Session3 | Ses03F_impro08 | overlap | 10 | 0.022 | 0.533 | 10 | -0.010 | 0.716 |
| Session3 | Ses03F_script01_2 | non_overlap | 10 | 0.111 | 0.000 | -3 | 0.004 | 0.592 |
| Session3 | Ses03F_script01_2 | overlap | -2 | 0.002 | 0.770 | -1 | 0.141 | 0.223 |
| Session3 | Ses03M_impro02 | non_overlap | -10 | -0.099 | 0.000 | 10 | -0.065 | 0.068 |
| Session3 | Ses03M_impro02 | overlap | -10 | 0.065 | 0.015 | -9 | 0.155 | 0.000 |
| Session3 | Ses03F_script03_1 | non_overlap | -5 | 0.080 | 0.002 | -9 | 0.087 | 0.012 |
| Session3 | Ses03F_script03_1 | overlap | 10 | 0.027 | 0.992 | 9 | 0.071 | 0.010 |
| Session3 | Ses03M_script01_3 | non_overlap | 1 | 0.029 | 0.052 | 10 | 0.052 | 0.013 |
| Session3 | Ses03M_script01_3 | overlap | 9 | 0.238 | 0.000 | -10 | 0.108 | 0.030 |
| Session3 | Ses03F_impro05 | non_overlap | 4 | 0.028 | 0.280 | 10 | 0.041 | 0.087 |
| Session3 | Ses03F_impro05 | overlap | 10 | -0.234 | 0.000 | 1 | 0.048 | 0.398 |
| Session3 | Ses03M_impro07 | non_overlap | -4 | 0.068 | 0.062 | 10 | -0.021 | 0.334 |
| Session3 | Ses03M_impro07 | overlap | -5 | 0.224 | 0.004 | 10 | 0.178 | 0.024 |
| Session3 | Ses03M_impro08a | non_overlap | -10 | 0.029 | 0.395 | 7 | 0.027 | 0.294 |
| Session3 | Ses03M_impro08a | overlap | 0 | 0.102 | 0.021 | -2 | 0.034 | 0.798 |
| Session3 | Ses03F_script02_1 | non_overlap | 7 | 0.133 | 0.000 | 9 | 0.080 | 0.007 |
| Session3 | Ses03F_script02_1 | overlap | 4 | 0.053 | 0.179 | -1 | 0.309 | 0.001 |
| Session3 | Ses03M_script01_1 | non_overlap | 4 | 0.029 | 0.132 | -6 | 0.013 | 0.288 |
| Session3 | Ses03M_script01_1 | overlap | -5 | 0.069 | 0.041 | 9 | 0.091 | 0.057 |
| Session3 | Ses03F_impro07 | non_overlap | -10 | 0.142 | 0.000 | 1 | 0.091 | 0.009 |
| Session3 | Ses03F_impro07 | overlap | 9 | 0.018 | 0.519 | -6 | -0.002 | 0.808 |
| Session3 | Ses03M_impro05b | non_overlap | 1 | 0.010 | 0.467 | -10 | 0.162 | 0.000 |
| Session3 | Ses03M_impro05b | overlap | -4 | 0.060 | 0.022 | 9 | 0.108 | 0.044 |
| Session3 | Ses03M_script03_2 | non_overlap | -3 | -0.029 | 0.510 | 10 | 0.049 | 0.020 |
| Session3 | Ses03M_script03_2 | overlap | -10 | -0.076 | 0.021 | -10 | -0.099 | 0.016 |
| Session3 | Ses03F_impro02 | non_overlap | -5 | 0.081 | 0.000 | 10 | 0.026 | 0.463 |
| Session3 | Ses03F_impro02 | overlap | 5 | 0.105 | 0.050 | 10 | 0.074 | 0.092 |
| Session3 | Ses03M_script02_2 | non_overlap | -10 | 0.057 | 0.001 | -10 | 0.010 | 0.476 |
| Session3 | Ses03M_script02_2 | overlap | 10 | -0.021 | 0.223 | 10 | 0.080 | 0.227 |
| Session4 | Ses04M_impro03 | non_overlap | 10 | 0.024 | 0.795 | 10 | 0.017 | 0.350 |
| Session4 | Ses04M_impro03 | overlap | 10 | 0.128 | 0.044 | -3 | 0.147 | 0.018 |
| Session4 | Ses04M_script02_1 | non_overlap | -4 | 0.029 | 0.322 | 0 | 0.018 | 0.380 |
| Session4 | Ses04M_script02_1 | overlap | -10 | 0.075 | 0.198 | -7 | 0.158 | 0.042 |
| Session4 | Ses04F_impro04 | non_overlap | 10 | 0.052 | 0.059 | -10 | 0.090 | 0.002 |
| Session4 | Ses04F_impro04 | overlap | -2 | 0.087 | 0.008 | 10 | 0.143 | 0.001 |
| Session4 | Ses04M_script03_1 | non_overlap | 3 | 0.115 | 0.000 | 3 | 0.146 | 0.000 |
| Session4 | Ses04M_script03_1 | overlap | -10 | 0.069 | 0.294 | -10 | -0.002 | 0.610 |
| Session4 | Ses04F_script01_3 | non_overlap | 0 | 0.067 | 0.001 | 4 | 0.061 | 0.006 |
| Session4 | Ses04F_script01_3 | overlap | 10 | 0.114 | 0.151 | -9 | 0.239 | 0.010 |
| Session4 | Ses04M_script01_2 | non_overlap | -1 | 0.150 | 0.000 | -6 | 0.210 | 0.000 |
| Session4 | Ses04M_script01_2 | overlap | -5 | 0.222 | 0.048 | 10 | 0.301 | 0.003 |
| Session4 | Ses04M_impro06 | non_overlap | -9 | 0.056 | 0.006 | -10 | 0.097 | 0.000 |
| Session4 | Ses04M_impro06 | overlap | 10 | -0.187 | 0.085 | 9 | -0.068 | 0.781 |
| Session4 | Ses04F_impro01 | non_overlap | 0 | 0.076 | 0.015 | 7 | 0.194 | 0.000 |
| Session4 | Ses04F_impro01 | overlap | -10 | 0.282 | 0.000 | -7 | 0.058 | 0.499 |
| Session4 | Ses04F_impro06 | non_overlap | -10 | 0.001 | 0.753 | 10 | -0.040 | 0.305 |
| Session4 | Ses04F_impro06 | overlap | 10 | 0.338 | 0.011 | 6 | 0.135 | 0.072 |
| Session4 | Ses04M_impro01 | non_overlap | -5 | 0.050 | 0.120 | -7 | 0.081 | 0.008 |
| Session4 | Ses04M_impro01 | overlap | -4 | 0.063 | 0.164 | -10 | 0.099 | 0.132 |
| Session4 | Ses04F_script02_2 | non_overlap | 2 | 0.087 | 0.000 | 0 | 0.089 | 0.000 |
| Session4 | Ses04F_script02_2 | overlap | -8 | 0.139 | 0.018 | 2 | 0.085 | 0.265 |
| Session4 | Ses04F_script03_2 | non_overlap | 5 | 0.069 | 0.005 | -1 | 0.077 | 0.001 |
| Session4 | Ses04F_script03_2 | overlap | 0 | 0.187 | 0.000 | 10 | 0.141 | 0.001 |
| Session4 | Ses04F_script01_1 | non_overlap | -9 | 0.082 | 0.000 | -7 | 0.058 | 0.006 |
| Session4 | Ses04F_script01_1 | overlap | 1 | 0.126 | 0.022 | -10 | 0.013 | 0.571 |
| Session4 | Ses04F_impro03 | non_overlap | -10 | 0.053 | 0.070 | -6 | 0.006 | 0.373 |
| Session4 | Ses04F_impro03 | overlap | 10 | -0.103 | 0.292 | 4 | 0.080 | 0.163 |
| Session4 | Ses04M_impro04 | non_overlap | -7 | 0.083 | 0.002 | -1 | 0.045 | 0.050 |
| Session4 | Ses04M_impro04 | overlap | 3 | 0.193 | 0.006 | -10 | 0.028 | 0.403 |
| Session4 | Ses04F_script01_2 | non_overlap | 5 | 0.064 | 0.007 | -10 | -0.004 | 0.946 |
| Session4 | Ses04F_script01_2 | overlap | -7 | 0.188 | 0.010 | -10 | 0.097 | 0.245 |
| Session4 | Ses04F_script03_1 | non_overlap | 9 | 0.082 | 0.001 | 10 | 0.015 | 0.749 |
| Session4 | Ses04F_script03_1 | overlap | 4 | 0.164 | 0.036 | 0 | 0.066 | 0.146 |
| Session4 | Ses04M_script01_3 | non_overlap | -5 | -0.017 | 0.873 | -10 | -0.063 | 0.002 |
| Session4 | Ses04M_script01_3 | overlap | -7 | 0.159 | 0.047 | 6 | 0.232 | 0.004 |
| Session4 | Ses04M_impro07 | non_overlap | -3 | -0.018 | 0.247 | 2 | 0.081 | 0.001 |
| Session4 | Ses04M_impro07 | overlap | -6 | 0.097 | 0.001 | -5 | 0.161 | 0.000 |
| Session4 | Ses04F_script02_1 | non_overlap | 7 | 0.209 | 0.000 | 10 | -0.005 | 0.783 |
| Session4 | Ses04F_script02_1 | overlap | -10 | 0.284 | 0.000 | -10 | 0.200 | 0.076 |
| Session4 | Ses04M_impro02 | non_overlap | 10 | 0.057 | 0.008 | 10 | 0.032 | 0.203 |
| Session4 | Ses04M_impro02 | overlap | 1 | 0.223 | 0.000 | -10 | 0.066 | 0.406 |
| Session4 | Ses04F_impro05 | non_overlap | -10 | -0.004 | 0.937 | 6 | -0.024 | 0.561 |
| Session4 | Ses04F_impro05 | overlap | -10 | 0.158 | 0.006 | 10 | -0.134 | 0.029 |
| Session4 | Ses04F_impro08 | non_overlap | 7 | 0.030 | 0.133 | 8 | 0.075 | 0.022 |
| Session4 | Ses04F_impro08 | overlap | -8 | 0.132 | 0.017 | 9 | -0.066 | 0.837 |
| Session4 | Ses04M_script01_1 | non_overlap | -10 | 0.026 | 0.310 | -3 | -0.004 | 0.991 |
| Session4 | Ses04M_script01_1 | overlap | -10 | 0.072 | 0.593 | 10 | -0.133 | 0.044 |
| Session4 | Ses04M_impro08 | non_overlap | 6 | 0.024 | 0.538 | -5 | -0.041 | 0.140 |
| Session4 | Ses04M_impro08 | overlap | -9 | 0.028 | 0.508 | -2 | 0.029 | 0.073 |
| Session4 | Ses04M_script03_2 | non_overlap | -5 | -0.026 | 0.266 | -6 | 0.048 | 0.010 |
| Session4 | Ses04M_script03_2 | overlap | -10 | 0.138 | 0.013 | 3 | -0.056 | 0.751 |
| Session4 | Ses04F_impro02 | non_overlap | -5 | -0.007 | 0.603 | -5 | 0.021 | 0.467 |
| Session4 | Ses04F_impro02 | overlap | -10 | 0.088 | 0.060 | 10 | -0.236 | 0.000 |
| Session4 | Ses04M_impro05 | non_overlap | 6 | 0.039 | 0.336 | 3 | 0.077 | 0.000 |
| Session4 | Ses04M_impro05 | overlap | -3 | 0.090 | 0.032 | 10 | 0.055 | 0.857 |
| Session4 | Ses04M_script02_2 | non_overlap | -3 | 0.002 | 0.445 | -8 | 0.024 | 0.160 |
| Session4 | Ses04M_script02_2 | overlap | -9 | 0.140 | 0.078 | -10 | 0.138 | 0.021 |
| Session4 | Ses04F_impro07 | non_overlap | -10 | -0.018 | 0.610 | -10 | 0.042 | 0.043 |
| Session4 | Ses04F_impro07 | overlap | -6 | -0.036 | 0.816 | 3 | 0.019 | 0.135 |
| Session5 | Ses05F_impro04 | non_overlap | 7 | -0.003 | 0.893 | -10 | 0.076 | 0.002 |
| Session5 | Ses05F_impro04 | overlap | 8 | 0.076 | 0.025 | 4 | 0.132 | 0.002 |
| Session5 | Ses05M_impro03 | non_overlap | -10 | -0.086 | 0.002 | 10 | -0.017 | 0.166 |
| Session5 | Ses05M_impro03 | overlap | -10 | -0.032 | 0.711 | -4 | 0.019 | 0.687 |
| Session5 | Ses05F_script01_1 | non_overlap | -5 | 0.056 | 0.004 | -3 | -0.021 | 0.583 |
| Session5 | Ses05F_script01_1 | overlap | -3 | 0.074 | 0.241 | 7 | 0.082 | 0.065 |
| Session5 | Ses05F_script03_2 | non_overlap | 3 | 0.190 | 0.000 | 10 | 0.181 | 0.000 |
| Session5 | Ses05F_script03_2 | overlap | -10 | 0.047 | 0.214 | -10 | 0.195 | 0.001 |
| Session5 | Ses05F_script02_2 | non_overlap | 10 | 0.027 | 0.486 | -10 | 0.101 | 0.000 |
| Session5 | Ses05F_script02_2 | overlap | -10 | 0.096 | 0.239 | -5 | 0.011 | 0.902 |
| Session5 | Ses05F_impro01 | non_overlap | 0 | -0.061 | 0.133 | 6 | 0.142 | 0.000 |
| Session5 | Ses05F_impro01 | overlap | 10 | -0.152 | 0.028 | -5 | 0.398 | 0.000 |
| Session5 | Ses05M_impro06 | non_overlap | 10 | -0.028 | 0.295 | -5 | 0.063 | 0.000 |
| Session5 | Ses05M_impro06 | overlap | -3 | 0.005 | 0.759 | -2 | -0.033 | 0.620 |
| Session5 | Ses05M_impro01 | non_overlap | -8 | 0.061 | 0.063 | -8 | 0.037 | 0.121 |
| Session5 | Ses05M_impro01 | overlap | 4 | 0.068 | 0.061 | 10 | 0.061 | 0.122 |
| Session5 | Ses05F_impro06 | non_overlap | -8 | -0.053 | 0.096 | 0 | 0.007 | 0.577 |
| Session5 | Ses05F_impro06 | overlap | -10 | 0.038 | 0.640 | 10 | -0.055 | 0.416 |
| Session5 | Ses05M_script01_2 | non_overlap | 5 | 0.098 | 0.009 | 0 | 0.094 | 0.025 |
| Session5 | Ses05M_script01_2 | overlap | -5 | 0.014 | 0.264 | 7 | 0.021 | 0.703 |
| Session5 | Ses05F_script01_3 | non_overlap | -5 | 0.021 | 0.306 | 10 | -0.009 | 0.862 |
| Session5 | Ses05F_script01_3 | overlap | -10 | -0.078 | 0.520 | -10 | 0.041 | 0.777 |
| Session5 | Ses05M_script03_1 | non_overlap | -10 | 0.013 | 0.256 | -10 | -0.019 | 0.207 |
| Session5 | Ses05M_script03_1 | overlap | -3 | 0.156 | 0.122 | -10 | -0.109 | 0.242 |
| Session5 | Ses05M_script01_1b | non_overlap | 10 | 0.004 | 0.486 | -10 | 0.037 | 0.029 |
| Session5 | Ses05M_script01_1b | overlap | -10 | 0.052 | 0.869 | -3 | 0.000 | 0.509 |
| Session5 | Ses05M_impro04 | non_overlap | 3 | 0.101 | 0.000 | -10 | 0.076 | 0.014 |
| Session5 | Ses05M_impro04 | overlap | 10 | 0.074 | 0.017 | -7 | 0.103 | 0.003 |
| Session5 | Ses05M_script02_1 | non_overlap | -10 | 0.013 | 0.098 | 9 | 0.017 | 0.035 |
| Session5 | Ses05M_script02_1 | overlap | 2 | 0.133 | 0.101 | -10 | 0.108 | 0.530 |
| Session5 | Ses05F_impro03 | non_overlap | -4 | 0.090 | 0.000 | -6 | 0.045 | 0.055 |
| Session5 | Ses05F_impro03 | overlap | 8 | 0.098 | 0.002 | 2 | 0.164 | 0.000 |
| Session5 | Ses05M_script02_2 | non_overlap | -9 | 0.099 | 0.000 | 7 | 0.087 | 0.000 |
| Session5 | Ses05M_script02_2 | overlap | 10 | -0.019 | 0.833 | 10 | -0.025 | 0.376 |
| Session5 | Ses05M_impro07 | non_overlap | 2 | -0.043 | 0.124 | -7 | 0.051 | 0.067 |
| Session5 | Ses05M_impro07 | overlap | 5 | 0.017 | 0.186 | -10 | 0.029 | 0.404 |
| Session5 | Ses05F_impro05 | non_overlap | -3 | 0.035 | 0.190 | 5 | 0.023 | 0.242 |
| Session5 | Ses05F_impro05 | overlap | -10 | 0.096 | 0.200 | 2 | 0.080 | 0.045 |
| Session5 | Ses05M_impro02 | non_overlap | 8 | 0.052 | 0.047 | -10 | -0.014 | 0.567 |
| Session5 | Ses05M_impro02 | overlap | 3 | 0.051 | 0.170 | -4 | 0.070 | 0.093 |
| Session5 | Ses05F_impro08 | non_overlap | 2 | 0.119 | 0.000 | 9 | 0.025 | 0.235 |
| Session5 | Ses05F_impro08 | overlap | -1 | 0.182 | 0.001 | -10 | 0.277 | 0.000 |
| Session5 | Ses05M_script03_2 | non_overlap | 10 | 0.165 | 0.000 | 9 | 0.124 | 0.000 |
| Session5 | Ses05M_script03_2 | overlap | 10 | 0.209 | 0.000 | -2 | 0.076 | 0.070 |
| Session5 | Ses05M_script01_1 | non_overlap | 3 | 0.034 | 0.044 | -3 | 0.037 | 0.010 |
| Session5 | Ses05M_script01_1 | overlap | 5 | 0.008 | 0.491 | 1 | 0.207 | 0.000 |
| Session5 | Ses05M_impro08 | non_overlap | -3 | 0.091 | 0.001 | -7 | 0.102 | 0.001 |
| Session5 | Ses05M_impro08 | overlap | 10 | 0.086 | 0.201 | 2 | 0.153 | 0.026 |
| Session5 | Ses05M_impro05 | non_overlap | 10 | -0.017 | 0.829 | -4 | 0.001 | 0.894 |
| Session5 | Ses05M_impro05 | overlap | 10 | 0.079 | 0.262 | 10 | 0.070 | 0.424 |
| Session5 | Ses05F_impro02 | non_overlap | 5 | -0.001 | 0.965 | 10 | 0.056 | 0.039 |
| Session5 | Ses05F_impro02 | overlap | 9 | 0.157 | 0.029 | 10 | 0.165 | 0.078 |
| Session5 | Ses05F_script02_1 | non_overlap | 0 | -0.131 | 0.002 | 6 | 0.177 | 0.000 |
| Session5 | Ses05F_script02_1 | overlap | 8 | 0.187 | 0.103 | 2 | 0.224 | 0.079 |
| Session5 | Ses05F_impro07 | non_overlap | -10 | 0.089 | 0.042 | 2 | 0.019 | 0.269 |
| Session5 | Ses05F_impro07 | overlap | 6 | 0.162 | 0.004 | -8 | 0.095 | 0.154 |
| Session5 | Ses05M_script01_3 | non_overlap | 6 | -0.079 | 0.000 | -3 | 0.005 | 0.877 |
| Session5 | Ses05M_script01_3 | overlap | 10 | 0.131 | 0.142 | 6 | 0.157 | 0.130 |
| Session5 | Ses05F_script03_1 | non_overlap | 1 | 0.052 | 0.105 | -4 | 0.012 | 0.509 |
| Session5 | Ses05F_script03_1 | overlap | 2 | 0.005 | 0.554 | -10 | -0.095 | 0.261 |
| Session5 | Ses05F_script01_2 | non_overlap | 10 | -0.001 | 0.879 | -10 | -0.078 | 0.038 |
| Session5 | Ses05F_script01_2 | overlap | 5 | -0.198 | 0.100 | -10 | -0.066 | 0.281 |

**This table reports the best lag values and the corresponding lag-adjusted Pearson correlation (smoothed) coefficients for arousal and valence, computed for each dyad and session under both non-overlapping and overlapping speech conditions. Each entry includes the correlation coefficient (Best corr) and the significance of the correlation (p-value).** Stronger correlations are highlighted in lighter shades, and weaker or negative correlations in darker shades. **Positive lag values indicate facial expressions lag behind speech, while negative values suggest the opposite.**

To illustrate variability in emotional synchrony, we examined individual dyads with the highest and lowest lag-adjusted correlations for arousal and valence under both non-overlapping and overlapping speech conditions (Figure 4a – 4h). In the non-overlapping condition, the highest arousal synchrony was observed in Ses01M_script01_2 at lag –6 with a correlation of $r = 0.28$ ($p < .001$) (Figure 4a), while the lowest was in Ses05F_script02_1 at lag 0 with $r = –0.13$ ($p = .002$), indicating negative alignment (Figure 4b). For valence, the strongest synchrony appeared in Ses04M_script01_2 at lag –6 with $r = 0.21$ ($p < .001$) (Figure 4c), whereas the weakest was in Ses03F_script03_2 at lag –10 with $r = –0.13$ ($p < .001$) (Figure 4d). These examples highlight that even in relatively well-aligned non-overlapping speech, some dyads exhibit notable desynchrony. In the overlapping condition, synchrony reached extreme levels. The highest arousal correlation was in Ses02F_impro06 at lag –5 with $r = 0.68$ ($p = .009$) (Figure 4e), whereas the lowest was in Ses02F_impro02 at lag –7 with $r = –0.42$ ($p < .001$) (Figure 4f), suggesting disrupted alignment. For valence, the strongest synchrony occurred in the same session (Ses02F_impro06) at lag –9 with an exceptionally high $r = 0.99$ ($p = .016$) (Figure 4g). Conversely, the weakest valence synchrony was found in Ses04F_impro02 at lag 10 with $r = –0.24$ ($p < .001$) (Figure 4h). These extreme examples demonstrate how overlapping speech can amplify both strong synchrony and deep misalignment, underscoring the dynamic and context-sensitive nature of emotional coordination.

## Non-overlap

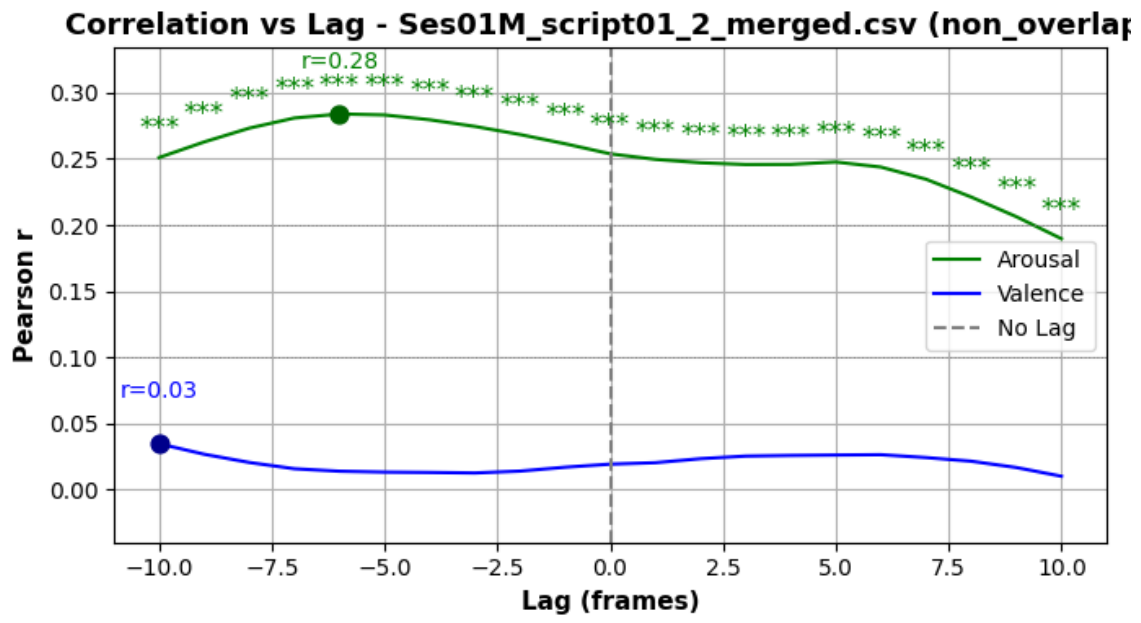

(a)

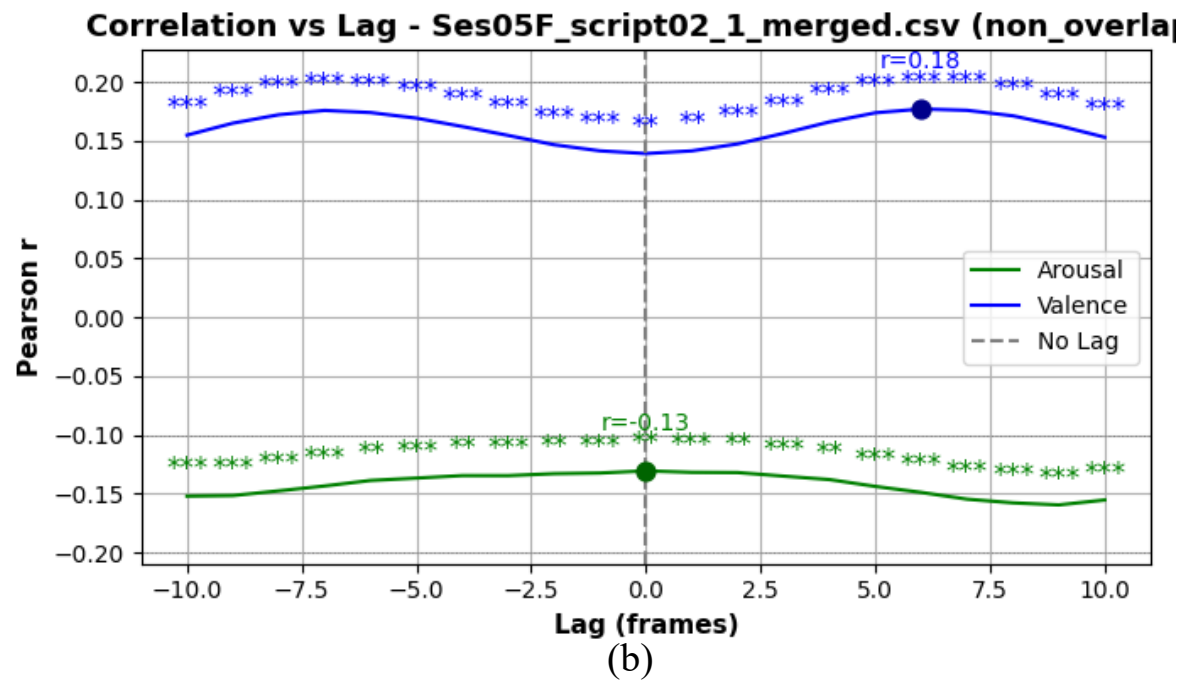

(b)

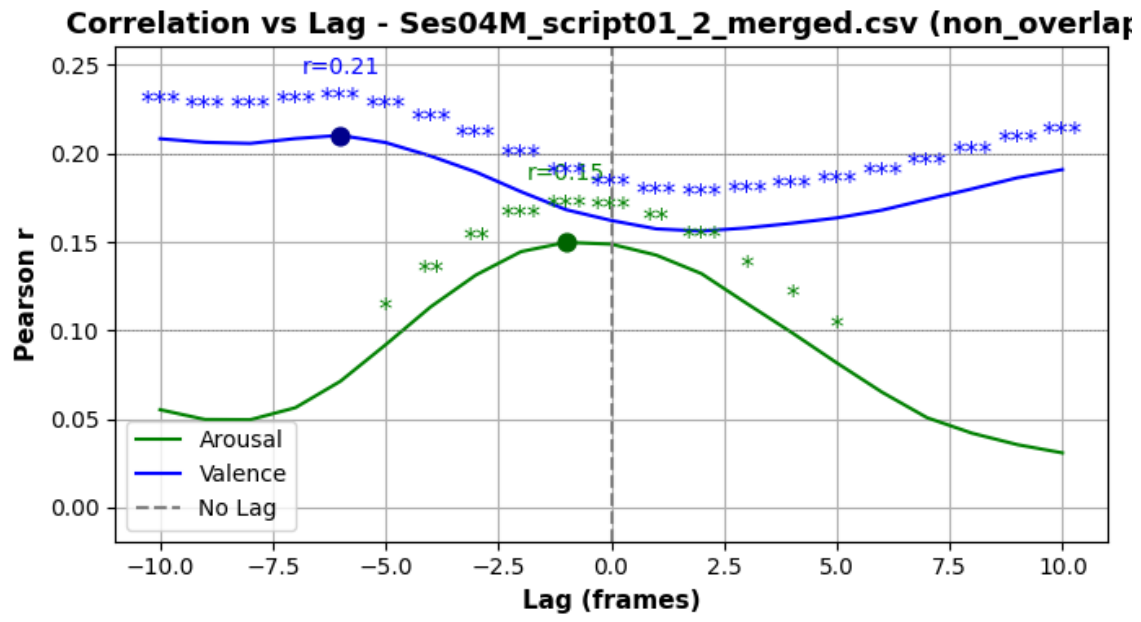

(c)

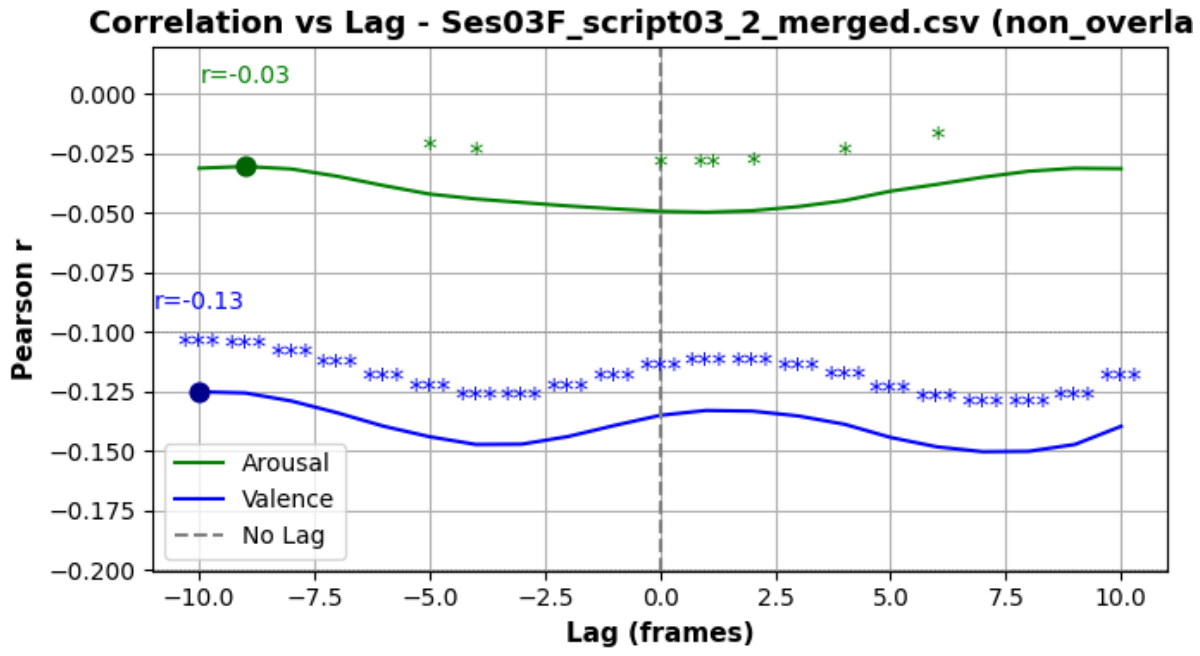

(d)

## Overlap

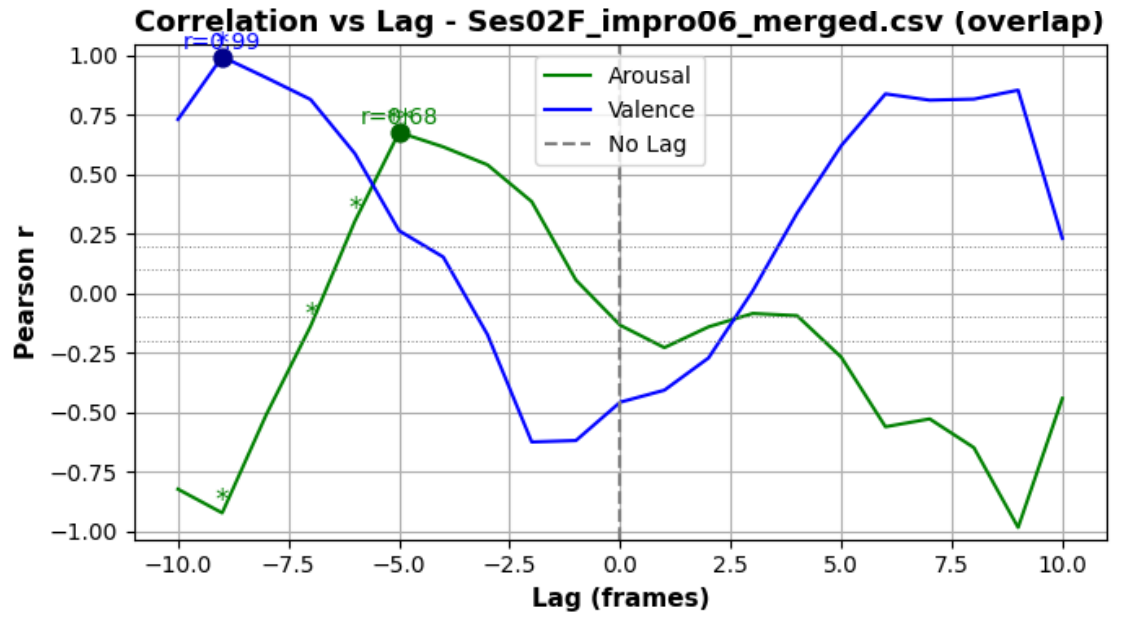

(e)

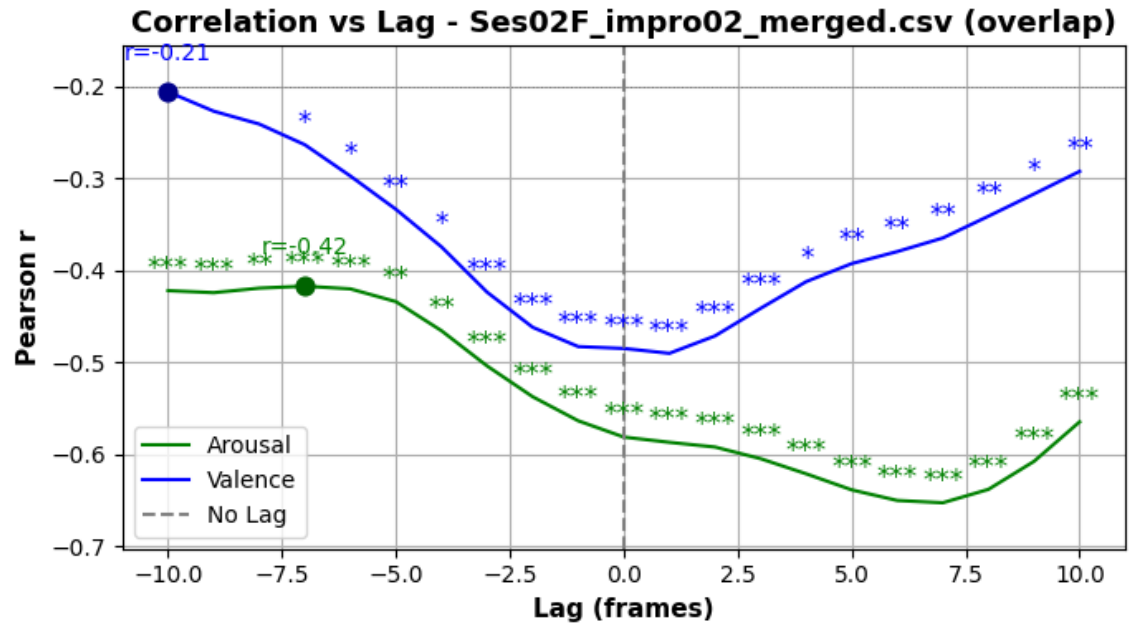

(f)

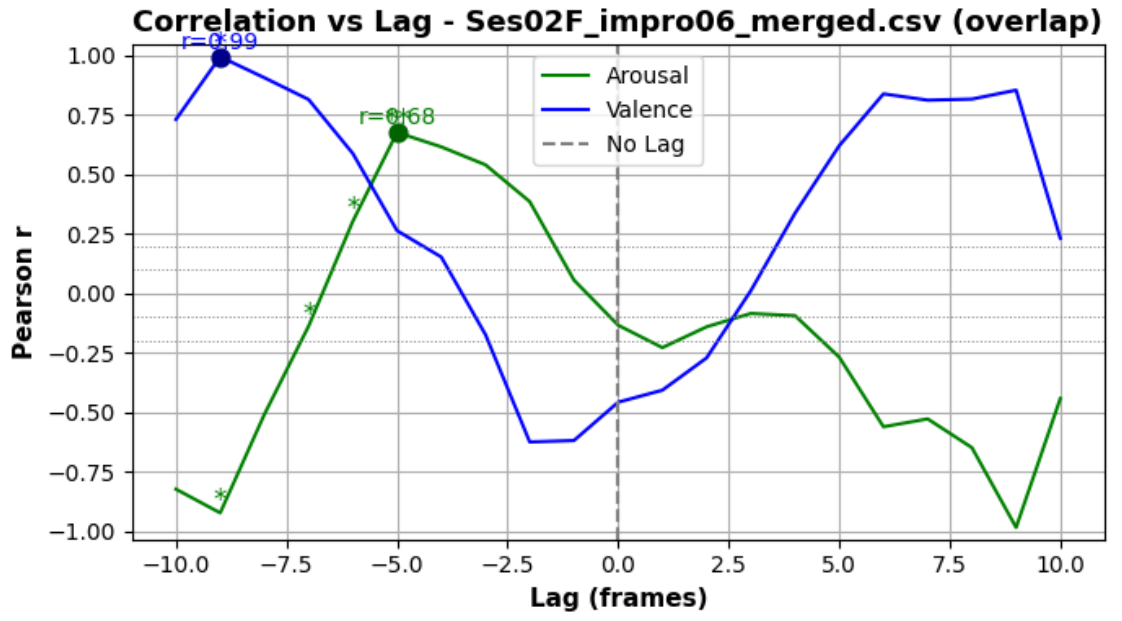

(g)

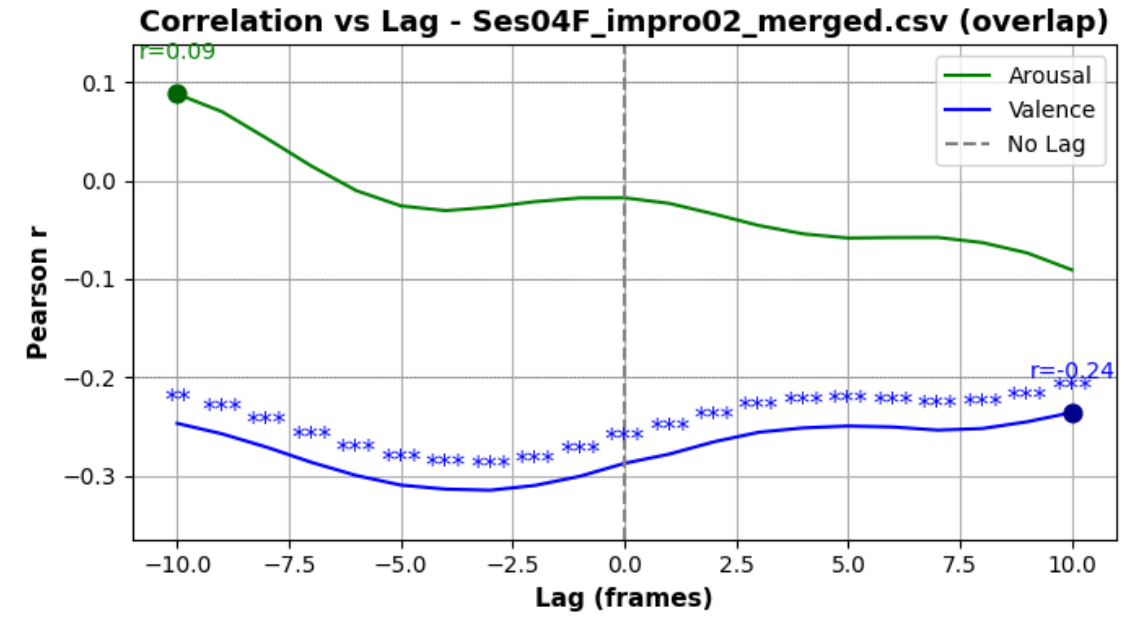

(h)

**Figure 4. Temporal correlation profiles across lag windows by speech condition illustrating how arousal (green) and valence (blue) synchrony between speech and facial expressions varies across lag windows (in frames) for selected dyads under non-overlapping (top) and overlapping (bottom) speech conditions. Solid lines indicate Pearson correlation values across lags, and dashed lines represent mean correlations. Markers denote the lag at which the highest correlation was observed.**

### c. Dynamic Time Warping (DTW) analysis

**Table 4. Dyad-Level Dynamic Time Warping (DTW) Metrics Across Conditions**

| | | | Arousal | | | Valence | | |
|---|---|---|---|---|---|---|---|---|
| Session | Dyads | Condition | Misalignment score | Lag direction | Mean shift | Misalignment score | Lag direction | Mean shift |
| Session1 | Ses01M_impro06 | non_overlap | 447.2 | speech | -457.3 | 467.8 | speech | -457.3 |
| Session1 | Ses01M_impro06 | overlap | 116.3 | speech | -91.8 | 92.2 | speech | -91.8 |
| Session1 | Ses01F_script02_2 | non_overlap | 701.9 | facial | 1014.7 | 1160.6 | facial | 1014.7 |
| Session1 | Ses01F_script02_2 | overlap | 67.3 | speech | -52.3 | 138.7 | speech | -52.3 |
| Session1 | Ses01F_impro01 | non_overlap | 76.2 | speech | -277.6 | 278.4 | speech | -277.6 |
| Session1 | Ses01F_impro01 | overlap | 25.3 | speech | -68.2 | 85.5 | speech | -68.2 |
| Session1 | Ses01M_impro03 | non_overlap | 218.7 | speech | -61.4 | 141.9 | speech | -61.4 |
| Session1 | Ses01M_impro03 | overlap | 114.3 | facial | 43.0 | 62.3 | facial | 43.0 |
| Session1 | Ses01F_impro04 | non_overlap | 146.5 | speech | -678.9 | 679.8 | speech | -678.9 |
| Session1 | Ses01F_impro04 | overlap | 83.9 | facial | 43.3 | 121.8 | facial | 43.3 |
| Session1 | Ses01F_script01_1 | non_overlap | 831.3 | speech | -872.1 | 872.2 | speech | -872.1 |
| Session1 | Ses01F_script01_1 | overlap | 79.0 | speech | -38.8 | 75.0 | speech | -38.8 |
| Session1 | Ses01F_script03_2 | non_overlap | 255.6 | facial | 287.6 | 373.4 | facial | 287.6 |
| Session1 | Ses01F_script03_2 | overlap | 108.1 | speech | -474.8 | 480.5 | speech | -474.8 |
| Session1 | Ses01F_impro03 | non_overlap | 231.1 | facial | 21.1 | 91.2 | facial | 21.1 |
| Session1 | Ses01F_impro03 | overlap | 35.0 | speech | -50.1 | 57.6 | speech | -50.1 |
| Session1 | Ses01M_script02_1 | non_overlap | 433.8 | facial | 321.0 | 463.6 | facial | 321.0 |
| Session1 | Ses01M_script02_1 | overlap | 154.5 | speech | -80.0 | 80.3 | speech | -80.0 |
| Session1 | Ses01M_impro04 | non_overlap | 177.9 | speech | -168.6 | 168.8 | speech | -168.6 |
| Session1 | Ses01M_impro04 | overlap | 95.1 | facial | 142.7 | 151.7 | facial | 142.7 |
| Session1 | Ses01F_impro06 | non_overlap | 179.9 | speech | -164.5 | 312.6 | speech | -164.5 |
| Session1 | Ses01F_impro06 | overlap | 117.9 | speech | -173.9 | 176.3 | speech | -173.9 |
| Session1 | Ses01M_impro01 | non_overlap | 135.9 | facial | 81.0 | 121.5 | facial | 81.0 |
| Session1 | Ses01M_impro01 | overlap | 90.2 | speech | -209.0 | 211.3 | speech | -209.0 |
| Session1 | Ses01M_script01_2 | non_overlap | 294.0 | speech | -272.6 | 273.3 | speech | -272.6 |
| Session1 | Ses01M_script01_2 | overlap | 81.0 | speech | -17.0 | 22.5 | speech | -17.0 |
| Session1 | Ses01M_script03_1 | non_overlap | 366.8 | speech | -79.5 | 260.8 | speech | -79.5 |
| Session1 | Ses01M_script03_1 | overlap | 100.5 | speech | -77.3 | 77.7 | speech | -77.3 |
| Session1 | Ses01F_script01_3 | non_overlap | 722.2 | facial | 204.9 | 411.9 | facial | 204.9 |
| Session1 | Ses01F_script01_3 | overlap | 19.0 | speech | -2.4 | 28.5 | speech | -2.4 |
| Session1 | Ses01M_impro02 | non_overlap | 589.8 | facial | 56.4 | 112.9 | facial | 56.4 |
| Session1 | Ses01M_impro02 | overlap | 74.3 | speech | -99.7 | 100.7 | speech | -99.7 |
| Session1 | Ses01F_impro05 | non_overlap | 216.9 | facial | 262.3 | 274.1 | facial | 262.3 |
| Session1 | Ses01F_impro05 | overlap | 78.7 | speech | -80.6 | 118.2 | speech | -80.6 |
| Session1 | Ses01M_script03_2 | non_overlap | 402.4 | speech | -1159.8 | 1160.0 | speech | -1159.8 |
| Session1 | Ses01M_script03_2 | overlap | 202.4 | speech | -79.3 | 82.6 | speech | -79.3 |
| Session1 | Ses01M_script01_1 | non_overlap | 396.5 | speech | -224.1 | 311.5 | speech | -224.1 |
| Session1 | Ses01M_script01_1 | overlap | 130.8 | speech | -81.8 | 81.8 | speech | -81.8 |
| Session1 | Ses01M_impro07 | non_overlap | 159.6 | facial | 246.8 | 249.7 | facial | 246.8 |
| Session1 | Ses01M_impro07 | overlap | 256.2 | facial | 102.9 | 115.9 | facial | 102.9 |
| Session1 | Ses01M_script02_2 | non_overlap | 860.2 | speech | -63.2 | 325.1 | speech | -63.2 |
| Session1 | Ses01M_script02_2 | overlap | 67.4 | speech | -76.0 | 79.5 | speech | -76.0 |
| Session1 | Ses01F_impro07 | non_overlap | 140.3 | facial | 199.0 | 222.9 | facial | 199.0 |
| Session1 | Ses01F_impro07 | overlap | 73.5 | facial | 152.8 | 157.1 | facial | 152.8 |
| Session1 | Ses01F_script03_1 | non_overlap | 423.3 | facial | 75.8 | 210.0 | facial | 75.8 |
| Session1 | Ses01F_script03_1 | overlap | 80.4 | speech | -132.5 | 134.0 | speech | -132.5 |
| Session1 | Ses01M_script01_3 | non_overlap | 622.3 | speech | -393.8 | 455.9 | speech | -393.8 |
| Session1 | Ses01M_script01_3 | overlap | 48.9 | speech | -28.3 | 33.1 | speech | -28.3 |
| Session1 | Ses01F_script01_2 | non_overlap | 220.8 | speech | -159.7 | 298.4 | speech | -159.7 |
| Session1 | Ses01F_script01_2 | overlap | 56.8 | speech | -61.1 | 166.7 | speech | -61.1 |
| Session1 | Ses01F_impro02 | non_overlap | 218.7 | facial | 52.3 | 255.0 | facial | 52.3 |
| Session1 | Ses01F_impro02 | overlap | 8.3 | speech | -15.4 | 24.0 | speech | -15.4 |
| Session1 | Ses01F_script02_1 | non_overlap | 222.7 | facial | 364.3 | 404.9 | facial | 364.3 |
| Session1 | Ses01F_script02_1 | overlap | 56.8 | facial | 29.9 | 35.5 | facial | 29.9 |
| Session1 | Ses01M_impro05 | non_overlap | 139.3 | speech | -76.5 | 159.7 | speech | -76.5 |
| Session1 | Ses01M_impro05 | overlap | 115.3 | speech | -97.5 | 114.3 | speech | -97.5 |
| Session2 | Ses02F_script01_1 | non_overlap | 211.2 | speech | -969.9 | 974.3 | speech | -969.9 |
| Session2 | Ses02F_script01_1 | overlap | 22.4 | facial | 78.8 | 93.8 | facial | 78.8 |
| Session2 | Ses02F_impro01 | non_overlap | 135.1 | facial | 118.5 | 512.0 | facial | 118.5 |
| Session2 | Ses02F_impro01 | overlap | 86.5 | speech | -68.7 | 83.9 | speech | -68.7 |
| Session2 | Ses02F_script03_2 | non_overlap | 314.0 | speech | -475.1 | 526.1 | speech | -475.1 |
| Session2 | Ses02F_script03_2 | overlap | 74.2 | speech | -92.9 | 99.0 | speech | -92.9 |
| Session2 | Ses02M_impro06 | non_overlap | 163.5 | facial | 251.9 | 284.4 | facial | 251.9 |
| Session2 | Ses02M_impro06 | overlap | 27.1 | speech | -7.0 | 12.4 | speech | -7.0 |
| Session2 | Ses02F_script02_2 | non_overlap | 335.7 | facial | 466.9 | 586.1 | facial | 466.9 |
| Session2 | Ses02F_script02_2 | overlap | 25.4 | speech | -37.7 | 50.0 | speech | -37.7 |
| Session2 | Ses02F_impro04 | non_overlap | 112.4 | speech | -615.3 | 643.9 | speech | -615.3 |
| Session2 | Ses02F_impro04 | overlap | 35.3 | speech | -4.1 | 35.5 | speech | -4.1 |
| Session2 | Ses02M_impro03 | non_overlap | 175.3 | speech | -137.3 | 149.3 | speech | -137.3 |
| Session2 | Ses02M_impro03 | overlap | 71.8 | facial | 23.9 | 80.7 | facial | 23.9 |
| Session2 | Ses02M_impro04 | non_overlap | 570.8 | facial | 362.9 | 386.7 | facial | 362.9 |
| Session2 | Ses02M_impro04 | overlap | 25.7 | speech | -2.3 | 20.6 | speech | -2.3 |
| Session2 | Ses02M_script01_2 | non_overlap | 96.8 | facial | 30.3 | 125.0 | facial | 30.3 |
| Session2 | Ses02M_script01_2 | overlap | 43.7 | facial | 40.9 | 53.8 | facial | 40.9 |
| Session2 | Ses02F_script01_3 | non_overlap | 164.7 | facial | 68.5 | 272.4 | facial | 68.5 |
| Session2 | Ses02F_script01_3 | overlap | 12.7 | facial | 11.1 | 11.4 | facial | 11.1 |
| Session2 | Ses02M_script03_1 | non_overlap | 153.8 | speech | -572.7 | 572.7 | speech | -572.7 |
| Session2 | Ses02M_script03_1 | overlap | 32.8 | facial | 44.7 | 59.4 | facial | 44.7 |
| Session2 | Ses02F_impro03 | non_overlap | 131.9 | facial | 396.8 | 401.3 | facial | 396.8 |
| Session2 | Ses02F_impro03 | overlap | 59.8 | speech | -159.1 | 183.3 | speech | -159.1 |
| Session2 | Ses02M_script02_1 | non_overlap | 411.3 | facial | 220.4 | 308.3 | facial | 220.4 |
| Session2 | Ses02M_script02_1 | overlap | 41.9 | speech | -17.8 | 46.0 | speech | -17.8 |
| Session2 | Ses02M_impro01 | non_overlap | 121.4 | speech | -23.8 | 154.5 | speech | -23.8 |
| Session2 | Ses02M_impro01 | overlap | 55.6 | facial | 40.2 | 52.9 | facial | 40.2 |
| Session2 | Ses02F_impro06 | non_overlap | 381.7 | facial | 317.1 | 336.9 | facial | 317.1 |
| Session2 | Ses02F_impro06 | overlap | 2.6 | facial | 4.7 | 4.7 | facial | 4.7 |
| Session2 | Ses02F_impro08 | non_overlap | 229.0 | speech | -270.1 | 270.9 | speech | -270.1 |
| Session2 | Ses02F_impro08 | overlap | 77.7 | facial | 69.9 | 77.9 | facial | 69.9 |
| Session2 | Ses02M_script02_2 | non_overlap | 181.8 | speech | -342.7 | 349.4 | speech | -342.7 |
| Session2 | Ses02M_script02_2 | overlap | 34.3 | speech | -152.5 | 152.5 | speech | -152.5 |
| Session2 | Ses02F_impro05 | non_overlap | 254.3 | facial | 458.4 | 467.2 | facial | 458.4 |
| Session2 | Ses02F_impro05 | overlap | 58.2 | facial | 54.4 | 58.3 | facial | 54.4 |
| Session2 | Ses02M_impro02 | non_overlap | 198.6 | speech | -297.6 | 297.7 | speech | -297.6 |
| Session2 | Ses02M_impro02 | overlap | 11.9 | facial | 17.5 | 17.5 | facial | 17.5 |
| Session2 | Ses02M_script03_2 | non_overlap | 363.6 | facial | 314.3 | 459.0 | facial | 314.3 |
| Session2 | Ses02M_script03_2 | overlap | 54.4 | speech | -113.1 | 147.9 | speech | -113.1 |
| Session2 | Ses02M_impro07 | non_overlap | 79.6 | facial | 312.3 | 317.6 | facial | 312.3 |
| Session2 | Ses02M_impro07 | overlap | 52.2 | speech | -60.1 | 109.9 | speech | -60.1 |
| Session2 | Ses02M_script01_1 | non_overlap | 696.3 | speech | -204.0 | 514.6 | speech | -204.0 |
| Session2 | Ses02M_script01_1 | overlap | 20.7 | facial | 104.4 | 119.1 | facial | 104.4 |
| Session2 | Ses02F_script02_1 | non_overlap | 172.3 | facial | 244.5 | 272.4 | facial | 244.5 |
| Session2 | Ses02F_script02_1 | overlap | 16.4 | facial | 9.2 | 14.6 | facial | 9.2 |
| Session2 | Ses02F_impro07 | non_overlap | 256.0 | facial | 334.3 | 364.5 | facial | 334.3 |
| Session2 | Ses02F_impro07 | overlap | 83.0 | speech | -161.7 | 201.8 | speech | -161.7 |
| Session2 | Ses02M_script01_3 | non_overlap | 748.0 | facial | 689.3 | 700.4 | facial | 689.3 |
| Session2 | Ses02M_script01_3 | overlap | 220.2 | facial | 201.7 | 203.4 | facial | 201.7 |
| Session2 | Ses02M_impro05 | non_overlap | 538.7 | facial | 339.3 | 341.8 | facial | 339.3 |
| Session2 | Ses02M_impro05 | overlap | 51.1 | speech | -38.1 | 39.6 | speech | -38.1 |
| Session2 | Ses02F_script03_1 | non_overlap | 147.8 | facial | 318.7 | 319.6 | facial | 318.7 |
| Session2 | Ses02F_script03_1 | overlap | 28.3 | facial | 61.7 | 65.3 | facial | 61.7 |
| Session2 | Ses02F_impro02 | non_overlap | 287.8 | speech | -596.0 | 596.4 | speech | -596.0 |
| Session2 | Ses02F_impro02 | overlap | 33.7 | speech | -30.0 | 30.3 | speech | -30.0 |
| Session2 | Ses02F_script01_2 | non_overlap | 86.2 | speech | -35.9 | 179.7 | speech | -35.9 |
| Session2 | Ses02F_script01_2 | overlap | 7.9 | speech | -30.6 | 32.6 | speech | -30.6 |
| Session2 | Ses02M_impro08 | non_overlap | 353.6 | facial | 100.5 | 131.4 | facial | 100.5 |
| Session2 | Ses02M_impro08 | overlap | 84.8 | facial | 9.8 | 93.2 | facial | 9.8 |
| Session3 | Ses03M_impro06 | non_overlap | 686.9 | speech | -291.2 | 293.8 | speech | -291.2 |
| Session3 | Ses03M_impro06 | overlap | 37.3 | speech | -14.2 | 40.8 | speech | -14.2 |
| Session3 | Ses03F_impro01 | non_overlap | 100.9 | speech | -50.4 | 138.6 | speech | -50.4 |
| Session3 | Ses03F_impro01 | overlap | 12.7 | facial | 47.3 | 67.4 | facial | 47.3 |
| Session3 | Ses03M_script02_1 | non_overlap | 222.2 | facial | 1304.2 | 1320.7 | facial | 1304.2 |
| Session3 | Ses03M_script02_1 | overlap | 141.0 | facial | 125.6 | 126.1 | facial | 125.6 |
| Session3 | Ses03M_impro03 | non_overlap | 147.4 | facial | 145.6 | 278.1 | facial | 145.6 |
| Session3 | Ses03M_impro03 | overlap | 78.2 | facial | 162.6 | 181.9 | facial | 162.6 |
| Session3 | Ses03M_script03_1 | non_overlap | 217.5 | speech | -47.7 | 270.6 | speech | -47.7 |
| Session3 | Ses03M_script03_1 | overlap | 26.5 | facial | 17.8 | 55.8 | facial | 17.8 |
| Session3 | Ses03F_script01_3 | non_overlap | 168.0 | facial | 824.9 | 890.7 | facial | 824.9 |
| Session3 | Ses03F_script01_3 | overlap | 22.4 | facial | 115.3 | 130.5 | facial | 115.3 |
| Session3 | Ses03M_script01_2 | non_overlap | 238.6 | speech | -19.3 | 277.1 | speech | -19.3 |
| Session3 | Ses03M_script01_2 | overlap | 55.3 | facial | 0.5 | 33.5 | facial | 0.5 |
| Session3 | Ses03F_impro04 | non_overlap | 210.4 | facial | 548.1 | 556.5 | facial | 548.1 |
| Session3 | Ses03F_impro04 | overlap | 50.6 | facial | 188.0 | 189.7 | facial | 188.0 |
| Session3 | Ses03M_impro05a | non_overlap | 168.1 | speech | -709.1 | 709.2 | speech | -709.1 |
| Session3 | Ses03M_impro05a | overlap | 34.3 | speech | -105.4 | 109.0 | speech | -105.4 |
| Session3 | Ses03F_impro03 | non_overlap | 461.2 | speech | -905.5 | 906.0 | speech | -905.5 |
| Session3 | Ses03F_impro03 | overlap | 102.1 | facial | 71.9 | 194.8 | facial | 71.9 |
| Session3 | Ses03M_impro08b | non_overlap | 446.3 | facial | 531.5 | 537.9 | facial | 531.5 |
| Session3 | Ses03M_impro08b | overlap | 57.6 | speech | -126.0 | 152.5 | speech | -126.0 |
| Session3 | Ses03M_impro04 | non_overlap | 231.3 | facial | 604.5 | 633.1 | facial | 604.5 |
| Session3 | Ses03M_impro04 | overlap | 142.1 | speech | -232.7 | 235.8 | speech | -232.7 |
| Session3 | Ses03F_script02_2 | non_overlap | 349.3 | speech | -1179.0 | 1444.3 | speech | -1179.0 |
| Session3 | Ses03F_script02_2 | overlap | 60.8 | facial | 255.6 | 259.3 | facial | 255.6 |
| Session3 | Ses03F_impro06 | non_overlap | 487.0 | facial | 1014.0 | 1151.9 | facial | 1014.0 |
| Session3 | Ses03F_impro06 | overlap | 58.9 | facial | 64.9 | 70.7 | facial | 64.9 |
| Session3 | Ses03F_script03_2 | non_overlap | 266.9 | speech | -559.0 | 568.1 | speech | -559.0 |
| Session3 | Ses03F_script03_2 | overlap | 55.5 | speech | -142.5 | 164.0 | speech | -142.5 |
| Session3 | Ses03M_impro01 | non_overlap | 189.6 | facial | 98.2 | 286.8 | facial | 98.2 |
| Session3 | Ses03M_impro01 | overlap | 44.1 | facial | 39.9 | 198.1 | facial | 39.9 |
| Session3 | Ses03F_script01_1 | non_overlap | 488.4 | speech | -1224.0 | 1482.6 | speech | -1224.0 |
| Session3 | Ses03F_script01_1 | overlap | 44.8 | speech | -208.0 | 284.2 | speech | -208.0 |
| Session3 | Ses03F_impro08 | non_overlap | 95.7 | facial | 824.4 | 858.2 | facial | 824.4 |
| Session3 | Ses03F_impro08 | overlap | 64.1 | facial | 150.3 | 222.3 | facial | 150.3 |
| Session3 | Ses03F_script01_2 | non_overlap | 205.7 | speech | -750.3 | 762.6 | speech | -750.3 |
| Session3 | Ses03F_script01_2 | overlap | 34.7 | speech | -57.6 | 65.8 | speech | -57.6 |
| Session3 | Ses03M_impro02 | non_overlap | 166.4 | speech | -37.1 | 482.8 | speech | -37.1 |
| Session3 | Ses03M_impro02 | overlap | 90.0 | facial | 113.2 | 135.2 | facial | 113.2 |
| Session3 | Ses03F_script03_1 | non_overlap | 217.5 | facial | 578.8 | 640.7 | facial | 578.8 |
| Session3 | Ses03F_script03_1 | overlap | 48.7 | speech | -192.5 | 282.3 | speech | -192.5 |
| Session3 | Ses03M_script01_3 | non_overlap | 970.3 | facial | 78.4 | 302.9 | facial | 78.4 |
| Session3 | Ses03M_script01_3 | overlap | 49.3 | speech | -115.4 | 122.2 | speech | -115.4 |
| Session3 | Ses03F_impro05 | non_overlap | 233.0 | speech | -234.7 | 296.6 | speech | -234.7 |
| Session3 | Ses03F_impro05 | overlap | 122.9 | facial | 192.3 | 198.9 | facial | 192.3 |
| Session3 | Ses03M_impro07 | non_overlap | 245.8 | facial | 156.5 | 176.9 | facial | 156.5 |
| Session3 | Ses03M_impro07 | overlap | 100.2 | facial | 62.0 | 67.5 | facial | 62.0 |
| Session3 | Ses03M_impro08a | non_overlap | 252.8 | speech | -21.2 | 200.1 | speech | -21.2 |
| Session3 | Ses03M_impro08a | overlap | 41.9 | speech | -489.5 | 493.4 | speech | -489.5 |
| Session3 | Ses03F_script02_1 | non_overlap | 42.8 | facial | 720.6 | 722.6 | facial | 720.6 |
| Session3 | Ses03F_script02_1 | overlap | 54.3 | speech | -96.0 | 111.4 | speech | -96.0 |
| Session3 | Ses03M_script01_1 | non_overlap | 400.3 | facial | 627.0 | 855.4 | facial | 627.0 |
| Session3 | Ses03M_script01_1 | overlap | 171.1 | facial | 315.2 | 341.2 | facial | 315.2 |
| Session3 | Ses03F_impro07 | non_overlap | 131.3 | speech | -167.9 | 286.7 | speech | -167.9 |
| Session3 | Ses03F_impro07 | overlap | 105.3 | facial | 229.3 | 232.9 | facial | 229.3 |
| Session3 | Ses03M_impro05b | non_overlap | 378.3 | speech | -196.1 | 253.0 | speech | -196.1 |
| Session3 | Ses03M_impro05b | overlap | 60.5 | speech | -85.8 | 170.7 | speech | -85.8 |
| Session3 | Ses03M_script03_2 | non_overlap | 341.4 | speech | -1131.8 | 1132.0 | speech | -1131.8 |
| Session3 | Ses03M_script03_2 | overlap | 111.7 | speech | -158.3 | 209.1 | speech | -158.3 |
| Session3 | Ses03F_impro02 | non_overlap | 183.2 | facial | 249.9 | 556.8 | facial | 249.9 |
| Session3 | Ses03F_impro02 | overlap | 21.2 | facial | 129.6 | 129.6 | facial | 129.6 |
| Session3 | Ses03M_script02_2 | non_overlap | 665.6 | facial | 779.8 | 781.0 | facial | 779.8 |
| Session3 | Ses03M_script02_2 | overlap | 48.9 | speech | -253.3 | 256.8 | speech | -253.3 |
| Session4 | Ses04M_impro03 | non_overlap | 550.5 | speech | -641.6 | 646.7 | speech | -641.6 |
| Session4 | Ses04M_impro03 | overlap | 37.5 | speech | -125.2 | 125.3 | speech | -125.2 |
| Session4 | Ses04M_script02_1 | non_overlap | 362.2 | facial | 80.4 | 280.1 | facial | 80.4 |
| Session4 | Ses04M_script02_1 | overlap | 26.4 | facial | 4.8 | 34.8 | facial | 4.8 |
| Session4 | Ses04F_impro04 | non_overlap | 144.5 | facial | 215.1 | 279.3 | facial | 215.1 |
| Session4 | Ses04F_impro04 | overlap | 85.5 | speech | -15.3 | 76.2 | speech | -15.3 |
| Session4 | Ses04M_script03_1 | non_overlap | 219.1 | speech | -116.1 | 177.6 | speech | -116.1 |
| Session4 | Ses04M_script03_1 | overlap | 22.7 | speech | -116.5 | 117.9 | speech | -116.5 |
| Session4 | Ses04F_script01_3 | non_overlap | 175.1 | speech | -108.3 | 163.0 | speech | -108.3 |
| Session4 | Ses04F_script01_3 | overlap | 34.2 | facial | 50.4 | 50.8 | facial | 50.4 |
| Session4 | Ses04M_script01_2 | non_overlap | 90.0 | speech | -167.6 | 323.7 | speech | -167.6 |
| Session4 | Ses04M_script01_2 | overlap | 27.7 | speech | -83.1 | 94.1 | speech | -83.1 |
| Session4 | Ses04M_impro06 | non_overlap | 551.6 | facial | 40.6 | 146.9 | facial | 40.6 |
| Session4 | Ses04M_impro06 | overlap | 26.3 | speech | -45.2 | 45.9 | speech | -45.2 |
| Session4 | Ses04F_impro01 | non_overlap | 116.4 | speech | -261.4 | 267.0 | speech | -261.4 |
| Session4 | Ses04F_impro01 | overlap | 60.5 | speech | -146.6 | 148.4 | speech | -146.6 |
| Session4 | Ses04F_impro06 | non_overlap | 79.1 | facial | 135.5 | 143.0 | facial | 135.5 |
| Session4 | Ses04F_impro06 | overlap | 43.4 | facial | 18.3 | 20.0 | facial | 18.3 |
| Session4 | Ses04M_impro01 | non_overlap | 68.3 | speech | -295.1 | 357.8 | speech | -295.1 |
| Session4 | Ses04M_impro01 | overlap | 173.1 | speech | -97.7 | 109.3 | speech | -97.7 |
| Session4 | Ses04F_script02_2 | non_overlap | 332.1 | facial | 1100.1 | 1101.7 | facial | 1100.1 |
| Session4 | Ses04F_script02_2 | overlap | 31.3 | facial | 35.9 | 42.7 | facial | 35.9 |
| Session4 | Ses04F_script03_2 | non_overlap | 456.9 | speech | -204.2 | 214.5 | speech | -204.2 |
| Session4 | Ses04F_script03_2 | overlap | 60.0 | speech | -75.1 | 84.1 | speech | -75.1 |
| Session4 | Ses04F_script01_1 | non_overlap | 300.3 | speech | -263.7 | 272.9 | speech | -263.7 |
| Session4 | Ses04F_script01_1 | overlap | 74.6 | speech | -78.4 | 89.9 | speech | -78.4 |
| Session4 | Ses04F_impro03 | non_overlap | 127.5 | speech | -349.6 | 350.0 | speech | -349.6 |
| Session4 | Ses04F_impro03 | overlap | 123.6 | speech | -106.5 | 136.0 | speech | -106.5 |
| Session4 | Ses04M_impro04 | non_overlap | 605.3 | facial | 7.7 | 187.4 | facial | 7.7 |
| Session4 | Ses04M_impro04 | overlap | 18.0 | facial | 27.4 | 50.1 | facial | 27.4 |
| Session4 | Ses04F_script01_2 | non_overlap | 143.5 | speech | -459.4 | 461.3 | speech | -459.4 |
| Session4 | Ses04F_script01_2 | overlap | 14.1 | speech | -55.7 | 62.5 | speech | -55.7 |
| Session4 | Ses04F_script03_1 | non_overlap | 97.0 | speech | -6.9 | 57.1 | speech | -6.9 |
| Session4 | Ses04F_script03_1 | overlap | 63.8 | speech | -82.2 | 84.2 | speech | -82.2 |
| Session4 | Ses04M_script01_3 | non_overlap | 681.3 | facial | 126.8 | 170.3 | facial | 126.8 |
| Session4 | Ses04M_script01_3 | overlap | 50.2 | speech | -6.3 | 21.4 | speech | -6.3 |
| Session4 | Ses04M_impro07 | non_overlap | 234.0 | facial | 83.5 | 219.2 | facial | 83.5 |
| Session4 | Ses04M_impro07 | overlap | 161.1 | facial | 38.5 | 133.5 | facial | 38.5 |
| Session4 | Ses04F_script02_1 | non_overlap | 61.1 | speech | -99.7 | 105.9 | speech | -99.7 |
| Session4 | Ses04F_script02_1 | overlap | 12.9 | facial | 29.2 | 36.7 | facial | 29.2 |
| Session4 | Ses04M_impro02 | non_overlap | 541.5 | facial | 368.7 | 426.1 | facial | 368.7 |
| Session4 | Ses04M_impro02 | overlap | 29.8 | speech | -25.9 | 38.0 | speech | -25.9 |
| Session4 | Ses04F_impro05 | non_overlap | 169.3 | speech | -183.4 | 183.7 | speech | -183.4 |
| Session4 | Ses04F_impro05 | overlap | 59.3 | facial | 123.5 | 125.6 | facial | 123.5 |
| Session4 | Ses04F_impro08 | non_overlap | 87.4 | facial | 232.0 | 268.6 | facial | 232.0 |
| Session4 | Ses04F_impro08 | overlap | 51.6 | speech | -248.6 | 248.6 | speech | -248.6 |
| Session4 | Ses04M_script01_1 | non_overlap | 545.6 | speech | -596.9 | 598.5 | speech | -596.9 |
| Session4 | Ses04M_script01_1 | overlap | 42.2 | speech | -172.6 | 172.6 | speech | -172.6 |
| Session4 | Ses04M_impro08 | non_overlap | 318.1 | speech | -289.0 | 289.1 | speech | -289.0 |
| Session4 | Ses04M_impro08 | overlap | 68.1 | facial | 11.7 | 70.7 | facial | 11.7 |
| Session4 | Ses04M_script03_2 | non_overlap | 517.4 | speech | -90.9 | 136.7 | speech | -90.9 |
| Session4 | Ses04M_script03_2 | overlap | 62.2 | facial | 53.7 | 121.3 | facial | 53.7 |
| Session4 | Ses04F_impro02 | non_overlap | 493.0 | facial | 291.7 | 292.3 | facial | 291.7 |
| Session4 | Ses04F_impro02 | overlap | 43.0 | facial | 168.5 | 168.7 | facial | 168.5 |
| Session4 | Ses04M_impro05 | non_overlap | 257.9 | speech | -594.2 | 596.0 | speech | -594.2 |
| Session4 | Ses04M_impro05 | overlap | 130.0 | speech | -108.8 | 115.7 | speech | -108.8 |
| Session4 | Ses04M_script02_2 | non_overlap | 271.6 | facial | 87.1 | 190.5 | facial | 87.1 |
| Session4 | Ses04M_script02_2 | overlap | 57.9 | speech | -61.7 | 70.2 | speech | -61.7 |
| Session4 | Ses04F_impro07 | non_overlap | 165.2 | facial | 213.4 | 244.9 | facial | 213.4 |
| Session4 | Ses04F_impro07 | overlap | 195.0 | speech | -240.2 | 240.8 | speech | -240.2 |
| Session5 | Ses05F_impro04 | non_overlap | 515.1 | facial | 639.6 | 807.4 | facial | 639.6 |
| Session5 | Ses05F_impro04 | overlap | 216.1 | speech | -25.9 | 47.0 | speech | -25.9 |
| Session5 | Ses05M_impro03 | non_overlap | 1086.0 | speech | -376.2 | 376.2 | speech | -376.2 |
| Session5 | Ses05M_impro03 | overlap | 439.7 | speech | -35.5 | 44.8 | speech | -35.5 |
| Session5 | Ses05F_script01_1 | non_overlap | 525.6 | speech | -259.1 | 741.8 | speech | -259.1 |
| Session5 | Ses05F_script01_1 | overlap | 140.6 | speech | -124.0 | 124.5 | speech | -124.0 |
| Session5 | Ses05F_script03_2 | non_overlap | 399.8 | speech | -800.9 | 806.2 | speech | -800.9 |
| Session5 | Ses05F_script03_2 | overlap | 38.1 | speech | -90.5 | 105.7 | speech | -90.5 |
| Session5 | Ses05F_script02_2 | non_overlap | 967.9 | facial | 1017.7 | 1156.8 | facial | 1017.7 |
| Session5 | Ses05F_script02_2 | overlap | 38.9 | speech | -44.4 | 72.1 | speech | -44.4 |
| Session5 | Ses05F_impro01 | non_overlap | 187.6 | speech | -308.3 | 312.7 | speech | -308.3 |
| Session5 | Ses05F_impro01 | overlap | 103.3 | speech | -112.3 | 118.7 | speech | -112.3 |
| Session5 | Ses05M_impro06 | non_overlap | 613.2 | speech | -162.4 | 292.2 | speech | -162.4 |
| Session5 | Ses05M_impro06 | overlap | 88.4 | speech | -26.4 | 51.5 | speech | -26.4 |
| Session5 | Ses05M_impro01 | non_overlap | 170.5 | speech | -125.0 | 130.5 | speech | -125.0 |
| Session5 | Ses05M_impro01 | overlap | 385.5 | speech | -145.8 | 147.5 | speech | -145.8 |
| Session5 | Ses05F_impro06 | non_overlap | 202.8 | speech | -931.7 | 966.0 | speech | -931.7 |
| Session5 | Ses05F_impro06 | overlap | 16.0 | speech | -10.6 | 32.9 | speech | -10.6 |
| Session5 | Ses05M_script01_2 | non_overlap | 327.5 | speech | -377.4 | 377.4 | speech | -377.4 |
| Session5 | Ses05M_script01_2 | overlap | 63.1 | facial | 10.6 | 27.9 | facial | 10.6 |
| Session5 | Ses05F_script01_3 | non_overlap | 277.2 | speech | -258.2 | 408.3 | speech | -258.2 |
| Session5 | Ses05F_script01_3 | overlap | 19.8 | speech | -32.8 | 33.6 | speech | -32.8 |
| Session5 | Ses05M_script03_1 | non_overlap | 263.5 | speech | -530.4 | 544.6 | speech | -530.4 |
| Session5 | Ses05M_script03_1 | overlap | 70.9 | facial | 17.5 | 44.1 | facial | 17.5 |
| Session5 | Ses05M_script01_1b | non_overlap | 615.6 | speech | -486.6 | 499.7 | speech | -486.6 |
| Session5 | Ses05M_script01_1b | overlap | 114.9 | speech | -56.3 | 56.3 | speech | -56.3 |
| Session5 | Ses05M_impro04 | non_overlap | 286.2 | speech | -235.7 | 297.6 | speech | -235.7 |
| Session5 | Ses05M_impro04 | overlap | 246.8 | speech | -173.8 | 174.2 | speech | -173.8 |
| Session5 | Ses05M_script02_1 | non_overlap | 593.7 | facial | 60.8 | 149.3 | facial | 60.8 |
| Session5 | Ses05M_script02_1 | overlap | 28.1 | speech | -14.6 | 22.5 | speech | -14.6 |
| Session5 | Ses05F_impro03 | non_overlap | 229.9 | facial | 29.3 | 118.3 | facial | 29.3 |
| Session5 | Ses05F_impro03 | overlap | 207.4 | facial | 32.1 | 77.0 | facial | 32.1 |
| Session5 | Ses05M_script02_2 | non_overlap | 315.6 | facial | 269.6 | 319.4 | facial | 269.6 |
| Session5 | Ses05M_script02_2 | overlap | 57.0 | facial | 24.0 | 53.1 | facial | 24.0 |
| Session5 | Ses05M_impro07 | non_overlap | 223.3 | speech | -2.4 | 87.2 | speech | -2.4 |
| Session5 | Ses05M_impro07 | overlap | 392.2 | speech | -214.4 | 216.8 | speech | -214.4 |
| Session5 | Ses05F_impro05 | non_overlap | 705.0 | speech | -735.1 | 735.9 | speech | -735.1 |
| Session5 | Ses05F_impro05 | overlap | 262.3 | speech | -13.4 | 149.5 | speech | -13.4 |
| Session5 | Ses05M_impro02 | non_overlap | 624.2 | speech | -164.0 | 208.0 | speech | -164.0 |
| Session5 | Ses05M_impro02 | overlap | 72.3 | speech | -80.1 | 80.1 | speech | -80.1 |
| Session5 | Ses05F_impro08 | non_overlap | 245.2 | speech | -65.2 | 216.9 | speech | -65.2 |
| Session5 | Ses05F_impro08 | overlap | 53.5 | speech | -24.2 | 55.9 | speech | -24.2 |
| Session5 | Ses05M_script03_2 | non_overlap | 142.7 | speech | -375.1 | 375.1 | speech | -375.1 |
| Session5 | Ses05M_script03_2 | overlap | 114.6 | speech | -158.6 | 158.9 | speech | -158.6 |
| Session5 | Ses05M_script01_1 | non_overlap | 313.1 | speech | -63.0 | 295.8 | speech | -63.0 |
| Session5 | Ses05M_script01_1 | overlap | 46.9 | speech | -99.7 | 99.7 | speech | -99.7 |
| Session5 | Ses05M_impro08 | non_overlap | 395.9 | speech | -63.5 | 139.0 | speech | -63.5 |
| Session5 | Ses05M_impro08 | overlap | 127.6 | speech | -26.6 | 32.8 | speech | -26.6 |
| Session5 | Ses05M_impro05 | non_overlap | 265.0 | speech | -181.0 | 181.2 | speech | -181.0 |
| Session5 | Ses05M_impro05 | overlap | 42.0 | facial | 54.5 | 54.6 | facial | 54.5 |
| Session5 | Ses05F_impro02 | non_overlap | 228.2 | speech | -645.2 | 771.6 | speech | -645.2 |
| Session5 | Ses05F_impro02 | overlap | 47.0 | speech | -89.9 | 96.1 | speech | -89.9 |
| Session5 | Ses05F_script02_1 | non_overlap | 210.5 | speech | -45.6 | 202.5 | speech | -45.6 |
| Session5 | Ses05F_script02_1 | overlap | 48.5 | speech | -70.3 | 71.9 | speech | -70.3 |
| Session5 | Ses05F_impro07 | non_overlap | 91.9 | speech | -8.1 | 54.9 | speech | -8.1 |
| Session5 | Ses05F_impro07 | overlap | 69.0 | speech | -102.8 | 118.9 | speech | -102.8 |
| Session5 | Ses05M_script01_3 | non_overlap | 854.7 | facial | 797.8 | 872.9 | facial | 797.8 |
| Session5 | Ses05M_script01_3 | overlap | 47.6 | speech | -57.8 | 62.8 | speech | -57.8 |
| Session5 | Ses05F_script03_1 | non_overlap | 280.3 | speech | -196.2 | 232.9 | speech | -196.2 |
| Session5 | Ses05F_script03_1 | overlap | 51.7 | speech | -36.4 | 54.0 | speech | -36.4 |
| Session5 | Ses05F_script01_2 | non_overlap | 207.5 | facial | 263.8 | 379.8 | facial | 263.8 |
| Session5 | Ses05F_script01_2 | overlap | 48.7 | speech | -1.2 | 32.9 | speech | -1.2 |

**This table presents dyad-level DTW misalignment scores for arousal and valence signals, comparing non-overlapping and overlapping speech conditions across sessions. Each entry includes the DTW misalignment score (lower values indicate stronger alignment), the direction of lag (which modality leads), and the mean shift between the time series.**

To complement the correlation-based analyses, we applied Dynamic Time Warping (DTW) to quantify the temporal alignment between speech- and facial-derived emotional signals. DTW computes the optimal alignment path between two time series by allowing nonlinear warping along the time axis, revealing how much one modality must stretch or compress to align with the other. Dyad-level DTW results provide further insight into variability across dyads, revealing not just the degree of synchrony but also the dynamic directionality of cross-modal emotional expression. To illustrate this, we identified sample sessions with the highest and lowest DTW misalignment scores for both arousal and valence, across non-overlapping and overlapping conditions (Figures 5a–5h). In the non-overlapping condition, Ses05M_impro03 had the highest arousal misalignment (lag score = 1085.99; Figure 5a), indicating strong desynchrony and a substantial temporal shift between modalities. Conversely, Ses03F_script02_1 showed minimal arousal misalignment (lag score = 42.79; Figure 5b), with an alignment path closely hugging the diagonal—suggesting near-synchronous expression. For valence, Ses03F_script01_1 showed the highest misalignment (lag score = 1482.60; Figure 5c), while Ses05F_impro07 demonstrated tight alignment (lag score = 54.93; Figure 5d), again reflecting strong multimodal coherence during structured turn-taking. In the overlapping condition, where simultaneous speech introduces higher coordination demands, we observed similar extremes. Ses05M_impro03 showed the highest arousal misalignment (lag score = 439.65; Figure 5e), whereas Ses02F_impro06 exhibited the lowest arousal misalignment (lag score = 2.61; Figure 5f), indicating almost perfect alignment even amid concurrent speech. For valence, Ses03M_impro08a had the highest misalignment (lag score = 493.40; Figure 5g), contrasting sharply with Ses02F_impro06, which again had the lowest valence lag score (4.70; Figure 5h), making it a rare case of dual low-lag synchrony across affective dimensions during overlap. These examples underscore the nuanced interplay between conversational structure and emotional coordination. While turn-taking generally supports stronger cross-modal synchrony, some dyads maintain alignment even during overlap. DTW thus not only captures the magnitude of alignment but also highlights interactional conditions under which synchrony may break down or flourish.

## Non-overlap

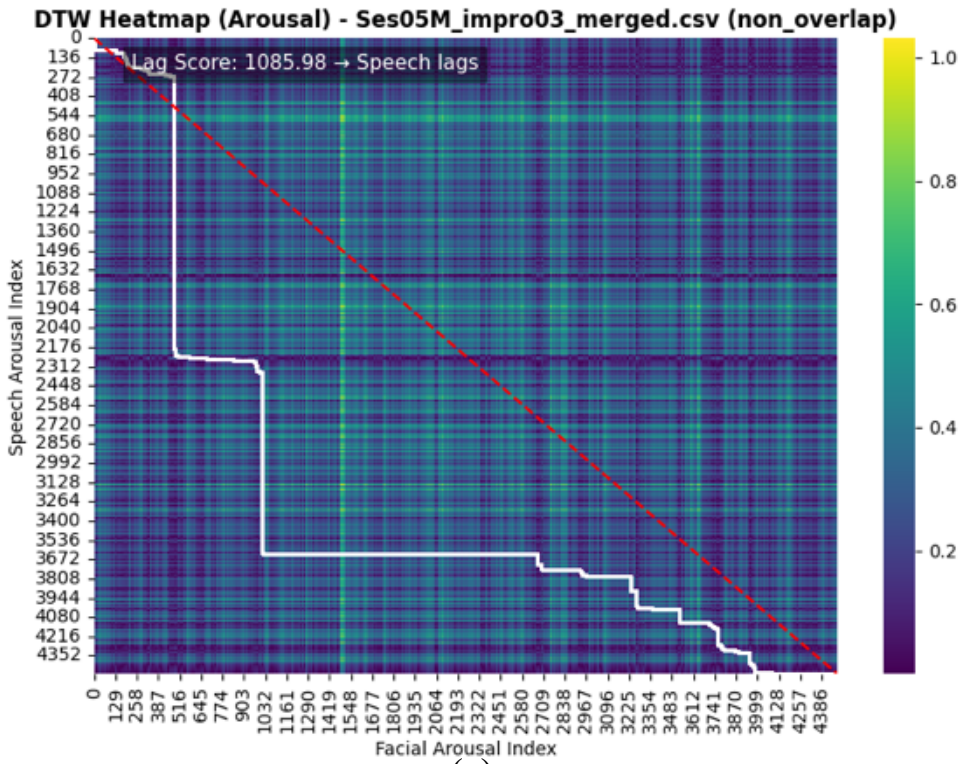

(a)

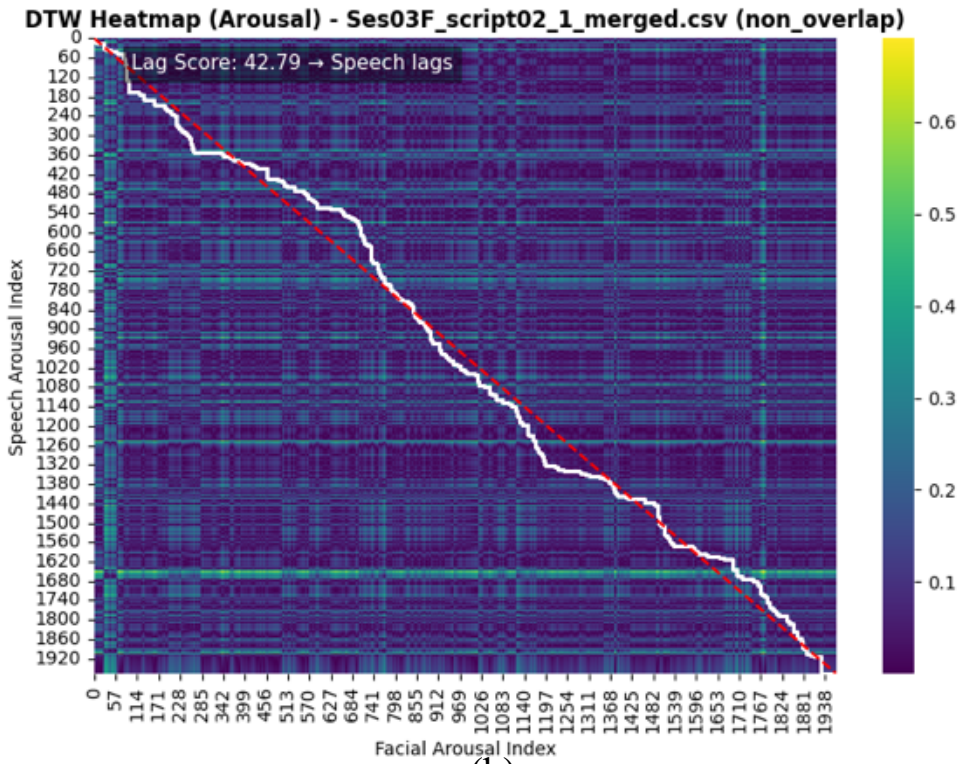

(b)

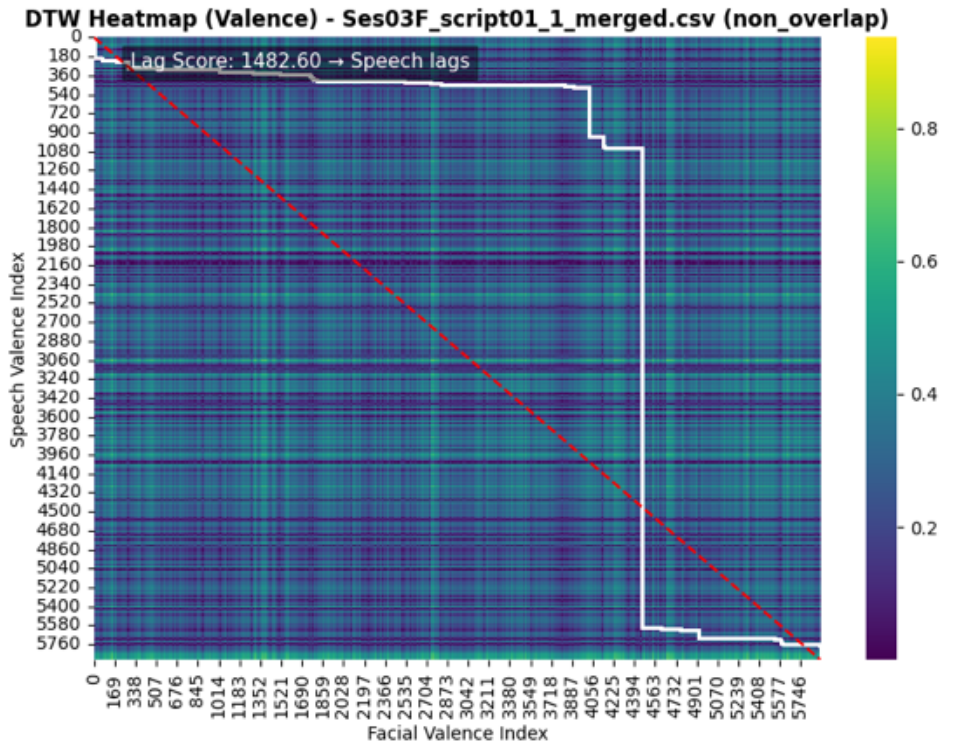

(c)

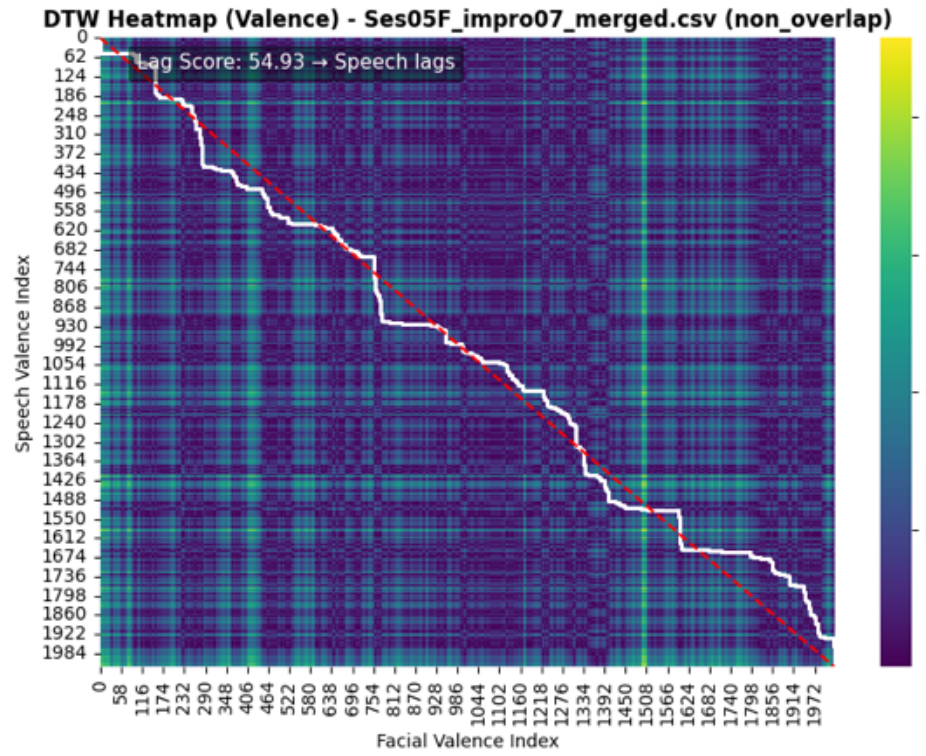

(d)

## Overlap

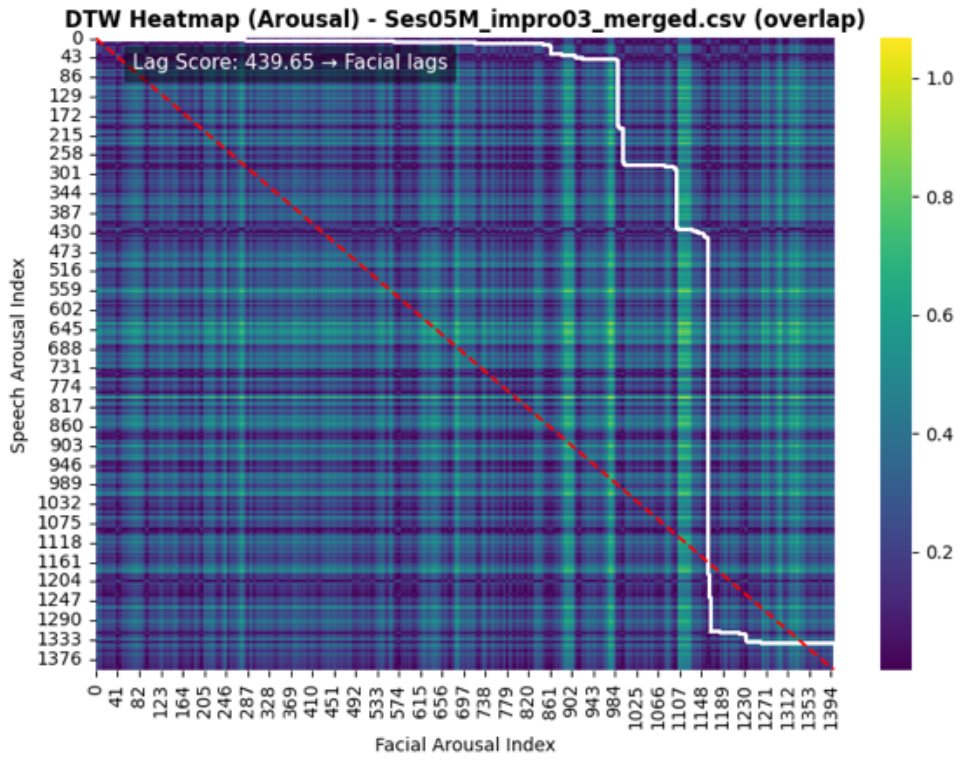

(e)

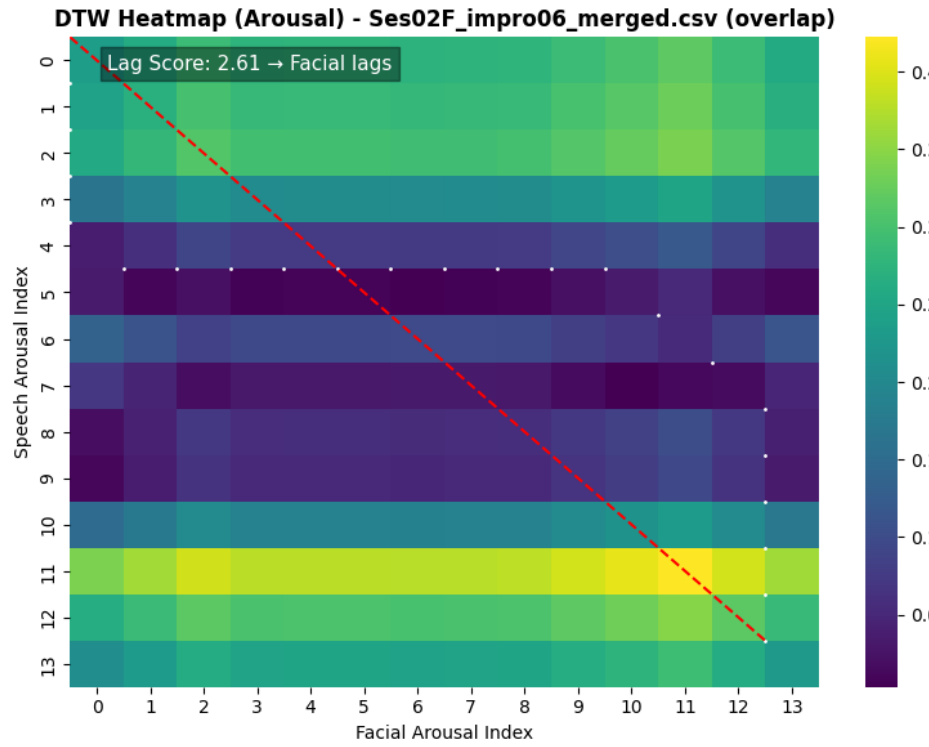

(f)

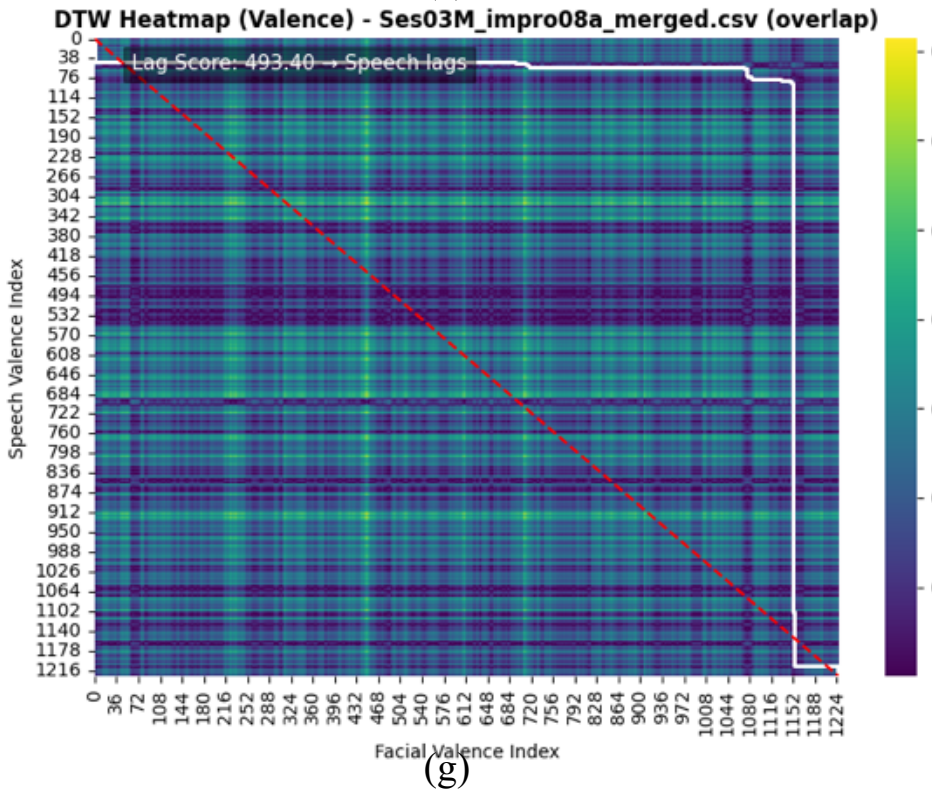

(g)

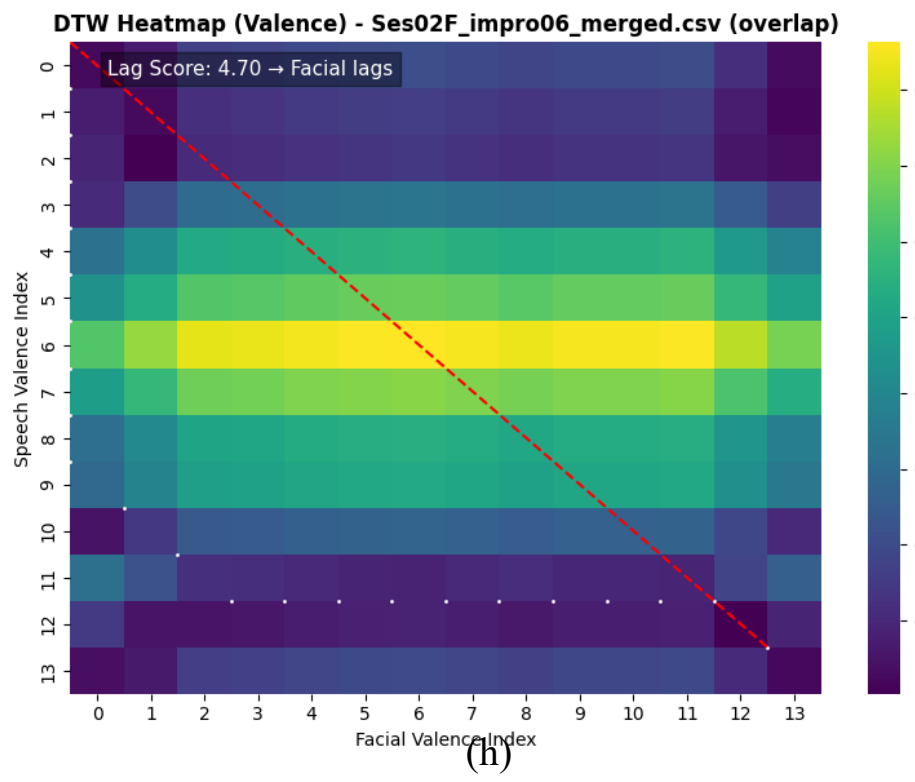

(h)

**Figure 5.** Dynamic Time Warping (DTW) misalignment heatmaps illustrating temporal alignment between speech- and facial-derived emotional signals (arousal and valence) across non-overlapping (top) and overlapping (bottom) speech conditions. Each panel shows the DTW cost matrix (color gradient), the optimal warping path (white line), and the ideal diagonal alignment (red dashed line).